\documentclass[apj,twocolumn, twocolappendix]{openjournal}
\usepackage{amsmath}
\usepackage{booktabs}
\usepackage{multirow}
\usepackage{color}
\usepackage{soul}
\usepackage{booktabs} 

\usepackage[dvipsnames]{xcolor} 

\usepackage[breaklinks,colorlinks,citecolor=blue,urlcolor=blue]{hyperref}

\newcommand{\HeI}{He~{\sc i}}
\newcommand{\OI}{O~{\sc i}}

\newcommand{\CII}{C~{\sc ii}}
\newcommand{\CI}{C~{\sc i}}

\newcommand{\MgII}{Mg~{\sc ii}}

\newcommand{\SiII}{Si~{\sc ii}}
\newcommand{\SiIII}{Si~{\sc iii}}

\newcommand{\SII}{S~{\sc ii}}
\newcommand{\CaII}{Ca~{\sc ii}}

\newcommand{\FeII}{Fe~{\sc ii}}
\newcommand{\FeIII}{Fe~{\sc iii}}
\newcommand{\CoII}{Co~{\sc ii}}

\newcommand{\Nifs}{$^{56}$Ni}

\newcommand{\Mch}{$M_{\rm ch}$}

\newcommand{\peakMJD}[0]{60406.6}

\usepackage{newunicodechar}
\DeclareRobustCommand{\okina}{%
  \raisebox{\dimexpr\fontcharht\font`A-\height}{%
    \scalebox{0.8}{`}%
  }%
}
\newunicodechar{ʻ}{\okina}

\renewcommand{\arraystretch}{1.2}  
\usepackage{listings}
\usepackage{color}
\definecolor{dkgreen}{rgb}{0,0.6,0}
\definecolor{gray}{rgb}{0.5,0.5,0.5}
\definecolor{mauve}{rgb}{0.58,0,0.82}
\definecolor{golden}{rgb}{0.86,0.65,0.01}
\usepackage{orcidlink}
\usepackage{hyperref}
\usepackage{amsmath}
\usepackage{amsthm}
\usepackage{amsfonts}
\usepackage{amssymb}

\begin{document}


\title{S\lowercase{eeing} \lowercase{the} O\lowercase{uter} E\lowercase{dge} \lowercase{of} \lowercase{the} I\lowercase{nfant} T\lowercase{ype} I\lowercase{a} S\lowercase{upernova} \lowercase{2024epr} \lowercase{in} \lowercase{the} O\lowercase{ptical} \lowercase{and} N\lowercase{ear} I\lowercase{nfrared}}

\author{\vspace{-1.0cm}
        W.~B.~Hoogendam$^{1*}$\orcidlink{0000-0003-3953-9532}}
\author{D.~O.~Jones$^{2}$\orcidlink{0000-0002-6230-0151}}
\author{C.~Ashall$^{1}$\orcidlink{0000-0002-5221-7557}}
\author{B.~J.~Shappee$^{1}$\orcidlink{0000-0003-4631-1149}}
\author{R.~J.~Foley$^{3}$\orcidlink{0000-0002-2445-5275}}

\author{M.~A.~Tucker$^{4,5,\dag}$\orcidlink{0000-0002-2471-8442}}
\author{M.~E.~Huber$^{1}$\orcidlink{0000-0003-1059-9603}}
\author{K.~Auchettl$^{6,3}$\orcidlink{0000-0002-4449-9152}}
\author{D.~D.~Desai$^{1}$\orcidlink{0000-0002-2164-859X}}
\author{A.~Do$^{7,8}$\orcidlink{0000-0003-3429-7845}}
\author{J.~T.~Hinkle$^{1\ddag}$\orcidlink{0000-0001-9668-2920}}
\author{S.~Romagnoli$^{6}$\orcidlink{0009-0003-8153-9576}}
\author{J.~Shi$^{6}$\orcidlink{0009-0008-3724-1824}}
\author{A.~Syncatto$^{1,9}$\orcidlink{0009-0000-6821-9285}}

\author{C.~R.~Angus$^{10,11}$\orcidlink{0000-0002-4269-7999}}
\author{K.~C.~Chambers$^{1}$\orcidlink{0000-0001-6965-7789}}
\author{D.~A.~Coulter$^{12}$\orcidlink{0000-0003-4263-2228}}
\author{K.~W.~Davis$^{3}$\orcidlink{0000-0002-5680-4660}}
\author{T.~de~Boer$^{1}$\orcidlink{0000-0001-5486-2747}}
\author{A.~Gagliano$^{13,14}$\orcidlink{0000-0003-4906-8447}}
\author{M.~Y.~Kong$^{1}$\orcidlink{0009-0005-5121-2884}}
\author{C.-C.~Lin$^{1}$\orcidlink{0000-0002-7272-5129}}
\author{T.~B.~Lowe$^{1}$}
\author{E.~A.~Magnier$^{1}$\orcidlink{0000-0002-7965-2815}}
\author{P. Mínguez$^{1}$\orcidlink{0009-0003-8803-8643}}
\author{Y.-C.~Pan$^{15}$\orcidlink{0000-0001-8415-6720}}
\author{K.C.~Patra$^{3}$\orcidlink{0000-0002-1092-6806}}
\author{S.~A.~Severson$^{16}$\orcidlink{0009-0001-0130-9734}} 
\author{K.~Taggart$^{3}$\orcidlink{0000-0002-5748-4558}}
\author{A.~R.~Wasserman$^{17,18}$\orcidlink{0000-0002-4186-6164}} 
\author{S.~K.~Yadavalli$^{13}$\orcidlink{0000-0002-0840-6940}}
\author{P.~Chen$^{19,20}$\orcidlink{0000-0003-0853-6427}}
\author{R.~S.~Post$^{21}$\orcidlink{0000-0003-3244-0337}}

\affiliation{$^1$Institute for Astronomy, University of Hawai\okina i, Honolulu, HI 96822, USA}
\affiliation{$^2$Institute for Astronomy, University of Hawai\okina i, 640 N.~Aʻohoku Pl., Hilo, HI 96720, USA}
\affiliation{$^3$Department of Astronomy and Astrophysics, University of California, Santa Cruz, CA 93105, USA}
\affiliation{$^4$Center for Cosmology \& Astroparticle Physics, The Ohio State University, Columbus, OH, USA}
\affiliation{$^5$Department of Astronomy, The Ohio State University, Columbus, OH, USA}
\affiliation{$^6$School of Physics, University of Melbourne, Parkville, VIC 3010, Australia}
\affiliation{$^7$Institute of Astronomy, Cambridge, CB3 0HA, UK}
\affiliation{$^8$Kavli Institute for Cosmology, Cambridge, CB3 0HA, UK}
\affiliation{$^9$Department of Physics \& Astronomy, University of Hawai'i at Hilo, Hilo, HI 96720, USA}
\affiliation{$^{10}$Astrophysics Research Centre, School of Mathematics and Physics, Queen’s University Belfast, Belfast BT7 1NN, UK}
\affiliation{$^{11}$DARK, Niels Bohr Institute, University of Copenhagen, Jagtvej 128, DK-2200 Copenhagen {\O} Denmark}
\affiliation{$^{12}$Space Telescope Science Institute, Baltimore, MD 21218, USA}
\affiliation{$^{13}$Center for Astrophysics \textbar{} Harvard \& Smithsonian, Cambridge, MA 02138, USA}
\affiliation{$^{14}$The NSF AI Institute for Artificial Intelligence and Fundamental Interactions}
\affiliation{$^{15}$Graduate Institute of Astronomy, National Central University, 300 Zhongda Road, Zhongli, Taoyuan 32001, Taiwan}
\affiliation{$^{16}$Department of Physics \& Astronomy, Sonoma State University, Rohnert Park, CA 94928, USA}
\affiliation{$^{17}$Department of Astronomy, University of Illinois at Urbana-Champaign, Urbana, IL 61801, USA}
\affiliation{$^{18}$Center for Astrophysical Surveys, National Center for Supercomputing Applications, Urbana, IL 61801, USA}
\affiliation{$^{19}$Institute for Advanced Study in Physics, Zhejiang University, Hangzhou 310027, China}
\affiliation{$^{20}$Institute for Astronomy, School of Physics, Zhejiang University, Hangzhou 310027, China}
\affiliation{$^{21}$Post Astronomy, Lexington, MA 02421}

\altaffiliation{$^*$NSF Fellow}
\altaffiliation{$^\dag$CCAPP Fellow}
\altaffiliation{$^\ddag$FINESST Fellow}

\email{Corresponding author: willemh@hawaii.edu}

\begin{abstract}
We present optical-to-near-infrared (NIR) photometry and spectroscopy of the Type Ia supernova (SN~Ia) 2024epr, including NIR spectra observed within two days of first light. The early-time optical spectra show strong, high-velocity Ca and Si features near rarely-observed velocities at $\sim$0.1$c$, and the NIR spectra show a \CI\ ``knee.'' Despite early-time, high-velocity features, SN~2024epr evolves into a normal SN~Ia, albeit with stronger peak-light Ca absorption than other SNe~Ia with the same light curve shape. Although we infer a normal decline rate, $\Delta m_{15}(B)=1.09\pm0.12$~mag, from the light-curve rise, SN~2024epr is a Branch ``cool'' object and has red early-time colors ($g-r\approx0.15$~mag at $-10$~days). The high velocities point to a density enhancement in the outer layers of the explosion, predicted by some models, but thick-shell He-detonation models do not match the smoothly rising light curve or apparent lack of He in our early-time NIR spectra. No current models (e.g., delayed detonation or thin He shell double detonation) appear to reproduce all observed properties, particularly the unusual early-time colors. Such constraints are only possible for SN~2024epr from the earliest optical and NIR observations, highlighting their importance for constraining SN~Ia models. Finally, we identify several literature SNe~Ia with intermediate mass elements at $\sim$30\,000~km~s$^{-1}$ within days after the explosion that evolve into otherwise normal SNe~Ia at peak light, suggesting the early-time spectra of SNe~Ia may hide a broad diversity of observational characteristics.
\end{abstract}


\maketitle  

\section{Introduction} \label{sec:intro}

Type Ia supernovae (SNe~Ia) are oft-occurring \citep[e.g.,][]{Desai24} astronomical transients associated with the thermonuclear explosions of carbon-oxygen white dwarf stars (CO WDs; \citealt{hoy60}). They produce a significant fraction of the intermediate mass and Fe-group elements in the Universe \citep[e.g.,][]{Raiteri96, Matteucci01} and enable precise distance determinations \citep{Phillips93} useful for constraining cosmological parameters \citep{riess98, Perlmutter99, Burns18, Jones22, Riess22}. Despite the many SN~Ia data sets (e.g., \citealp{Hicken09, Ganeshalingam10, Brown14b, Krisciunas17, Holoien17, Holoien17B, Holoien17C, Foley18, Holoien19, Jones19, Phillips19, Tucker20, Fausnaugh21, Jones21, Scolnic22, Fausnaugh23, Neumann23, Peterson23, Morrell24, Do25}) and theoretical models (e.g., \citealt{Nomoto84, Thielemann86, Khokhlov91, Woosley94, Hoeflich95, Hoeflich96, Iwamoto99, Ropke07, Shen10, Woosley11, Pakmor12, Seitenzahl13, Fink14}), the progenitor systems and explosion mechanisms of SNe~Ia are not yet fully understood, and fundamental, qualitative questions remain unanswered (reviews include \citealt{Maoz14, Jha19, Liu23}). 

While there is broad consensus and strong evidence \citep{hoy60, Nugent11} that SNe~Ia originate from exploding CO WDs, many disparate models exist. Each model includes a progenitor scenario (i.e., the constituent objects before any explosion occurs) and an explosion mechanism (i.e., how the CO WD ignites thermonuclear runaway). Two oft-invoked progenitor scenarios are the single-degenerate (SD) scenario with a CO WD and a non-degenerate companion such as a main sequence or red giant star (e.g., \citealp{Whelan73, Nomoto82b, Nomoto97}) and the double-degenerate (DD) scenario with at least two CO WDs or a CO WD and a He WD (e.g., \citealp{Nomoto80, iben84, Webbink84}).  Other scenarios exist as well, such as the core-degenerate (CD) scenario involving a CO WD and the degenerate CO core of an asymptotic giant branch star (AGB star; e.g., \citealp{Hoeflich96}), but are most often invoked only to explain highly peculiar SNe~Ia (e.g., 2003fg-like SNe~Ia \citealp{Lu21, Ashall21}).

As with the progenitor scenario, several explosion mechanisms may produce a SN~Ia. Historically, SNe~Ia were thought to detonate (i.e., explode with a supersonic propagation speed) when they near the Chandrasekhar mass ($M_{\rm ch}\approx1.4$\,M$_\odot$, \citealp{Chandrasekhar31}). However, subsequent studies have found different explosion mechanisms may produce SNe~Ia below, at, or above this mass. For example, material from either a degenerate (DD scenario) or non-degenerate (SD scenario) companion can accrete onto the CO WD, triggering an explosion through central carbon ignition as the CO WD approaches the Chandrasekhar mass \citep{hoy60, Whelan73, Nomoto82a, Piersanti03} or through a He detonation on the surface of a CO WD below the Chandrasekhar mass \citep{Nomoto80, Nomoto82b, Livne90, Woosley94, Hoeflich96, Shen14b, Hoeflich17, Maeda18, Shen18, Polin19, Gronow20, Shen24}. Alternatively, a merger \citep{iben84, Webbink84, vanKerkwijk10, Scalzo10, Pakmor10, Pakmor13, Kromer13, Kromer16, Tucker25} or a third/fourth-body induced collision of two CO WDs \citep{Rosswog09, Raskin09, Thompson11, Kushnir13, Pejcha13} may occur in the DD scenario causing explosions with a wide range of masses less than, near, or exceeding \Mch. Finally, a dynamical merger between a CO WD and the CO core of an AGB star in the CD scenario may also produce a SN~Ia near \Mch\ \citep{Hoeflich96, Kashi11, Ilkov13, AznarSiguan15, Noebauer16, Maeda23}.

Comparing the plethora of SNe~Ia models is difficult, as many models can reproduce the broad homogeneity of SNe~Ia at peak light. However, at phases before peak light, especially those shortly after the explosion, model predictions diverge significantly because of differences in the initial conditions of a model most strongly reflected in the outermost layers of the ejecta. The outermost layers of the expanding ejecta are the first to become optically thin. Thus, observing these outermost layers of the ejecta requires early observations shortly after the explosion. For example, the origin of high-velocity ($\sim$20\,000~km~s$^{-1}$, \citealp{Benetti04, Benetti05, Wang09_HV, Harvey25}) Si and Ca features in early-time optical spectra is still not well understood, with potential explanations including ionization effects \citep[e.g.,][]{Blondin13}, an increased amount of ejecta at high velocities from the explosion \citep[e.g.,][]{Mazzali05, Tanaka06, Kato18},  material swept up by the ejecta after the explosion  \citep[e.g.,][]{Gerardy04}, or interaction with circumstellar material \citep[e.g.,][]{Mulligan19}. 

In addition to potentially depending on the progenitor scenario, high-velocity features may also arise from differences in the explosion mechanism. For example, a later transition from deflagration (subsonic flame propagation) to detonation in the delayed-detonation mechanism increases the ejecta velocities \citep{Iwamoto99}. Additionally, He detonation models may also produce high-velocity ejecta, depending on the viewing angle \citep{Polin19, Boos21, Collins23, Collins24, Boos24b}. Distinguishing these two models is difficult, but early-time data, in particular NIR spectra, have strong discriminatory power \citep{Hsiao19}.  

In the NIR, spectroscopic features have shallower optical depths than in the optical \citep{Wheeler98, Hoeflich02} and thus probe material deeper in the ejecta than the optical regime at the same epoch. While optical features such as \CaII\ H\&K and NIR triplet and \SiII\ $\lambda$6355 easily saturate \citep[e.g.,][]{Hachinger08}), NIR features do not, making them important diagnostics of SN~Ia physics \citep[see, e.g.,][and references therein]{Hsiao19, Hoogendam25b, Muller-Bravo25}. One example is tracing the distribution of unburnt carbon in the explosion, which differs between the deflagration (more unburnt carbon) and detonation (less unburnt carbon) models \citep[e.g.,][]{Hoeflich02, Kasen09}.  Carbon is more easily measured with the less blended NIR \CI\ features than optical \CII\ features, which suffer severe blending with \SiII\ (e.g., \citealp{Hsiao13, Hsiao15, Marion15, Wyatt21}). Additionally, there are two He features in the NIR (\HeI~1.083~and~2.058~$\mu$m) that can be used to test predictions of He absorption made by some He-detonation models \citep[e.g.,][although other models do not predict He features, e.g., \citealp{Boos21, Boos24b}]{Collins23, Callan24, Collins24}.  Unfortunately, few SNe~Ia have optical and NIR spectral time series data within two days of the explosion (e.g., \citealp{Hsiao13, Hsiao15, Pearson24}). 

In this manuscript, we present a detailed analysis of SN~2024epr, a nearby SN~Ia in the galaxy NGC~1198. Both optical and NIR spectra were obtained within two days of the estimated time of first light, $\sim$17~days before maximum light. The \CaII\ NIR triplet of SN~2024epr is particularly notable because it is extremely high velocity at early times and remains strong throughout the photospheric phases. We obtained further optical and NIR spectroscopic epochs, with dense spectroscopic sampling in the earliest phases after discovery. Our photometric follow-up observations have a cadence of approximately one to two days during the rise. \S \ref{sec:data} presents our data set and the data reduction methods. We present the photometric data and analysis in \S \ref{sec:photometry} and the spectroscopic data and analysis in \S \ref{sec:spectroscopy}. We discuss the implications for the explosion mechanism for SN~2024epr in \S \ref{sec:discussion}. Lastly, we offer a summary and concluding remarks in \S \ref{sec:conclusion}. Throughout this work, we use $H_0 = 72$~km~s$^{-1}$~Mpc$^{-1}$, $\Omega_m = 0.30$, and $\Omega_{vac} = 0.70$.

\section{Data} \label{sec:data}
\begin{figure*}
    \centering
    \includegraphics[width=0.49\linewidth]{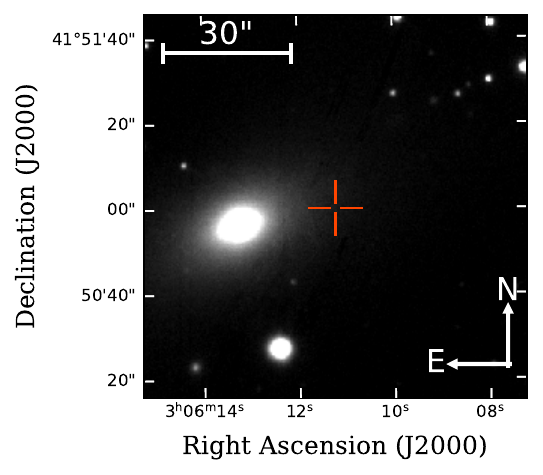}
    \includegraphics[width=0.49\linewidth]{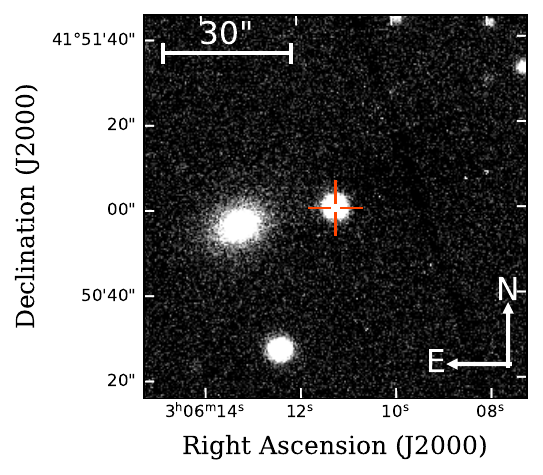}\\
    \caption{
        Finder charts for SN~2024epr. \emph{Left:} Archival Pan-STARRS $r$-band image before explosion \citep{Chambers16,Flewelling20,Magnier20}. \emph{Right:} Pan-STARRS $r$-band image on UT~2024-04-08~at~05:45:08 (MJD~=~60408.2) by YSE.
    }
    \label{fig:finder}
\end{figure*}

The Zwicky Transient Facility (ZTF; \citealp{Bellm19}) discovered SN~2024epr at 
$(\alpha,\delta) = (03^{\mathrm{h}}06^{\mathrm{m}}11\fs224,+41\degr51\arcmin 00\farcs18)$
on UT~2024-03-19 at 05:10:26 (MJD~=~60388.2) with a $g$-band magnitude of 19.4~mag \citep{2024epr_discovery} and classified it as a pre-peak SN~Ia \citep{2024epr_classification}\footnote{Note that \citet{Karambelkar24} report an earlier spectrum within a day of discovery, but this spectrum was not uploaded as part of an official TNS classification report and remains proprietary.}. The ZTF last non-detection was on UT~2024-03-10 at 03:26:47 (MJD~=~60379.1) with a limiting $g$-band magnitude of 18.9~mag. Figure \ref{fig:finder} shows the location of SN~2024epr in NGC 1198 within the Pan-STARRS \citep{Chambers16} $r$-band template image and a near-peak image taken by the Young Supernova Experiment \citep[YSE;][]{Jones21} as part of our photometric follow-up (see \S \ref{sec:phot-data}).

\subsection{Host Galaxy}\label{sec:host-galaxy}
SN~2024epr exploded in NGC~1198, an elliptical (S0\^-, \citealp{deVaucouleurs91}) galaxy with a redshift $z = 0.0053$ \citep{Huchra99}. The Milky Way line-of-sight extinction toward NGC~1198 is $E(B-V)=0.113$~mag \citep{Schlafly11}. To derive an approximate distance to NGC~1198, we use our assumed cosmological parameters, correct the redshift from the heliocentric to the cosmic microwave background frame, and apply a peculiar velocity correction using a density field derived from the 2M$++$ redshift catalog \citep{Lavaux11}. Our derived distance is $22.7\pm3.5$~Mpc ($\mu = 31.76\pm0.34$~mag), including a velocity error of 250~km~s$^{-1}$ in the error budget. We adopt this distance. 

We use the \texttt{Blast} \citep{Jones24} tool\footnote{\url{https://blast.scimma.org/}.} to acquire archival host-galaxy photometry and estimate the properties of NGC~1198. \verb|Blast| measures galaxy photometry from the Two Micron All Sky Survey \citep[2MASS;][]{Skrutskie06}, the Wide-field Infrared Survey Explorer \citep{Wright10}, the Sloan Digital Sky Survey \citep{Fukugita96, York00}, and the Panoramic Survey Telescope and Rapid Response System (Pan-STARRS; \citealp{Chambers16}) using PSF-matched elliptical apertures and runs \verb|Prospector| SED fits on the data \citep{Leja19, Johnson21} with speed engagements from a simulation-based inference (SBI++) from \citet{Wang23_SBIpp}.  {\tt Blast} uses the \verb|Prospector|-$\alpha$ model with a non-parameteric star-formation history and parameters/priors \citep[for more information, see Appendix A in][]{Jones24}.

For NGC~1198, we find the following host-galaxy parameters:
a mass of $\log_{10}M/M_\odot=9.9^{+0.1}_{-0.1}$,
a star-formation rate of $\log_{10}\mathrm{SFR/yr} = -2.2_{-1.5}^{+0.6}$,
a specific star-formation rate of $\log_{10}\mathrm{sSFR/yr} = -12.1_{-1.3}^{+0.6}$, and 
a mass-weighted mean stellar age of $8.4_{-1.3}^{+3.5}$~Gyr, 
confirming visual inspection that NGC~1198 is a quiescent, low-mass, and reasonably compact elliptical galaxy. 

We find an angular separation of 22.4\arcsec\ from the host galaxy, corresponding to a projected separation of $\sim$2.5~kpc at the distance of NGC~1198 (22.7~Mpc). 

     
\subsection{Photometric Data}\label{sec:phot-data}

Our optical photometry of SN~2024epr is from the Asteroid Terrestrial-impact Last Alert System \citep[ATLAS;][]{Tonry18}, YSE, and ZTF surveys, and additional follow-up observations were taken using the Nickel telescope and the Lulin 1m Telescope. Table \ref{tab:phot-data} presents a log of our photometric observations. 

ATLAS observes the entire sky multiple times per night with four telescopes in Chile, South Africa, and Hawaiʻi \citep{Tonry18}. It uses ``cyan" and ``orange" broadband filters that are approximately equivalent to $g$+$r$ and $r$+$i$ bandpasses, respectively. Most observations of SN~2024epr were taken with the orange ($o$) filter, which has a wavelength coverage of 560-820~nm \citep{Tonry18}. Data were retrieved from the ATLAS Transient Science Server \citep{smith20, Shingles21}, and same-night data were stacked using a weighted average, excluding data affected by clouds.

YSE is a time-domain survey on the Pan-STARRS telescopes surveying $\sim$1500~deg$^2$ of sky at a time with a three-day cadence in the $griz$ filters. For particular targets of interest, including SN~2024epr, observations at 1- or 2-day cadence are obtained in $grizy$. The Pan-STARRS observations were reduced using the {\tt Photpipe} package \citep{Rest05} as implemented by the YSE team \citep{Jones21, Aleo23}. We used the {\tt YSE-PZ} target and observation manager \citep{Coulter22_YSEPZ, Coulter23} to compile the data for this SN.

ZTF is a time-domain survey currently providing public, two-day cadence $g$- and $r$-band data for the northern extragalactic sky. We obtained ZTF photometry using the ZTF forced photometry service \citep{Masci19}\footnote{\href{https://irsa.ipac.caltech.edu/Missions/ztf.html}{https://irsa.ipac.caltech.edu/Missions/ztf.html}.}.

Follow-up observations from the Nickel and Lulin telescopes used CCD imagers. Point-spread-function photometry is measured from the images and calibrated by comparing field stars to the Pan-STARRS photometric catalog \citep{Flewelling20}, using standard bias and sky-flat-field procedures. 

After uploading to arXiv, two of us (R.P. and P.C.) contacted the other authors with additional images taken with the Corrected Dall-Kirkham 24'' and Ritchey-Chretien 32'' telescopes by R.P. at the Post Observatory. The images were taken with Sloan filters produced by Astrodon with Pan-STARRS archival images as the template for image subtraction. The photometric calibration is against the SDSS magnitudes transformed from the Pan-STARRS photometric catalog \citep{Flewelling20}. 

NIR $Y\!J\!H$ photometry was obtained with the Wide Field Camera (WFCAM; \citealt{Casali07, Hodgkin09}) mounted on the University of Hawai'i-owned and operated UKIRT 3.8m telescope. We again processed these data with {\tt Photpipe} and calibrated them using 2MASS following \citet{Hodgkin09} and \citet{Peterson23}. We used unforced, non-difference-imaged photometry from \texttt{DOPHOT} \citep{Schechter93}. Because SN~2024epr was bright and located near the outskirts of a smooth elliptical galaxy, a local background estimation and centroid determinations from individual images were considered sufficient for reliable photometry.

\begin{table}[]
    \centering
    \caption{Log of the photometric observations of SN~2024epr. The full table is available on the arXiv listing.}\label{tab:phot-data}
    \begin{tabular}{ccccccc}
        \hline
        MJD & Filter & $m$ & $\sigma_m$ & Source \\ 
        \hline \hline 
            60374.19 & $g$     & 22.64 &    99.99 & ZTF         \\
            60375.30 & $o$     & 18.46 &    99.99 & ATLAS       \\
            \vdots   & \vdots  & \vdots& \vdots   & \vdots  \\
            60652.12 & $g$     & 19.52 &     0.18 & ZTF         \\
        \hline \\
    \end{tabular}
\end{table}

\subsection{Spectroscopic Data}\label{sec:Spec_data}

We obtained 16 optical and 10 NIR spectra of SN~2024epr. The first public spectrum (from \citealp{2024epr_discovery}) was taken by the Nordic Optical Telescope (NOT) using the Alhambra Faint Object Spectrograph and Camera (ALFOSC)\footnote{Accessed via \href{https://www.wis-tns.org/object/2024epr}{https://www.wis-tns.org/object/2024epr}. }. Seven further optical spectra were taken on the UH2.2m telescope using the SuperNova Integral Field Spectrograph (SNIFS; \citealt{Lantz04}) as part of the Spectroscopic Classification of Astrophysical Transients (SCAT; \citealt{Tucker22c}) survey, with reductions using the standard SCAT pipeline \citep{Tucker22c}. One spectrum was taken with the Keck Cosmic Web Imager (KCWI; \citealp{Morrissey18, McGurk24}). We extracted the 1D spectrum using a custom pipeline from the automatic KCWI data reduction pipeline-produced data cubes\footnote{The documentation for which is available \href{https://kcwi-drp.readthedocs.io/en/latest/index.html}{online}.}. Seven optical spectra were observed by the Shane 3m Telescope using the Kast Double Spectrograph \citep{Miller94}. The Kast data were reduced using the UCSC Spectral Pipeline\footnote{See the \href{https://ucsc-spectral-pipeline.readthedocs.io/en/latest/}{online} documentation.} \citep{Siebert20}. Table \ref{tab:opt-spec} compiles the optical spectroscopic observations.

\begin{table}[]
    \centering
    \caption{Log of the optical spectroscopic observations of SN~2024epr. The reported rest-frame phases are with respect to the time of $B$-band maximum on MJD \peakMJD.}\label{tab:opt-spec}
    \begin{tabular}{cccccc}
        \hline
        UT Date & MJD & Phase &  Telescope & Spectrograph & $R$ \\ 
                &     & [days]&            &              &     \\
        \hline \hline
        2024-03-20 &	60389.8	      & $ -16.6$    &  NOT	   &   ALFOSC & 710  \\
        2024-03-23 &	60391.2	      & $ -15.3$    &  UH88    &   SNIFS  & 2000 \\
        2024-03-24 &	60392.2	      & $ -14.2$    &  UH88    &   SNIFS  & 2000 \\
        2024-03-27 &	60396.2	      & $ -10.3$    &  Lick    &   KAST   & 1800 \\
        2024-04-06 &	60406.2	      & $  -0.3$    &  UH88    &   SNIFS  & 2000 \\
        2024-06-27 &    60488.6       & $ +81.6$    &  UH88    &   SNIFS  & 2000 \\
        2024-07-05 &    60496.6       & $ +89.6$    &  UH88    &   SNIFS  & 2000 \\
        2024-07-07 &    60498.5       & $ +91.4$    &  Lick    &   KAST   & 1800 \\
        2024-07-19 &    60510.6       & $+103.5$    &  UH88    &   SNIFS  & 2000 \\
        2024-08-06 &    60528.4       & $+121.2$    &  Lick    &   KAST   & 1800 \\
        2024-08-09 &    60531.6       & $+124.4$    &  UH88    &   SNIFS  & 2000 \\
        2024-08-27 &    60549.5       & $+142.1$    &  Lick    &   KAST   & 1800 \\
        2024-09-05 &    60558.6       & $+151.3$    &  Keck II &   KCWI   & 900  \\
        2024-09-13 &    60566.5       & $+159.0$    &  Lick    &   KAST   & 1800 \\ 
        2024-09-24 &    60577.4       & $+169.9$    &  Lick    &   KAST   & 1800 \\ 
        2024-10-03 &    60586.5       & $+179.0$    &  Lick    &   KAST   & 1800 \\ 
        \hline
    \end{tabular}
\end{table}

Seven NIR spectra were taken on the Gemini North 8m telescope using the Gemini Near-InfraRed Spectrograph (GNIRS; \citealt{Elias06a, Elias06b}). Reductions were performed using standard \textsc{PyPeit} procedures \citep{Prochaska20}. An additional spectrum was taken using NASA's InfraRed Telescope Facility (IRTF) with SpeX \citep{Rayner03}. Finally, the Keck Infrared Transients Survey \citep{Tinyanont24} obtained two post-peak transitional-phase NIR spectra with the Near-InfraRed Echellette Spectrometer on Keck II (NIRES; \citealp{Wilson04_NIRES}). The GNIRS, NIRES, and SpeX data were reduced with telluric corrections from an A0V star using the standard PypeIt (GNIRS and NIRES; \citealp{PypeIt_JOSS}) or Spextool (SpeX; \citep{Cushing04}) procedures. Table \ref{tab:nir-spec} catalogs our NIR spectroscopic observations.

\begin{table}[]
    \centering
    \caption{Log of the NIR Spectroscopic Observations of SN~2024epr. The reported rest-frame phases are relative to the $B$-band maximum time on MJD \peakMJD.}\label{tab:nir-spec}
    \begin{tabular}{cccccc}
        \hline
        UT Date & MJD & Phase &  Telescope & Spectrograph & $R$ \\ 
                &     & [days]&            &              &     \\
        \hline \hline 
            2024-03-21 &	60390.1	      & $-16.5   $     &  Gemini   &   GNIRS & 1500   \\
            2024-03-22  &	60391.2	      & $-15.3   $     &  IRTF     &   SpeX  & 80     \\
            2024-03-23 &	60392.0	      & $-14.5   $     &  Gemini   &   GNIRS & 1500   \\
            2024-03-30 &	60398.8	      & $-7.8    $     &  Gemini   &   GNIRS & 1500   \\
            2024-04-06 &	60406.2	      & $-0.4    $     &  Gemini   &   GNIRS & 1500   \\
            2024-07-05 &    60496.6       & $+89.6   $     &  Keck II  &   NIRES & 2700   \\
            2024-07-06 &    60497.5       & $+90.4   $     &  Gemini   &   GNIRS & 1500   \\
            2024-07-27 &    60518.5       & $+111.4  $     &  Gemini   &   GNIRS & 1500   \\ 
            2024-09-04 &    60557.5       & $+150.2  $     &  Gemini   &   GNIRS & 1500   \\ 
            2024-09-12 &    60565.5       & $+158.1  $     &  Keck II  &   NIRES & 2700   \\ 
            \hline
    \end{tabular}
\end{table}

\section{Photometry} \label{sec:photometry}
\subsection{Light Curves}\label{sec:light_curves}

\begin{figure*}
    \centering
    \includegraphics[width=0.99\linewidth]{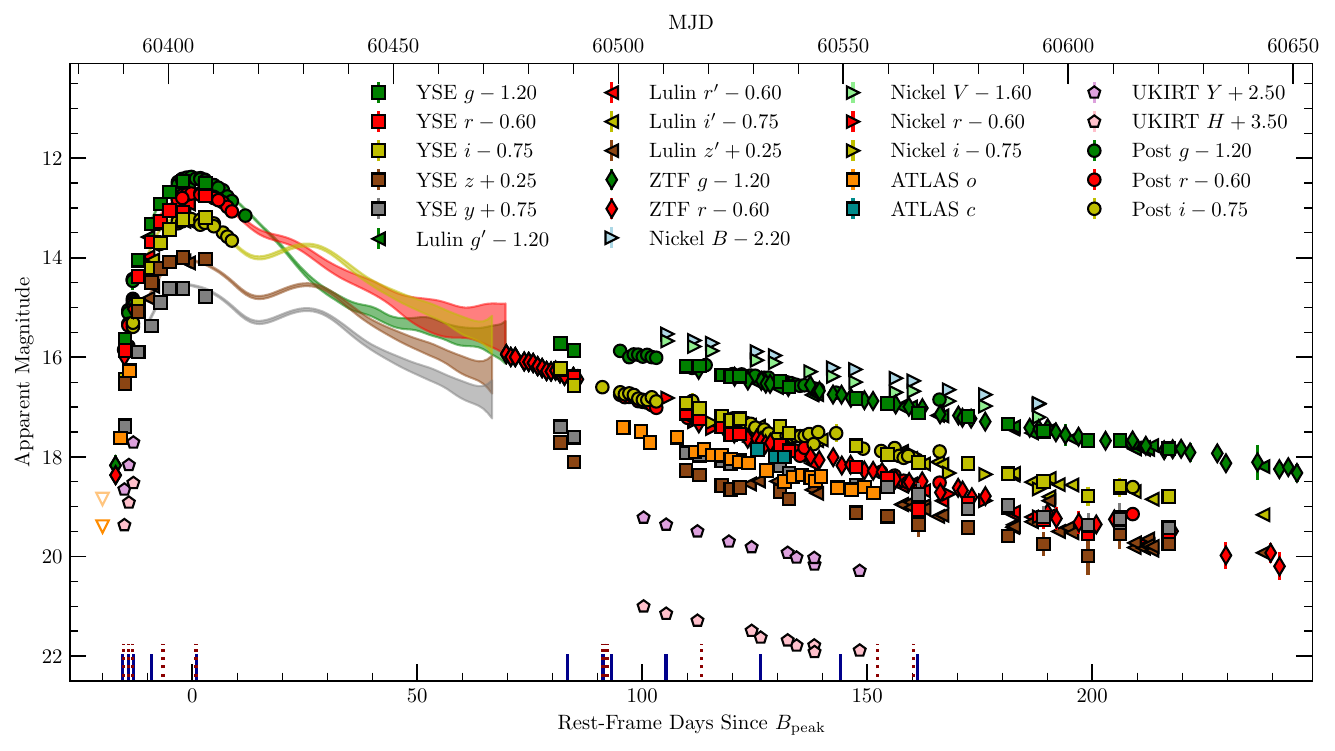} \\
    \caption{
        The optical and NIR photometry of SN~2024epr. Unfilled triangles represent photometric upper limits. The shaded regions denote model fits described in \S \ref{sec:light_curves}. Spectroscopic observations are shown as solid blue (optical) and dotted red (NIR) lines on the bottom of the figure. The ATLAS non-detection is 5$\sigma$. 
        }
    \label{fig:lightcurves}
\end{figure*}

Figure \ref{fig:lightcurves} shows the optical and NIR photometry for SN~2024epr, along with SuperNovae in object-oriented Python (\texttt{SNooPy}; \citealp{Burns11, Burns14}) fits (using the \texttt{max\_model}, see following discussion) to the Post Observatory and Pan-STARRS data. While we do not have $B$-band data near maximum light, \texttt{SNooPy} computes SED fits whose synthetic photometry is used to return parameters with respect to the $B$ band. Using {\tt EBV\_model2} with $\Delta m_{15}(B)=1.09$~mag as a prior (from the \texttt{max\_model} fit described below), the \texttt{SNooPy} fit finds a host-galaxy extinction of $E(B-V)=0.12\pm0.07$~mag and peak $B$-band magnitude of 13.80$\pm$0.01~mag.

The \texttt{SNooPy} \verb|EBV_model2| model fits the light curve of SN~2024epr reasonably well in the $gzy$ bands, with worse fits in the $ri$ bands. The issues in the $ri$ bands may be due to the significant Si and Ca absorption, respectively, which could cause band-dependent shifts in color relative to a typical reddening law (see Section \ref{sec:optical-spectra}).  We consider the high extinction measurement to be most likely due to the unusual colors of 2024epr or the fact that \texttt{SNooPy} may not fully model variation in SN intrinsic colors \citep{Burns14}.

Since SN~2024epr has peculiar color evolution (see \S \ref{sec:photometry_colors}), we report the \verb|max_model| values for 
$t_{\rm peak}^B$, which is on MJD $60406.6\pm0.4$, 
$\Delta m_{15}(B)=1.09\pm0.07$~mag, and 
$s_{BV} = 1.189\pm0.04$. 
The $gri$ peak magnitudes are $13.23\pm0.04$~mag, $13.07\pm0.04$~mag, and $13.76\pm0.03$~mag, respectively. 

Finally, the ATLAS last non-detection is notably deep. The 3$\sigma$ upper limit is 19.40~mag and the 5$\sigma$ upper limit is 18.85~mag. Correcting for the Milky Way extinction ($A_o=0.27$~mag) and host-galaxy distance, this yields 3$\sigma$ and 5$\sigma$ absolute magnitude limits of $-12.64$~mag and $-12.08$~mag, respectively.

\begin{figure}
    \centering
    \includegraphics[width=0.98\columnwidth]{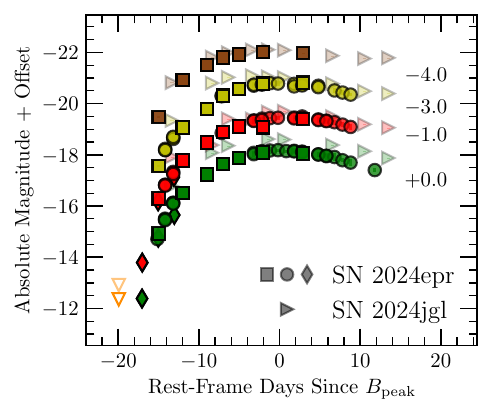} \\
    \caption{
        Comparison of the rising YSE light curves of SN~2020jgl (right-facing triangles; from YSE DR1 \citealp{Aleo23}) and SN~2024epr. Circles, squares, and diamonds denote Post, YSE, and ZTF photometry, respectively. The ATLAS 3$\sigma$ (bold) and 5$\sigma$ (light) last nondetections are orange, downward triangles; there are no ZTF non-detections a week before discovery. Photometric bands are, from the bottom up, $griz$. When shifted by peak time, the light curves line up, confirming that the \texttt{SNooPy} time of $B$-band maximum is reliable. Offsets (not including the 0.67~mag difference in the host-galaxy distance moduli) are reported next to each band.
    }
    \label{fig:peak_validation}
\end{figure}

We emphasize that we do not directly measure $\Delta m_{15}(B)$ or $s_{BV}$ because our data do not extend to the necessary time scales to do so (15 days and 20-40 days, respectively; \citealp{Burns14}). Rather, we infer it from fitting the rising light curve with \texttt{SNooPy}. To ensure the accuracy of our computed $\Delta m_{15}(B)$ (and of the time of maximum light), we compare our phase-shifted light curves with SN~2020jgl \citep{Galbany25}, another SN~Ia with strong and high-velocity \CaII\ absorption at early times (see \S \ref{sec:spectroscopy} for spectral comparison). Figure \ref{fig:peak_validation} shows the overlaid light curves of SNe~2020jgl~and~2024epr shifted to the rest-frame phase with respect to the epoch of peak $B$-band magnitude and shifted by the difference in distance modulus between the two host galaxies. The light curves agree well using the \texttt{SNooPy} \verb|max_model| peak times for SN~2024epr. SN~2020jgl has a similar $\Delta m_{15}(B)$ value ($\Delta m_{15}(B)=1.111\pm0.018$~mag, \citealp{Galbany25}) to our inferred $\Delta m_{15}(B)$ value for SN~2024epr.

We also fit the data with the SALT3 model \citep{Kenworthy21, Taylor23} and see similar color discrepancies in both the early-time and near-maximum data, with the $i$-band flux in particular lower than the best-fit model. Again, we interpret this as an indication of stronger-than-normal Ca absorption affecting this band. The best-fit SALT parameters for SN~2024epr are a relatively red $c = 0.19 \pm 0.02$~mag, a slightly fast (but uncertain) decline rate of $x_1 = -0.39 \pm 0.33$, and a somewhat underluminous $B$-band absolute magnitude that is approximately 0.7~mag fainter than the mean SN~Ia at that redshift; however, adopting a peculiar velocity uncertainty of 250 km/s at the redshift of SN~2024epr gives an additional magnitude uncertainty $\sim$0.34 mag. Using the SALT nuisance parameters from \citet{Brout22} gives a distance modulus to NGC~1198 of $31.90 \pm 0.11$~mag, closer than the \texttt{SNooPy} distance. For the rest of this work, we adopt the host-galaxy redshift distance from \S \ref{sec:host-galaxy}

Our inferred $\Delta m_{15}(B)$ value is somewhat inconsistent with SN~2024epr being a \citet{Branch06} ``cool'' object (see \ref{sec:optical-spectra}) and the similarities in the NIR spectra to subluminous SNe Ia (see \ref{sec:nir-comp}). In light of this, we also estimate $\Delta m_{15}(B)$ based on the relationship between $\mathcal{R}(\mathrm{Si})$ (the ratio of the \SiII~$\lambda$5972 and $\lambda$6355 pEWs at peak light; see \citealt{Nugent95}) and $\Delta m_{15}(B)$ presented in \citet{Benetti05}, which gives $\Delta m_{15}(B)\approx1.3$~mag\footnote{\citet{Benetti05} do not fit the data, so we estimate by eye.}. 

\begin{figure}
    \centering
    \includegraphics[width=0.98\columnwidth]{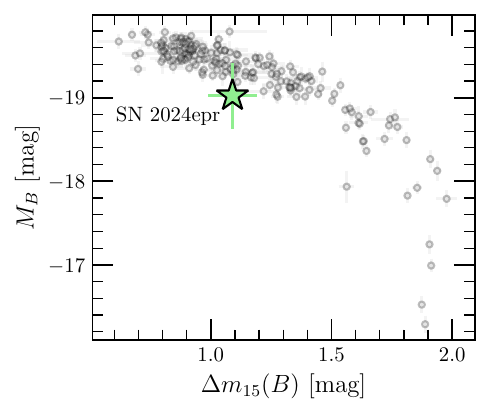} \\
    \caption{
        SN~2024epr (light green star) on the luminosity-width relationship \citep{Phillips93}. SN~2024epr is a normal SN~Ia when corrected for Milky Way extinction and SN color as described in the text. We infer the distance from the host-galaxy redshift (there is no independent distance to NGC~1198), assuming a peculiar velocity error of 250~km~s$^{-1}$. The uncertainty in distance is the dominant source of luminosity uncertainty. The $B$-band magnitude is computed using SED fit from \texttt{SNooPy}.  
    }
    \label{fig:LWR}
\end{figure}

Correcting for the distance and the Milky Way extinction, we find that the $B$-band absolute magnitude of SN~2024epr is $M_B\approx-18.4\pm0.4$~mag. 
We correct for SN color by adopting the \verb|EBV_model2| host-galaxy extinction and $R_B=R_V+1=3.2$, since the median $R_V$ for SNe~Ia is known to be lower than that of the Milky Way \citep[e.g.,][]{Burns14}. This is likely not from actual host-galaxy extinction but rather reflects the known empirical correlation between SN color and luminosity \citep[e.g.,][]{Burns14}.
After this correction, SN~2024epr is consistent within uncertainties with normal SNe~Ia (Figure \ref{fig:LWR}). Its slightly subluminous location could be explained by a variety of factors, including a poor distance estimate to the host and the lack of a $B$-band peak measured directly from the data, rather than inferred from an SED that may only superficially match that of SN~2024epr. 
The peculiar-velocity-induced distance uncertainty is also a non-negligible factor in the absolute magnitude uncertainty. We estimate this uncertainty assuming a peculiar velocity of 250~km~s$^{-1}$ \citep{Peterson22}. 

\subsection{Color Curves}\label{sec:photometry_colors}
\begin{figure}
    \centering
    \includegraphics{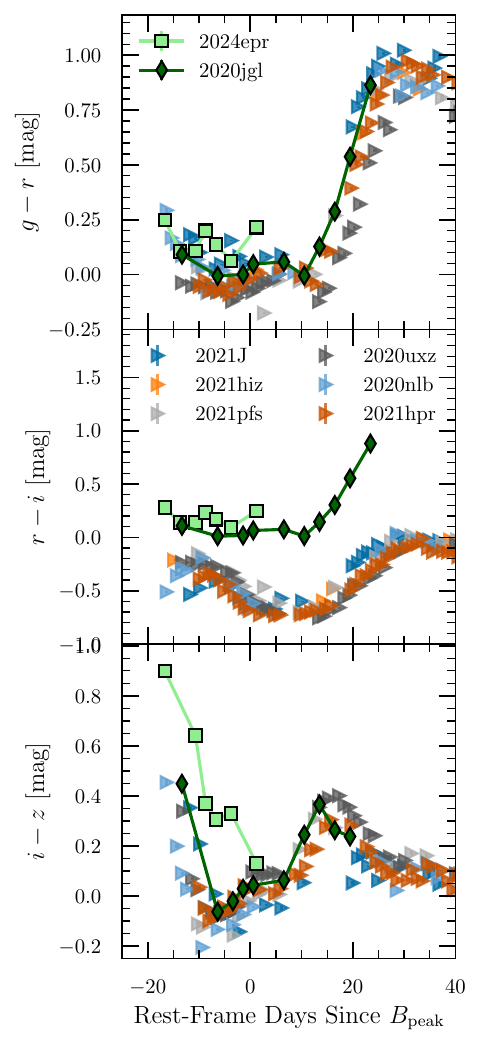} \\
    \caption{
        Milky-Way extinction-corrected color curves of SNe~Ia with strong \CaII\ NIR features (SNe 2020jgl and 2024epr) compared to normal SNe~Ia from YSE DR1 \citep{Aleo23}. The two SNe~Ia with strong \CaII, SNe 2020jgl and 2024epr, are redder than the normal SNe~Ia. 
    }
    \label{fig:colorcurves}
\end{figure}

Figure \ref{fig:colorcurves} shows the Milky Way extinction-corrected $g-r$, $r-i$, and $i-z$ color curves for SN~2024epr compared to the Milky Way extinction-corrected colors for SN~2020jgl, another SN~Ia with strong \CaII\ features, and well-observed normal SNe~Ia from the YSE DR1 sample (SNe 2020nlb, 2020opy, 2020uxz, 2021J, 2021hiz, 2021hpr, 2021pfs; \citealp{Aleo23}). 

SNe~2020jgl and 2024epr are similar to the comparison SNe~Ia in the $g-r$ color throughout their evolution. Interestingly, despite the similarity in the $g-r$ and $i-z$ colors, SNe~2020jgl and 2024epr differ from the other SNe~Ia in their $r-i$ color curve, being much flatter ($r-i\approx0$~mag from pre-peak to maximum light) and not showing any color evolution until 10~days past $B$-band maximum. This implies the color difference is not solely due to reddening. One possible explanation is that, at early times, the high-velocity Ca feature is so blue-shifted that the Ca absorption trough is solely in the $i$ band (see \S \ref{sec:optical-spectra}). This difference may explain why the early-time $i-z$ color is much redder than that of the comparison SNe~Ia, similar to what is seen for SN~2021aefx, which exhibits a high-velocity \CaII\ H\&K feature \citep{Ashall22}. 

Another SN Ia with similarly low luminosity for its measured $\Delta m_{15}(B)$ is SN~2006bt \citep{Foley10}, which also has peculiar colors; however, the colors shown by SN~2020jgl and 2024epr ($r-i\approx+0.3$~mag) at maximum light are much redder than SN~2006bt ($r-i = -0.3$~mag), suggesting that SN~2024epr is unlike SN~2006bt, which also had significantly slower velocities. 

\section{Spectroscopy} \label{sec:spectroscopy}

\subsection{Optical Spectra}\label{sec:optical-spectra}

\begin{figure*}
    \centering
    \includegraphics{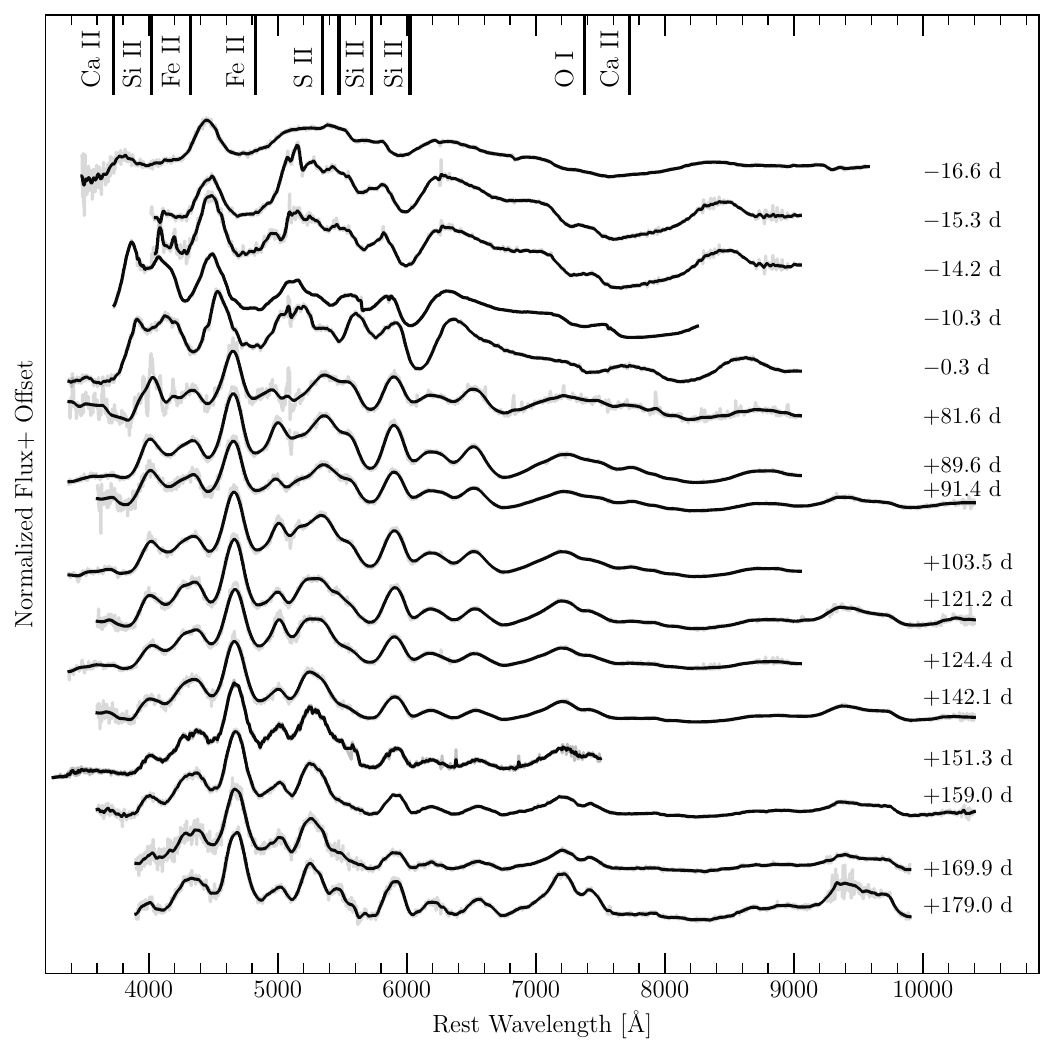} \\
    \caption{
        Time series of optical spectra for SN~2024epr (grey unsmoothed; black smoothed). Epochs are with respect to $B$-band maximum on MJD \peakMJD. Spectrographs are listed in Table \ref{tab:opt-spec}.
    }
    \label{fig:opt_spectra}
\end{figure*}

Figure \ref{fig:opt_spectra} presents the optical spectroscopic time series of SN~2024epr. The spectra include some of the earliest optical spectra of a SN~Ia, with rest-frame phases ranging from $-16.6$ to $+159.0$~days relative to the $B$-band peak. The optical spectra look like a typical SN~Ia but with high velocities at early times and a much stronger \CaII\ NIR triplet. 

\subsubsection{Early-Time Comparison}

\begin{figure*}
    \centering
    \includegraphics[width=\textwidth]{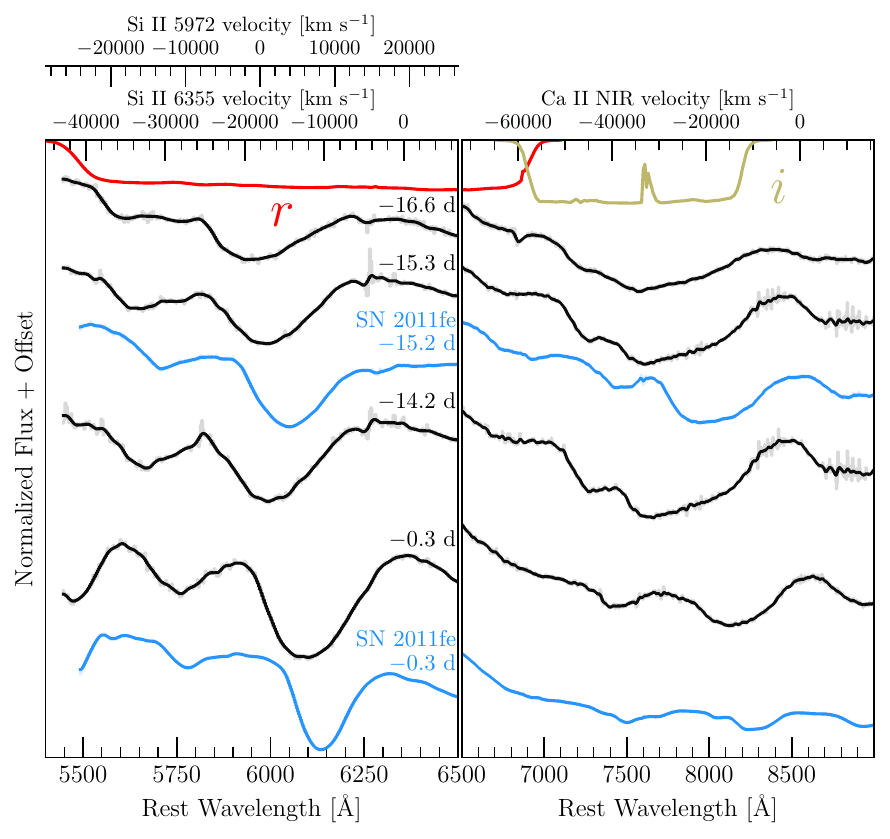} \\
    \caption{
        Comparison of the early-time spectral time series data of SN~2024epr (unsmoothed grey, smoothed black) with that of SN~2011fe (blue). \emph{Left:} the spectral region around \SiII~$\lambda6355$. \emph{Right:} spectral region around the Ca NIR triplet. Epochs are with respect to $B$-band maximum on MJD \peakMJD, and the $i$-band transmission function is shown in dark khaki on the top of the left panel.  
    }
    \label{fig:flux_comp_11fe}
\end{figure*}

We compare our photospheric spectra to the earliest ($-15.3$~days) and peak-light ($-0.3$~days) SNIFS spectra of SN~2011fe \citep{Pereira13} in Figure \ref{fig:flux_comp_11fe}. The most notable difference is the strength and high-velocity ($\sim$35\,000~km~s$^{-1}$ versus $\sim$20\,000~km~s$^{-1}$) of \CaII\ in SN~2024epr versus SN~2011fe at $\sim-16$~days. On the other hand, the \SiII\ feature has a similar velocity to SN~2011fe, albeit slightly faster, and the \CII\ $\lambda$6580 feature is at the same velocity as SN~2011fe.

\begin{figure*}
    \centering
    \includegraphics{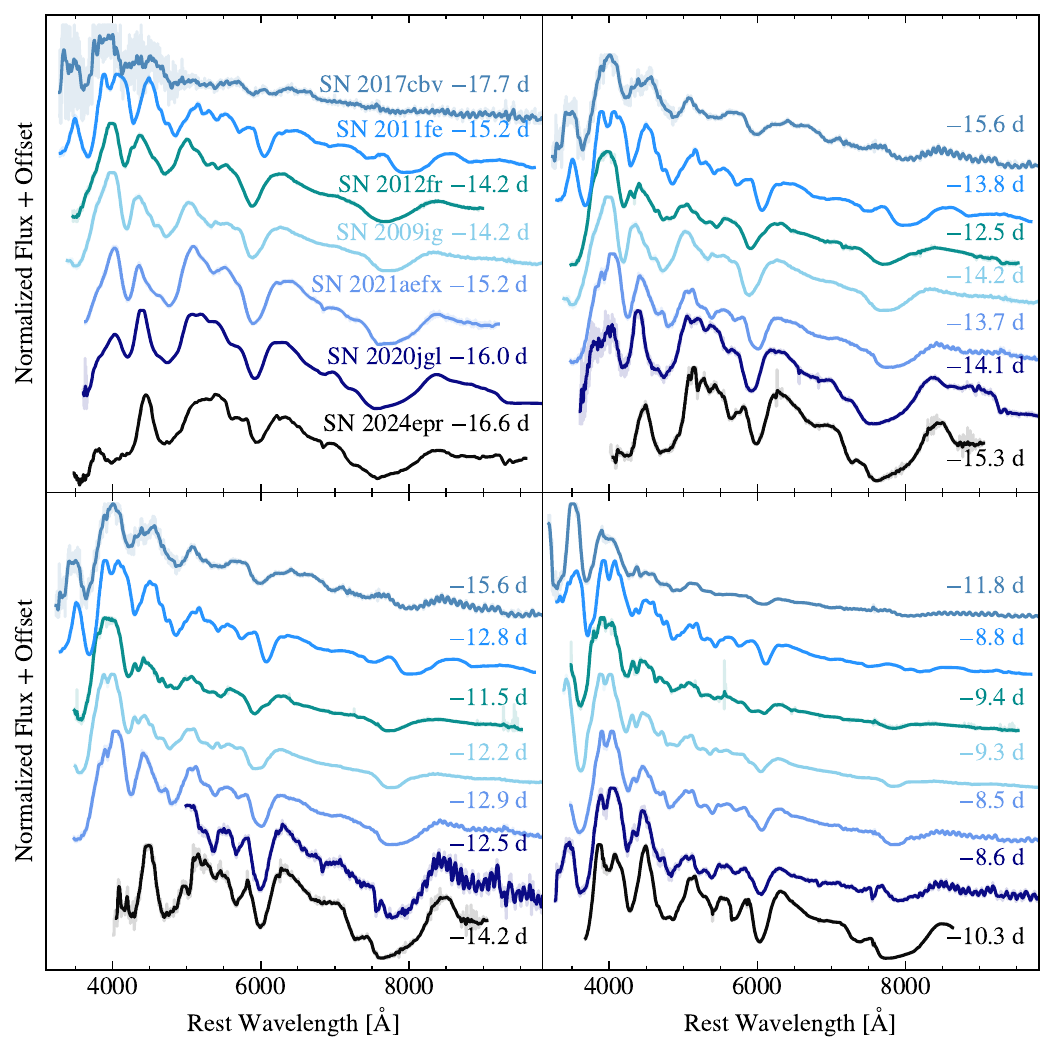} \\
    \caption{
        Comparison of SN~2024epr (black, bottom) to other SNe~Ia with early-time spectroscopic time series data. The plotted order is SN~2017cbv, SN~2011fe, SN~2012fr, SN~2009ig, SN~2021aefx, and SN~2020jgl from top to bottom in each panel.
    }
    \label{fig:four_panel_optical_spectra_comp}
\end{figure*}

In Figure \ref{fig:four_panel_optical_spectra_comp}, we broaden our comparison sample to include the following SNe~Ia: SN~2009ig \citep{Foley12}, SN~2011fe \citep{Pereira13}, SN~2012fr \citep{Childress13}, SN~2017cbv \citep{Hosseinzadeh17}, SN~2020jgl \citep{Galbany25}, and SN~2021aefx \citep{Ashall22, Hosseinzadeh22}. These SNe were all classified as normal SNe~Ia and were chosen due to the availability of their early-time optical spectra, and some, such as SNe~2020jgl and 2021aefx, are known to have high-velocity \CaII\ features \citep{Ashall22, Galbany25}.

In their earliest spectra, SNe~2020jgl and 2024epr have stronger and faster \CaII\ NIR than the comparison SNe~Ia, with the extremely high velocity causing the feature to blend with the \OI\ $\lambda$7774 feature. The other events with high-velocity Ca features (2009ig, 2012fr and 2021aefx) do not show strong \OI\ $\lambda$7774, with a weak, $\sim$20\,000~km~s$^{-1}$ \OI\ feature in SN~2021aefx and no clear \OI\ in SNe~2009ig or 2012fr. At slightly later epochs, 14~and~13~days before maximum light, the spectra are largely similar to the $-15$~day spectra, and even at $-9$~days the \OI\ and \CaII\ features are still slightly blended.
 
Like the \OI\ and \CaII\ features, the \SiII\ $\lambda\lambda$5972, 6355 features blend in SNe~2020jgl~and~2024epr, with some \SiII\ $\lambda$6355 at $\sim$35\,000~km~s$^{-1}$ in the earliest spectrum. Unlike \OI\ and \CaII\ features, the Si features become less blended a day later, and by $-13$~days the two features have only minimal blending. At 9~days before maximum light, the \SiII\ features are no longer blended, and SN~2024epr has the strongest \SiII\ $\lambda$5972 out of all the plotted SNe~Ia. 

On the red edge of the \SiII\ $\lambda$6355, each SN~Ia shows a potential flattening or notch feature in their earliest spectrum, which \citet{Thomas11} attribute to \CII\ $\lambda6580$. The strongest feature is seen in SN~2017cbv at early times, with the \CII\ feature equaling the \SiII\ $\lambda$6355 feature in depth. A clear notch is seen in SN~2011fe, with tentative \CII\ features seen in SNe 2020jgl and 2021aefx. SN~2024epr may have a notch similar to SN~2011fe; however, the feature's location also corresponds to an $\rm{O}_2$ skyline. Given the other $\rm{O}_2$ skyline at 6800 in our $-16.6$~day spectrum and evidence of sky subtraction issues in the subsequent two, we cannot confidently claim a \CII\ $\lambda6580$ feature in SN~2024epr. 

As discussed in \S \ref{sec:nir-spectra}, we also find evidence for the presence of C in SN~2024epr at early times by identifying a \CI\ feature in the NIR spectra at the same epoch (see Figs. \ref{fig:NIR_He_vel} and \ref{fig:NIR_comp}), highlighting the power of contemporary optical and NIR spectroscopic coverage. Except for SNe~2011fe and 2017cbv, the \CII\ feature disappears from the SNe~Ia spectra within a few days after the first spectrum implying that the C may be located in the outermost layers. 

Finally, the morphology of the \FeII, \SiII\ and \FeIII\ blend between 4600 \AA\ and 5000 \AA\ shows SNe 2020jgl and 2024epr are very similar to each other and have approximately the same features at the same velocities as SNe 2009ig, 2011fe, and 2021aefx. This explains the early-time color curves of SNe 2020jgl and 2024epr when compared to the normal YSE sample: colors using bluer bands (e.g., $g-r$) match the colors of normal SNe~Ia better than colors using redder filters (e.g., $i-z$), as discussed in \S \ref{sec:photometry_colors}. 

\subsubsection{Peak-Light Comparison}

\begin{figure}
    \centering
    \includegraphics{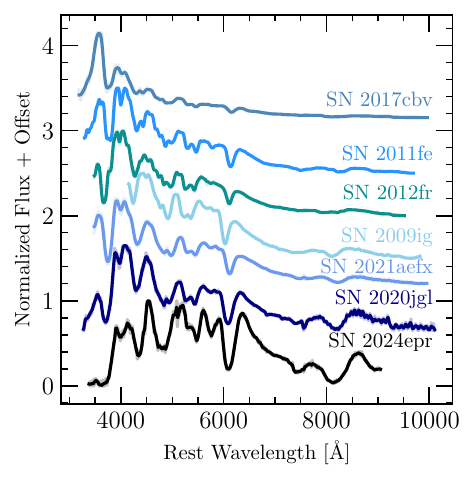} \\
    \caption{
        Near maximum light comparison of SN~2024epr to the same SNe~Ia as Figure \ref{fig:four_panel_optical_spectra_comp}. 
    }
    \label{fig:peak_comparison}
\end{figure}

Figure \ref{fig:peak_comparison} shows the spectra at the time of maximum light. Here, SNe~Ia with faster early-time velocities (e.g, SNe~2009ig, 2020jgl, 2021aefx, and 2024epr) generally have faster peak-time velocities. Still, peak velocities are within the standard range of SNe~Ia velocities. Of the high-velocity SNe~Ia, SN~2024epr has the strongest secondary \SiII\ feature; SNe~Ia with high-velocity early-time features generally show weaker secondary \SiII, in contrast to SN~2024epr. 

\subsubsection{Pseudo-Equivalent Widths and Velocities}
To quantify spectral differences, we use the Python package  \verb|misfits|\footnote{\href{https://github.com/sholmbo/misfits}{https://github.com/sholmbo/misfits}} to measure the pseudo-equivalent width (pEW) and velocity for each feature in each spectrum (for a description of \verb|misfits| and the routines used, see \citealp{Hoogendam22} and \citealp{Holmbo23}). We note that when fitting the high-velocity \CaII\ triplet, it blends with the \OI\ $\lambda$7774 feature. Thus, we fit three features in this case: one for the \OI\ $\lambda$7774 feature and two for the \CaII\ triplet feature, which has a high-velocity and photospheric-velocity component. 

\begin{figure}
    \centering
    \includegraphics[width=\linewidth]{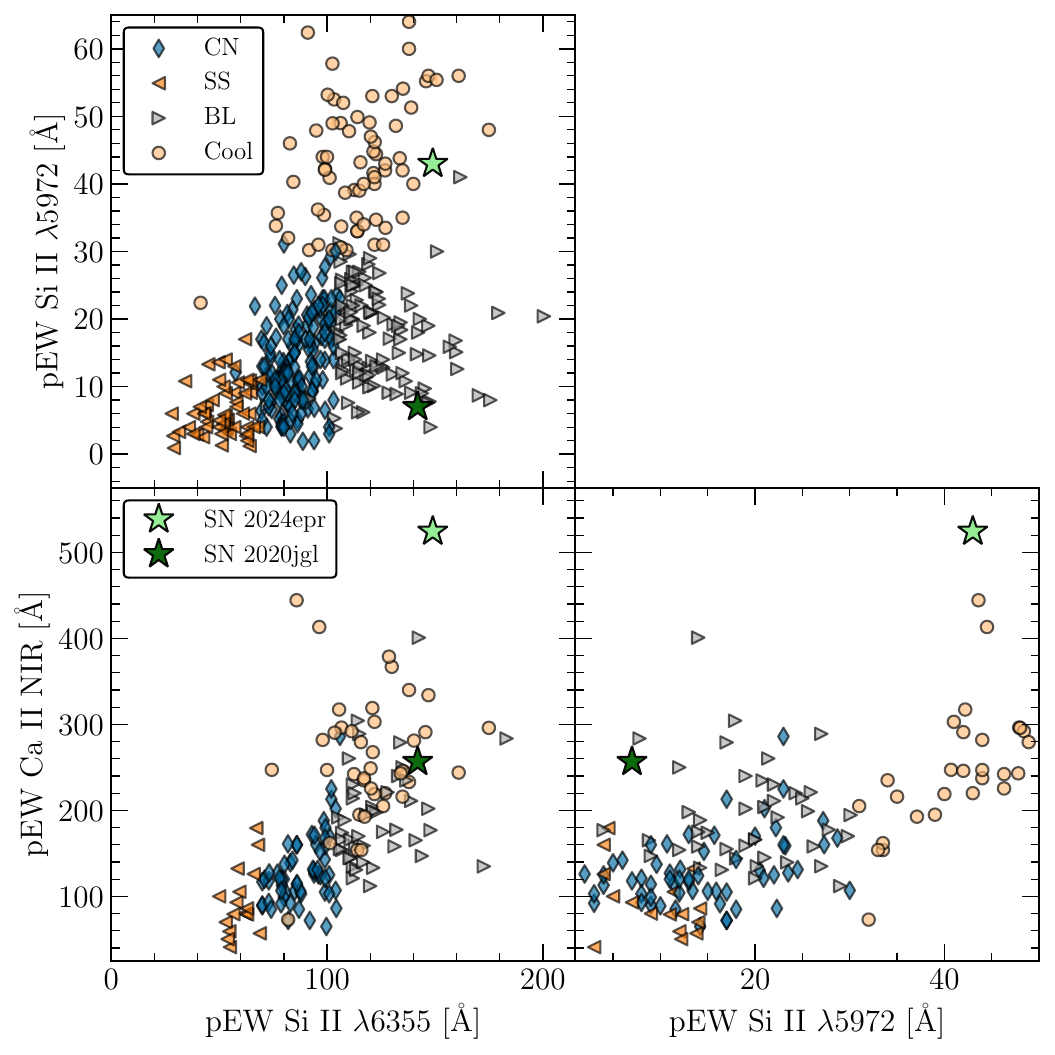} \\
    \caption{
        Branch diagram (comparison of SN~Ia spectral features at maximum light; classifications based on \citealp{Branch06}) with SN~2024epr as the light green star and SN~2020jgl as the dark green star. The other SNe~Ia data are from \citet{Blondin12}, \citet{Folatelli13}, \citet{Hoogendam22}, and \citet{Morrell24}. \emph{Top:} Canonical \citet{Branch06} diagram with the two \SiII\ lines as diagnostics. \emph{Bottom:} Pseudo-Branch diagram with the \CaII\ NIR feature replacing the \SiII\ $\lambda$5972 feature. While SN~2024epr is an unremarkable ``cool'' SN~Ia in the canonical Branch scheme, when using the \CaII\ line as a diagnostic, SN~2024epr is more extreme than any other SN~Ia plotted.
    }
    \label{fig:branch}
\end{figure}

Figure \ref{fig:branch} shows the \citet{Branch06} classification scheme, which classifies SNe~Ia into four groups based on the pEW of the \SiII\ $\lambda\lambda$5972 and 6355 features. The ``core normal'' SNe~Ia are a highly homogeneous group with similar spectral features. For ``broad line'' SNe~Ia, the \SiII\ $\lambda$6355 feature is broader than the core normal SNe~Ia, perhaps indicating a high-velocity component to the feature. Alternatively, ``shallow silicon'' SNe~Ia have shallower \SiII\ features, especially the weaker \SiII\ $\lambda$5972 feature, which may be entirely absent from the SN~Ia spectrum. As shallow silicon SNe~Ia tend to be overluminous, a likely explanation is the explosion energy is high enough to doubly ionize Si, resulting in \SiIII\ features instead of \SiII. Finally, the ``cool'' SNe~Ia have a stronger \SiII\ $\lambda$5972 features due to a lower temperature compared to the core normal SNe~Ia.  In the canonical \citet{Branch06} scheme, SN~2024epr is a ``cool'' SN~Ia, with a strong secondary \SiII\ feature at max (pEWs of $\sim$43~\AA\ and $\sim$149~\AA\ for \SiII\ $\lambda\lambda$5972 and 6355, respectively). This is dissimilar to SN 2020jgl (pEWs of  7~\AA\ and 142~\AA, respectively), which has a weaker \SiII\ $\lambda$5972 at peak. 

We also construct a pseudo-\citet{Branch06} diagram in the bottom panel of Figure \ref{fig:branch} to compare the \CaII\ pEW ($\sim$525~\AA\ for SN~2024epr, and $\sim$257~\AA\ for SN~2020jgl) to that of \SiII\ $\lambda$6355. In general, we see that core normal and shallow-silicon SNe~Ia tend to have smaller Ca pEW and are generally distinguishable from the broad line and cool SNe~Ia as in the standard \citet{Branch06} diagram. However, unlike the standard \citet{Branch06} diagram, the broad line and cool SNe~Ia overlap in the \CaII\ NIR diagram. This is explained by the \CaII\ NIR triplet saturating more easily than \SiII\ $\lambda$5972. Additionally, there is a weak linear correlation for all SNe~Ia between the \SiII\ $\lambda$6355 and \CaII\ NIR pEWs at maximum light. SN~2020jgl is consistent with this trend despite its strong early-time \CaII\ NIR feature, but SN~2024epr is not. Of the plotted SNe~Ia, SN~2024epr has the largest \CaII\ pEW, indicative of a higher-than-usual abundance of Ca. 

\subsection{NIR Spectra}\label{sec:nir-spectra}

\begin{figure*}
    \centering
    \includegraphics{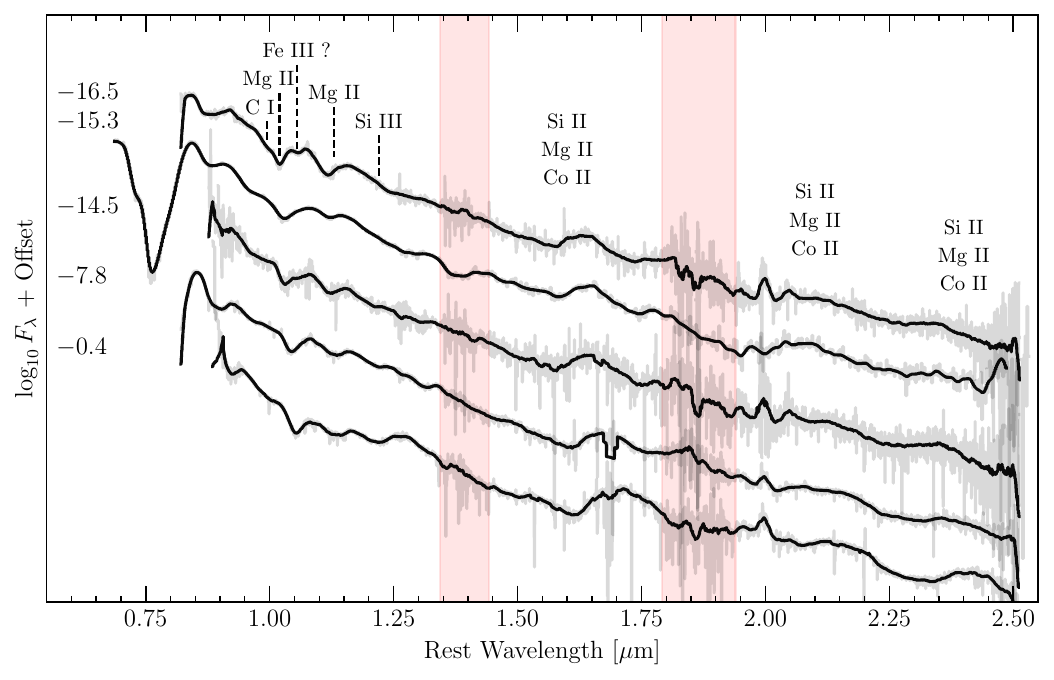} \\
    \caption{
        Pre-peak NIR spectroscopic time series for SN~2024epr. Light red shading denotes telluric regions.  
    }
    \label{fig:nir_spectra_plot}
\end{figure*}

Figure \ref{fig:nir_spectra_plot} presents our pre-peak NIR spectroscopic time series data for SN~2024epr. The spectra show common SN~Ia NIR features such as \MgII\ 1.0092~$\mu$m, \CI\ 1.0693~$\mu$m, and \MgII\ 1.0927~$\mu$m \citep{Hsiao13,Hsiao15,Marion15,Hsiao19,Lu23}. 

The absorption feature at 1.06~$\mu$m in the first spectrum may be high-velocity ($\sim$20\,000~km~s$^{-1}$) \OI~1.1286,~1.1302~$\mu$m, high-velocity ($\sim$40\,000~km~s$^{-1}$) \CI~1.1754~$\mu$m, or \FeIII~1.1323~$\mu$m. There are difficulties with each of these interpretations for the 1.06~$\mu$m feature. First, the optical \OI\ is present in every optical spectrum, whereas this feature is not seen in subsequent spectra of SN~2024epr. Second, if it is \CI~1.1754~$\mu$m, it would be higher velocity than \CI~1.0693~$\mu$m. Finally, if the feature is from \FeIII\ recombining into \FeII, the presence of \FeIII\ in the outermost layers of the ejecta within a day after the explosion requires higher temperatures \citep[see, e.g.,][]{DerKacy20} for a short time after the explosion before cooling from expansion. However, a higher temperature at such early times contradicts the presence of \SiII~$\lambda$5972 at early times, a hallmark spectral feature of cooler explosions.

The next redward absorption feature is at 1.124~$\mu$m, which \citet{Hoeflich02} attribute to \MgII~1.1620 and 1.1600~$\mu$m. If so, the velocity of this feature is much slower than the bluer \MgII\ 1.0927~$\mu$m feature. Given the paucity of features in this region (see, e.g., the line list of \citealp{Marion09}), \MgII\ may be the best option despite this velocity inconsistency. 

\citet{Hsiao19} identify the minimum at 1.23~$\mu$m as a blend of \SiIII~1.2523 and 1.2601~$\mu$m. Other \SiIII\ features include \SiIII~1.3395, 1.3497, and 1.3644~$\mu$m, which we do not observe in SN~2024epr at early times, but may appear near peak. Alternatively, the models of \citet{Hoeflich02} predict weak \OI~1.3164~$\mu$m features near 1.23~$\mu$m, which we may see given the strong \OI\ $\lambda$7774 feature in the optical. 

Si/Mg/Co blends dominate the $H$ and $K$-band regions of the spectra \citep{Hoeflich02, Marion09}. In the $H$ band, the bluest edge is determined by \MgII~1.6787 ~$\mu$m and the red emission from the P Cygni profile is from \CoII~1.5759, 1.6064, 1.6361, 1.7772, 1.7462~$\mu$m \citep{Marion09, Ashall19a, Ashall19b}. The $K$-band has \MgII~2.1569~$\mu$m, \MgII~2.1369, 2.1432, 2.4041, 2.4044, 2.4125~$\mu$m, \SiII~2.1920 and 2.1990~$\mu$m \citep{Hoeflich02}, and \CoII~2.1350, 2.2205, 2.3613, 2.4596~$\mu$m \citep{Marion09} features. 

After the seasonal break, we obtained additional NIR spectra. Spectra starting $\sim$90~days after peak light trace the transition between the end of the photospheric phase with an optically thick, emitting photosphere and the beginning of the nebular phase when the ejecta are optically thin. These spectra are dominated by the permitted \FeII\ and \CoII\ features \citep{Marion09, Gall12, Friesen14} which fade as the SN~Ia evolves and eventually forbidden emission dominates the spectrum $\sim$1 year after explosion. 

\begin{figure*}
    \centering
    \includegraphics{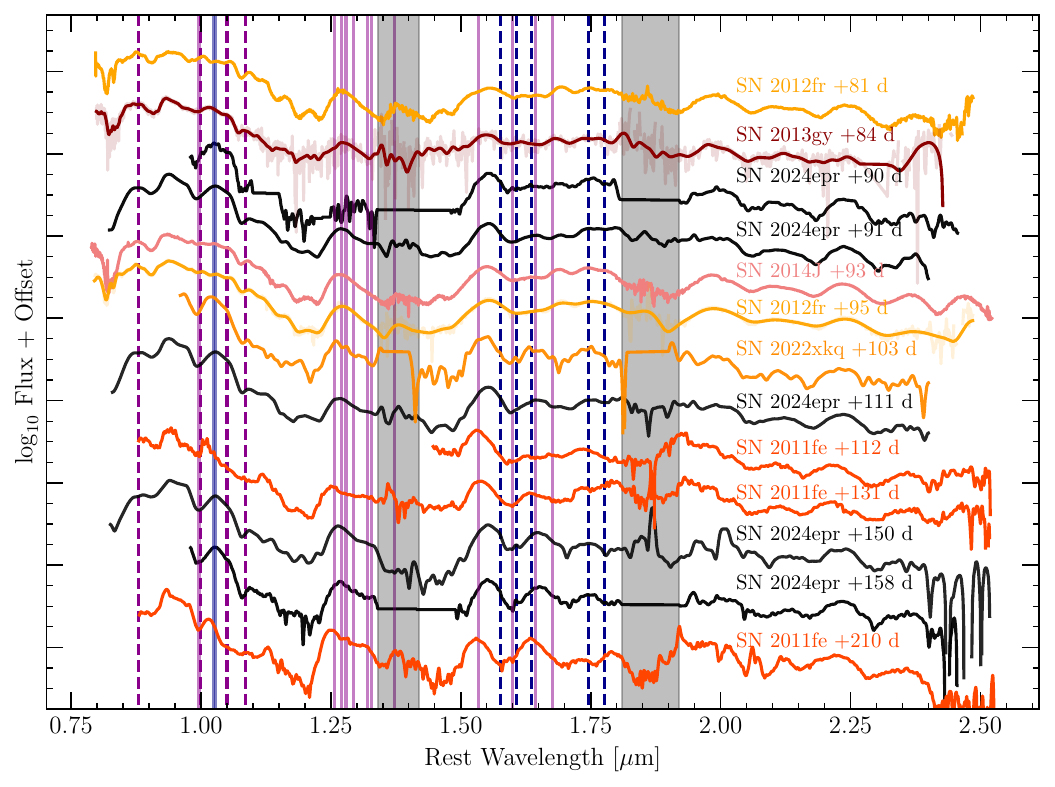} \\
    \caption{SN~2024epr compared to other SNe~Ia with NIR data in the $+81$~to~$+210$~days phase range. Solid lines denote forbidden Fe (magenta) and Co (blue) transitions, respectively, and dashed lines denote permitted Fe and Co transitions, respectively. Grey-shaded regions denote telluric regions. 
    } \label{fig:NIR_late-time_comp}
\end{figure*}

Figure \ref{fig:NIR_late-time_comp} compares the late-time NIR spectra of SN~2024epr with other SNe~Ia during the photospheric/nebular transition phase, and it includes several never-before-published spectra of SN~2011fe (PI: Phillips). These spectra were taken by Gemini/GNIRS and accessed through the Gemini Science Archive \citep{Gemini_Archive}. The three spectra were observed on MJDs 55927.6, 55945.5, and 56038.5. We reduced these spectra in the same manner as the spectra of SN~2024epr (see \S \ref{sec:Spec_data}). 

Additional spectra come from \citet{Sand16}, \citet{Lu23}, and \citet{Pearson24}. This sample of SNe~Ia spans a variety of peak absolute magnitudes and early-time NIR diversity (see \S \ref{sec:nir-comp}), yet despite that, the spectra are relatively similar in the +80~day to +210~day window.  

We identify a mix of permitted Fe and Co blends (from the \citealt{Marion09} line list) and forbidden Fe and Co blends (from the \citealt{Hoeflich21} line list). Figure \ref{fig:NIR_late-time_comp} includes these lines, denoting permitted emission with dashed lines and forbidden emission with solid lines. 

The 1.25~$\mu$m and 1.55~$\mu$m features during the transition from photospheric to nebular phases are ubiquitous in SNe~Ia. While $+90$~days is early to transition to the nebular phase in the optical, the NIR becomes nebular and shows forbidden emission lines earlier than the optical. 
The broad emission feature at $\sim$1.25~$\mu$m is a blend of [\FeII]~1.257,~1.271,~1.279,~and~1.294~$\mu$m. At 1.55~$\mu$m, the feature is likewise a blend of [\FeII] 1.534, 1.600, 1.644, and 1.677 $\mu$m.The strongest permitted features are \FeII~0.8801, 0.9998, 1.0500, 1.0863~$\mu$m and \CoII~1.5759, 1.6064, 1.6361, 1.7772, 1.7462~$\mu$m \citep{OHora24}. 

Every SN~Ia except SN~2014J shows absorption features at 0.90~$\mu$m and 1.00~$\mu$m. At first glance, this may be attributable to \MgII~0.9227 and 1.0092, respectively, but \MgII~1.0927 is the stronger feature at lower temperatures \citep[see Table 5 in][]{Marion09}.
In the $H$~band, the $+$80--100-day spectra have emission at 1.7~$\mu$m, decreasing in strength over time. Different blends of Fe/Co begin to appear after $+$100~days.  

Finally, the $K$-band features are generally similar for all SNe~Ia in this phase range. Longer wavelengths measure deeper into the ejecta; thus, the $K$~band may already be optically thin and start to show blends of forbidden Fe/Co features, or there could still be permitted \CoII\ emission. 

\section{Implications for SNe~I\lowercase{a} Models}\label{sec:discussion}
The early-time observations of SN~2024epr provide helpful constraints to evaluate proposed models for SNe~Ia. Below, we discuss the rising light curve morphology (\S \ref{sec:rising_lc_discussion}), the early-time optical spectra (\S \ref{sec:opt_model_disc}), and the NIR spectroscopic features at early times (\S \ref{sec:NIR_model_disc}).

We use these observational diagnostics to qualitatively discuss predictions made by various physical models (see sections below). While we are unable to link SN~2024epr to a single SN~Ia model definitively, the high-velocity features challenge some delayed-detonation models, and we rule out He detonation models with a thick ($M_{\mathrm{shell}}\geq0.05M_\odot$) He shell from the fits to the multiband photometry (\S \ref{sec:rising_lc_discussion}), and the apparent lack of He features in the NIR spectra (\S \ref{subsec:no-He}). 

\subsection{Model Implications from the Rising Light Curve}\label{sec:rising_lc_discussion}
Some SNe~Ia models may differ in their predictions for the shape of the rising light curve (e.g., \citealt{Kasen10, Piro13, Piro14, Piro16, Maeda18, Polin19, Magee20b, Maeda23}). Such differences may arise from intrinsic differences in the SN Ia progenitor scenario (e.g., companion interaction) or explosion mechanism (e.g., double detonation), or from an extrinsic effect like viewing angle. 
Observationally, there are three categories of early-time rising light curve: ``single'' power law rises in which the rising light curve is well-described by a single power law, ``double'' or ``excess'' rises, which are best described by a power law with an additional component (e.g., another power law or a Gaussian), and ``bump'' rising light curves which have non-monotonic light curves exhibiting a decline in flux before rising to the main light curve peak \citep[][and potentially correlate with negative Hubble residuals, e.g., \citealp{Ye24}]{Hoogendam24}. 

\begin{figure}
    \centering
    \includegraphics{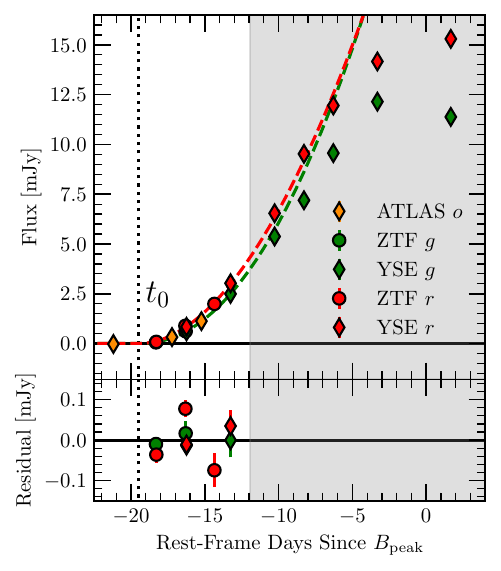} \\
    \caption{
        Rising light curve fits and residuals for SN~2024epr. The grey shaded region denotes the data range excluded from the fit, and the vertical dotted line denotes the time of explosion $t_0$ as determined from a single power law fit. \emph{Top:} The combined YSE (diamonds) and ZTF (circles) $g$- and $r$-band light curves (green and red) and best-fit lines (dashed lines) along with ATLAS $o$-band data (orange diamonds). \emph{Bottom:} Fit residuals for the single component fit. 
    }
    \label{fig:rising_lc_fits}
\end{figure}

For SN~2024epr, we fit the combined rest-frame Pan-STARRS and ZTF $g$-band light curve with a single power-law model, masking data after $-10$~days. \citet{Vallely19, Vallely21, Fausnaugh21} find that the choice of the fitting range of a light curve can significantly change the fit results and recommend fitting only up to 40\% of the SN peak flux, which, for SN~2024epr is at $\sim$10~days. Other studies \citep[e.g.,][]{Fausnaugh23, Hoogendam24b, Hoogendam25b} use a double power-law to fit transients through maximum light, negating this potential systematic. Despite having photometric coverage at this time, there is an offset in the Post light curves compared to those from ZTF and YSE, which have a negligible difference. This offset is likely due to differences in the filter functions, and it ultimately prevents us from using this more robust fitting technique. Thus, we opt to fit only at early times, with a model defined by

\begin{equation}
    f(t) = 
    \begin{cases} 
        0 & t < t_0 \\
        a\times\left(x-t_0\right)^b & t_0 \leq t, 
    \end{cases}\label{eq:1pl}
\end{equation}
where $a$ is a multiplicative scaling factor, $b$ is the power law index, and $t_0$ is the time of first light from the model. The best values from the fit are 
$a=0.11\pm0.01$~mJy
$b=1.80\pm0.01$, and
$t_0=-18.9\pm0.3$~days, or MJD $60387.7\pm0.3$. Figure \ref{fig:rising_lc_fits} shows the data and fit.

A single power-law model fits the rising light curve in this phase range well ($\chi^2/n=2.3$; with an additional 2\% uncertainty added to the photometry to account for potential underestimation of the photometric precision). There is no trend in the residuals as a function of time. For those reasons, we do not perform a double power law or power law and Gaussian fit. Based on this simple power-law fit, the first optical and NIR spectra of SN~2024epr are $2.1\pm0.1$/$2.4\pm0.1$~days after first light, respectively, making them remarkably early spectra for a SN~Ia. Additionally, given the constraining last non-detection from ATLAS, SN~Ia models that require additional flux beyond a simple single power law at early times may not accurately match the rising light curve of SN~2024epr. 

\begin{figure*}
    \centering
    \includegraphics{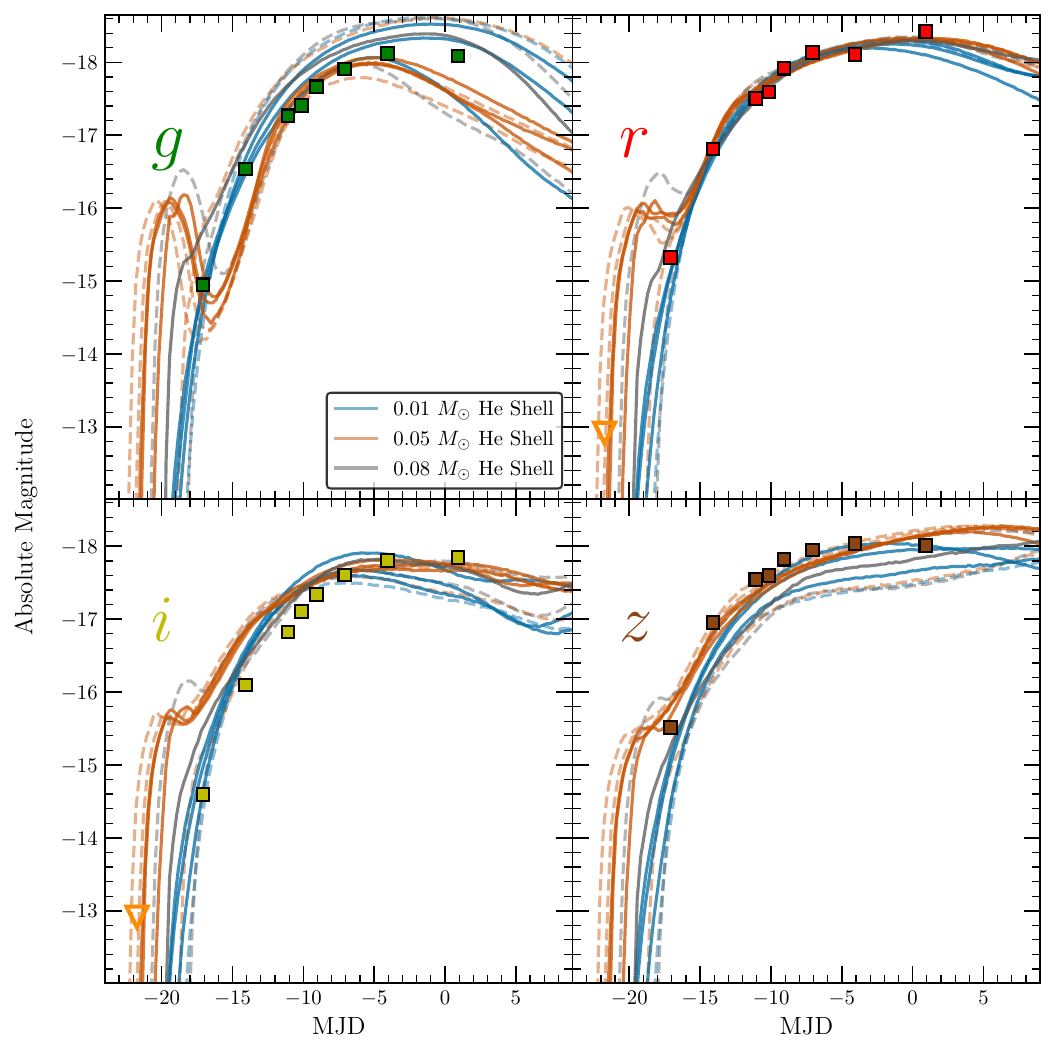} \\
    \caption{
        Comparison of selected He detonation model light curves from \citet{Polin19} to our Pan-STARRS $griz$-band photometry (squares) that have the lowest reduced $\chi^2$ values after fitting for optimal vertical (accommodating uncertainty in the redshift-derived distance) and horizontal (accommodating uncertainty in the time of peak) offsets between the model and data. The ATLAS last non-detection is shown in orange but not included in the $\chi^2$ fitting. The models are colored by He-shell mass, with different models of the same color varying in WD mass. Solid lines denote models with $\chi^2/\mathrm{N}\leq6$, and dashed lines denote models with $6<\chi^2/\mathrm{N}\leq9$. An additional 2\% uncertainty is included in the fit uncertainties to account for the potential underestimation of photometric uncertainties. 
    }
    \label{fig:he-det_comp}
\end{figure*}

We also compare the multi-band light curves of SN~2024epr with those of He detonation models. Figure \ref{fig:he-det_comp} shows model He detonation absolute magnitudes from \citet{Polin19} and our Pan-STARRS $griz$ photometry. Model fits are highly sensitive to the time of first light and the distance to the SN~Ia, so we perform $\chi^2$ minimization fits to determine the horizontal and vertical shifts to the model light curves to obtain the best fit for our data. Even with the phase and distance offsets (i.e., horizontal and vertical shifts to minimize the reduced $\chi^2$ value), the \citet{Polin19} model light curves struggle to reproduce the observed light curves. 

Generally, the smaller 0.01~$\mathrm{M}_\odot$ He shells match the $gri$ light curves well at 10~days before peak, but the larger 0.05~$\mathrm{M}_\odot$ He shell light curves match better at peak. No single model light curve can explain our observed data, particularly the unusual colors near peak. Given the last non-detection from ATLAS for SN~2024epr of 18.8~mag ($\sim$$-13$~absolute~magnitude using the redshift-inferred distance, or $\sim$$-13.5$~mag using the distance from SN~2024epr) at $\sim$$-21$~days before maximum light, we rule out the scenarios with larger He shells ($M_\mathrm{shell} \geq0.05$~$\mathrm{M}_\odot$) that would produce excess flux in the early-time observations. Thus, thinner He shells (e.g., \citealp{Boos21, Boos24a}), which do not produce excess flux in the rising light curve, may be preferential to thicker He shells to explain SN~2024epr and other SNe~Ia with strong, high-velocity Ca features and smoothly rising light curves. This is not to suggest that SN~2024epr definitively originates from a He-detonation. Non-He-detonation models may also be able to replicate the observed properties of SN~2024epr (a thorough analysis of every model is beyond the scope of this work).

\subsection{Model Implications from Early-Time Optical Spectra}\label{sec:opt_model_disc}

There are apparent differences between extremely high-velocity SNe~Ia, like SNe 2009ig, 2012fr, 2020jgl, 2021aefx, and 2024epr, and the canonically normal SNe~2011fe and 2017cbv. We find a qualitative trend between the \CaII\ NIR feature strength and the \SiII\ 5972 feature strength at peak, as shown in Figure \ref{fig:branch}. SN~2009ig has the weakest \CaII\ NIR and \SiII\ 5972 features, whereas SN~2024epr has the strongest \CaII\ NIR and \SiII\ 5972; the other two SNe~Ia have \SiII\ 5972 and \CaII\ strengths between those of SNe~2009ig and 2024epr.

The presence of Si and Ca in the outermost layers of the SN~2024epr ejecta (at $\sim$38\,000~km~s$^{-1}$) also has modeling implications. High-velocity Ca features may be from primordial Ca \citep{Lentz00}, so such objects may be explosions of higher metallicity CO WDs. This would agree with the generally redder colors of SN~2020jgl \citep{Galbany25} and SN~2024epr (Section \ref{sec:photometry_colors}). Alternatively, the Ca may be the product of SN~Ia nucleosynthesis; higher-density deflagration-to-detonation transitions in the delayed detonation models of  \citet{Iwamoto99} produce higher Ca velocities. The \citet{Iwamoto99} models match the apparent lack of He and the smoothly rising light curve of SN~2024epr, but it is unclear whether they can produce Ca with a velocity of 35\,000~km~s$^{-1}$. 

Alternatively, double detonation models may have high-velocity features, especially if the line-of-sight is aligned with the location of the He detonation (e.g., \citealp{Boos21}). A thin He shell burns only to Si or Ca and could also produce high-velocity features \citep[e.g.,][]{Moore13, Shen14b, Townsley19, Gronow20, Boos21}. This is also consistent with less line blanketing, which is predicted to occur in thick-shell double detonation models. These models have significant line blanketing for wavelengths less than 5000~\AA\ due to He burned to Ni in the initial detonation \citep[e.g.,][]{Polin19}. This is inconsistent with SN~2024epr, which does not have significant flux suppression in that spectral region.

\subsection{Model Implications from Early-Time NIR Spectra}\label{sec:NIR_model_disc}

\subsubsection{Implications from No Obvious He Features}\label{subsec:no-He}

He may form spectral features in the optical, but potential He features would blend with stronger features typical of SNe~Ia, such as \SII\ and \SiII, making them difficult to detect confidently. In the NIR, however, the strongest He features, at \HeI~1.083~$\mu$m and \HeI~2.058~$\mu$m, suffer less severe blending. 

\begin{figure}
    \centering
    \includegraphics{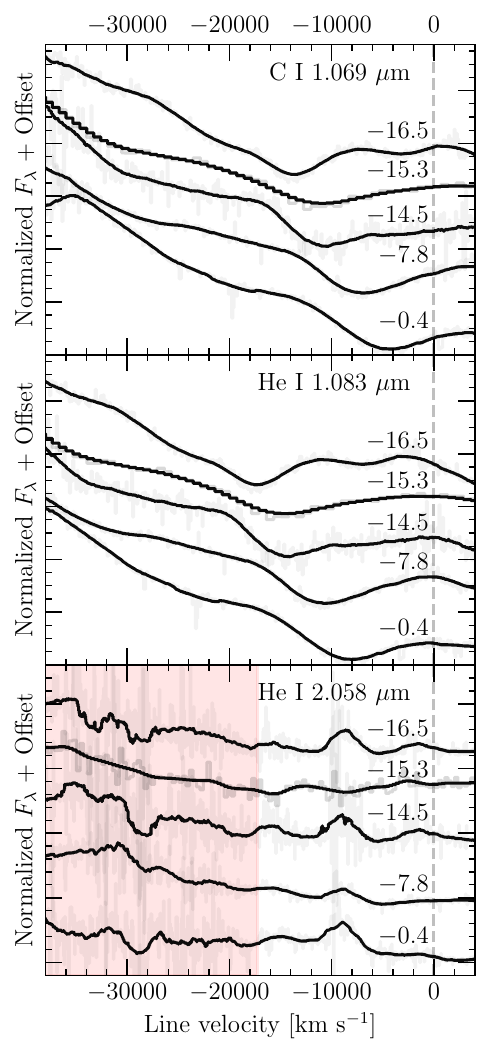} \\
    \caption{The photospheric NIR spectral evolution of SN~2024epr in regions near potential He features. 
    \emph{Top:} Velocities with respect to \CI\ 1.0693~$\mu$m. 
    \emph{Middle:} Velocities with respect to \HeI\ 1.083~$\mu$m
    \emph{Bottom:} Velocities with respect to \HeI\ 2.058~$\mu$m. 
    We rule out the presence of \HeI\ given the lack of an obvious feature at $\sim2$~$\mu$m to complement any feature in the 1.083~$\mu$m region. The emission feature at $\sim$$-10\,000$~km~s$^{-1}$ in the bottom panel is a reduction artifact and not physical. The shaded red area in the bottom panel denotes the telluric region. 
    }
    \label{fig:NIR_He_vel}
\end{figure}

We do not observe evidence for He in our NIR spectral time series (Figure \ref{fig:NIR_He_vel}). For the \HeI\ 1.083~$\mu$m feature, several absorption features could be \HeI, but their velocities are too slow to be consistent with the outer layers of the ejecta of SN~2024epr. We would expect the early-time He velocity to be similar to Ca because it must be the most exterior ejecta layer; however, He may be optically thin at these velocities. The most likely \HeI\ feature in this wavelength region is between 20\,000~km\,s$^{-1}$ and 10\,000~km\,s$^{-1}$, but we do not see a similarly evolving feature in the $K$-band at 2.058~$\mu$m. Thus, we prefer the interpretation of this ``knee'' feature as \CI\ 1.0693~$\mu$m, which has been previously identified in SNe~Ia \citep{Hsiao13, Marion15, Hsiao15, Hsiao19, Pearson24}. If it is \CI, this feature has a velocity of $\sim$23\,000~km~s$^{-1}$. While this is almost twice as fast as the \CI\ velocity in iPTF13ebh and SN~2022xkq, it is not inconsistent since iPTF13ebh and SN~2022xkq are both low-luminosity SNe~Ia with slower ejecta velocities. The detection of \CI\ in SNe~Ia has been associated with high ejecta masses and delayed detonation explosions \citep{Hoeflich02, Hsiao15, Hsiao19}.

Double detonation models with thin (e.g., \citealp{Boos21}) or thick (e.g., \citealp{Collins23}) shells may have He in the outer layers of the ejecta. Thick-shell models may have enough He to produce a detectable absorption feature \citep{Callan24, Collins24}, which is disfavored by our observations. In contrast, the He from thin-shell models is optically thin \citep{Boos24b} and leaves no spectroscopic imprint. However, double-detonation models tend to have less C \citep[e.g.,][]{Fink10}, which may be in contradiction to the NIR spectra.  

\subsubsection{Implications from Mg Features}

\begin{figure*}
    \centering
    \includegraphics{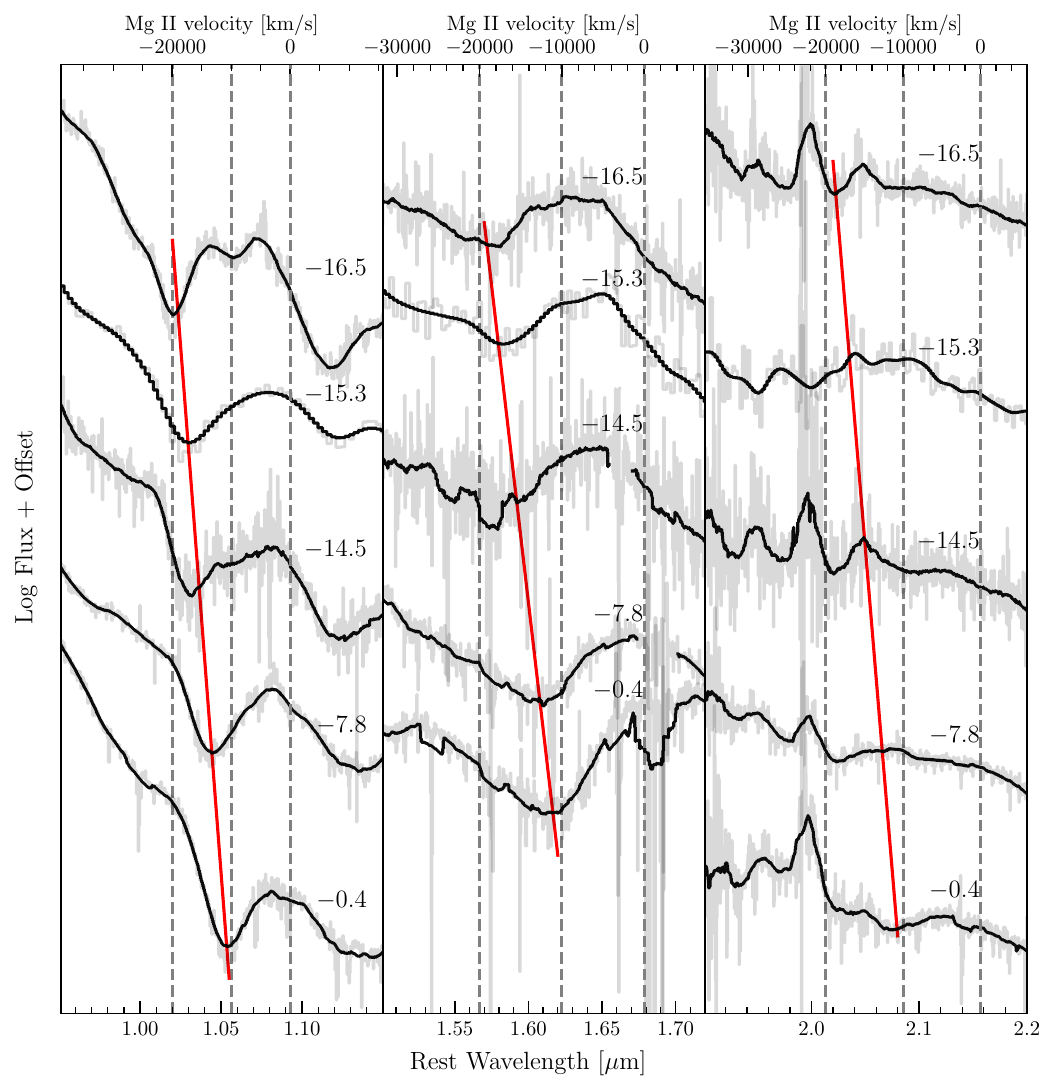} \\
    \caption{
        The velocity evolution of \MgII\ NIR spectral features in SN~2024epr. The fiducial red line is to guide the eye along the velocity evolution of the \MgII\ features and is not physically meaningful. The dashed grey lines indicate \MgII\ velocities of 20\,000, 10\,000, and 0~km~s$^{-1}$, respectively.
    }
    \label{fig:NIR_Mg_vel}
\end{figure*}

Similarly to Figure \ref{fig:NIR_He_vel}, Figure \ref{fig:NIR_Mg_vel} shows the velocity evolution of prominent \MgII\ features (the location of \MgII\ was determined by \citealt{Hsiao13, Marion15, Hsiao15} using \verb|SYNAPPS|; \citealp{Thomas11_SYNAPPS}). Mg traces the extent of explosive carbon burning, and the epoch where the Mg velocity becomes constant in time indicates the boundary in velocity space between explosive C and explosive O burning \citep{Wheeler98}.

In SN~2024epr, \MgII\ evolves from $\sim$20\,000~km~s$^{-1}$ shortly after the explosion to $\sim$10\,000~km~s$^{-1}$ near peak.  
The delayed detonation models presented in \citet{Wheeler98} have the C/O boundary at 15\,000 to 16\,000~km~s$^{-1}$, which is higher than the velocity seen in SN~2024epr, especially given that our spectral time series stops at peak where the \MgII\ velocities may still be slowing.
This suggests that the boundary between explosive C and explosive O burning is deeper in the ejecta than the \citet{Wheeler98} delayed detonation models predict. Variations in transition density,  metallically, and C/O ratio in delayed detonation models are able to produce Mg velocities more similar to those seen in SN~2024epr \citep{Hoeflich17, Cain18}.

\subsubsection{Comparison to Other SNe~Ia}\label{sec:nir-comp}

\begin{figure*}
    \centering
    \includegraphics{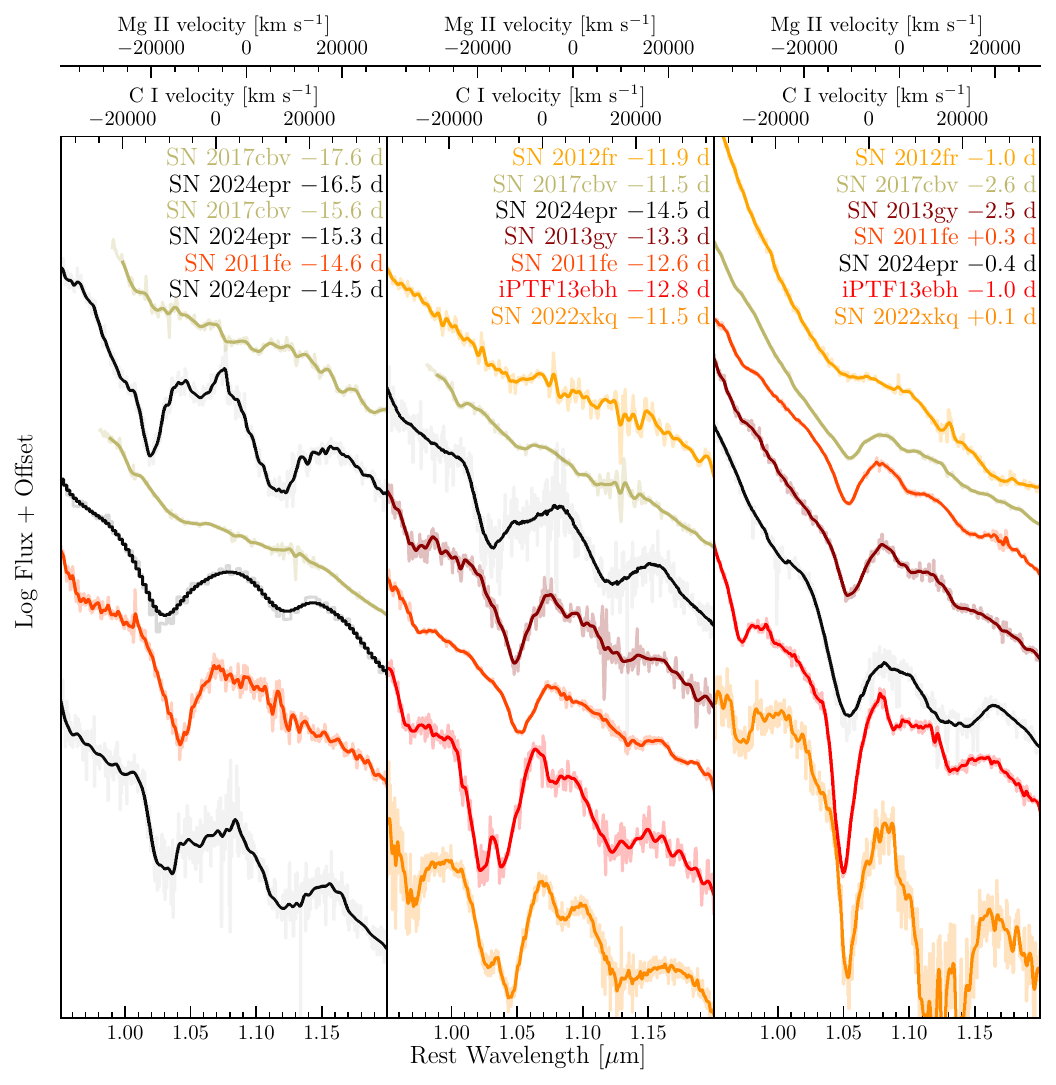} \\
    \caption{
        Comparison of other early-time NIR spectra to other SNe~Ia with early-time NIR spectra.
    }
    \label{fig:NIR_comp}
\end{figure*}

We compare our NIR spectra of SN~2024epr to SN~2011fe \citep{Hsiao13}, SN~2012fr \citep{Lu23}, iPTF13ebh \citep{Hsiao15}, SN~2013gy \citep{Lu23}, SN~2017cbv \citep{Wang20}, and SN~2022xkq \citep{Pearson24} in Figure \ref{fig:NIR_comp}. Our comparison sample consists of all SNe~Ia whose NIR spectral time series begins 10~days before peak light and contains at least two epochs. 

In the earliest SNe~Ia NIR spectra, we see three different NIR behaviors: SN~2017cbv is a featureless continuum, SN~2011fe has only \MgII, and SN~2024epr has \MgII\ and 1.06, and 1.12~$\mu$m features. The underlying physics driving the diversity in the early-time NIR spectral features is unclear, and a larger observational sample is needed.

There is tentative evidence that the presence of features in the NIR correlates with the luminosity of the SNe~Ia themselves. The most luminous SNe in our comparison sample, SNe~2012fr and 2017cbv, have the fewest absorption features between $-14$~to~$-10$~days. The other SNe~Ia show at least one strong absorption feature, \MgII\ 1.0972~$\mu$m. This \MgII\ feature is stronger in the subluminous iPTF13ebh and SN 2022xkq than in the normal SNe 2011fe and 2013gy. Additionally, the subluminous SNe~Ia have a second absorption feature to the blue, which is claimed to be unburnt C \citep{Hsiao15, Pearson24}, and an additional \MgII\ feature observed at $\sim$1.12~$\mu$m; we also claim these features are present in SN~2024epr. This trend may arise from composition or excitation differences.  

The $\sim$1.12~$\mu$m \MgII\ feature strength also appears to depend on luminosity. This feature is seen throughout the spectral time series of iPTF13ebh, SN~2022xkq, and SN~2024epr, and it is present at maximum light, albeit with differing strengths and width, in the aforementioned SNe~Ia and SNe~2011fe and 2013gy. Both SNe~2022xkq and 2024epr have a single feature centered roughly at $\sim$1.12~$\mu$m, whereas iPTF13ebh has two absorption troughs, similar to the weaker troughs potentially seen in SNe 2011fe and 2013gy. SN~2017cbv shows a weak \MgII\ feature, whereas SN~2012fr is still featureless, even at maximum light. This trend has several potential causes, including ionization differences, variations in the outer layers from differences in the progenitor C to O ratio or metallicity, or a difference in the explosion mechanism.  

Finally, the ``notch'' feature at $\sim$1.07~$\mu$m seen in iPTF13ebh, SN~2022xkq, and SN~2024epr at earlier times may be \FeIII, which would suggest the presence of ionizing photons, and thus plausibly \Nifs\ as well, in the outermost ejecta layers. Relative to the time of the explosion, the notch appears within approximately the first two~days. Fe-group elements in the outermost layers of the ejecta could arise from either enhanced \Nifs\ mixing or the presence of surface \Nifs\ \citep[e.g.,][]{Magee20a, Magee20b}. While not confirmed, understanding the origins of this notch feature may help constrain the extent of mixing or surface burning in SNe~Ia. One caveat to this is that a high-metallicity progenitor could also have primordial Fe that produces this feature (and may naturally also explain the redder colors), and its short lifetime could be due to an ionization change rather than a surface-heavy distribution. 

\section{Conclusions}\label{sec:conclusion}
Our observations of SN~2024epr are some of the earliest-ever SN~Ia observations, especially in the NIR. When measured in days before $B$-band maximum, only the 0.9–1.8~$\mu$m spectrum of SN~2017cbv from \citet{Wang20} is earlier than our first 0.8-2.5~$\mu$m spectrum of SN~2024epr. Such early-time NIR spectroscopic time series data within a couple of days of the explosion exist for only a few SNe~Ia \citep{Hsiao13, Hsiao15, Wang20, Pearson24} and are strong diagnostics of the underlying physics \citep[see, e.g.,][]{Hsiao19}.

The early-time optical spectra show high velocity \SiII\ $\lambda$6355 and \CaII\ NIR features at speeds of $\sim$0.1$c$, yet there are only photospheric velocity features by maximum light. SN~2024epr is in the ``cool'' region of the \citet{Branch06} classification scheme and has the strongest \CaII\ NIR pEW in our comparison sample.

 The early-time NIR spectra show strong Mg features at a velocity of $\sim$20\,000~km~s$^{-1}$, and a fast-evolving \CI\ feature at $\sim$23\,000~km~s$^{-1}$. Despite obtaining these spectra two days after the estimated time of first light, we do not find any evidence for He features in our spectral time series. Compared to other SNe~Ia with early-time NIR spectra, SN~2024epr is normal, albeit with higher Mg velocities. Generally, there is growing evidence for a diversity of early-time NIR spectra, and more early-time NIR spectra will improve our understanding of the observational and, ultimately, physical differences between SNe~Ia. 

The transitional phase NIR spectra show clear Fe group elements in emission (\FeII\ and \CoII), and we tentatively propose \SiIII\ to explain the feature at 1.25~$\mu$m. We also identify what may be \CaII\ and \MgII\ features in the transitional phase spectra, suggesting that inner layers of the ejecta also undergo incomplete burning. Lastly, these later time spectra do not show evidence for H or He from a non-degenerate companion or surrounding H- or He-rich circumstellar medium. 

Because our photometric coverage begins promptly after discovery, we searched for excess flux in the rising light curve of SN~2024epr. We do not find evidence for excess flux in SN~2024epr, ruling out thick He shells, which may produce excess flux at early times. We also compare the Pan-STARRS photometry to model He-detonation light curves from \citet{Polin19}, and we find that no single model fits the data well. In particular, the $i$ band, which hosts the extremely strong \CaII\ NIR feature, tends to be fainter than the model predicts. The best-fitting double detonation models favor smaller He shells, which do not produce as much early-excess flux as larger He shells. Thin shell He donations do not burn completely to Fe group elements; instead, they produce intermediate mass elements in the outermost layers consistent with the early-time spectra of SN~2024epr. Alternatively, it may be plausible that a higher transition density between deflagration and detonation may also produce such extremely high-velocity features in a delayed detonation explosion. 

Finally, we find several other SNe~Ia (namely, SNe 2009ig, 2012fr, 2020jgl, and 2021aefx) also have high velocity ($v\approx30\,000$~km~s$^{-1}$) intermediate mass elements shortly after explosion. Given the paucity of SNe~Ia with prompt spectroscopic observations after the explosion and that these objects all evolve to be consistent with normal SNe~Ia at peak light, a significant fraction of SNe~Ia potentially may have unobserved high-velocity features at early times, which require prompt discovery and rapid spectroscopic classification to reveal. 

We find evidence for several trends in the early-time behavior of SNe~Ia. First, SNe~Ia with high-velocity features at early times generally evolve to be normal SNe~Ia at peak. Second, the substantial NIR diversity between SNe~Ia at early times is also seen at maximum light, contrasting the optical where most early-time differences between SNe~Ia become less stark near maximum light. 
This work highlights the utility of early-time optical and NIR observations to probe the composition of the outermost ejecta layers most sensitive to differences in the progenitor scenario and/or explosion mechanism, making it essential to grow the sample of SNe~Ia with early-time data to understand their explosive origins better.

\section*{Acknowledgments}
We thank the anonymous referee for providing helpful feedback that improved this manuscript. 
We thank Sahana Kumar, Jing Lu, Abigail Polin, Ken Shen, Eddie Baron, and Peter Hoeflich for their helpful discussions. 
This material is based upon work supported by the National Science Foundation Graduate Research Fellowship Program under Grant Nos.\ 1842402 and 2236415. Any opinions, findings, conclusions, or recommendations expressed in this material are those of the authors and do not necessarily reflect the views of the National Science Foundation.
D.O.J.\ acknowledges support from NSF grants AST-2407632 and AST-2429450, NASA grant 80NSSC24M0023, and HST/JWST grants HST-GO-17128.028, HST-GO-16269.012, and JWST-GO-05324.031, awarded by the Space Telescope Science Institute (STScI), which is operated by the Association of Universities for Research in Astronomy, Inc., for NASA, under contract NAS5-26555.
The Shappee group at the University of Hawai`i is supported with funds from NSF (grants AST-1908952, AST-1911074, \& AST-1920392) and NASA (grants HST-GO-17087, 80NSSC24K0521, 80NSSC24K0490, 80NSSC24K0508, 80NSSC23K0058, \& 80NSSC23K1431).
The UCSC team is supported in part by NASA grants 80NSSC23K0301 and 80NSSC24K1411; and a fellowship from the David and Lucile Packard Foundation to R.J.F.
A.D.\ is supported by the European Research Council (ERC) under the European Union’s Horizon 2020 research and innovation programme (Grant Agreement No.\ 101002652).
A.R.W.\ is supported by funding through AST-2421845 and funding from the Simons Foundation for the NSF-Simons SkAI Institute, the LSST Discovery Alliance through the LINCC Incubator Program and the Catalyst Program, supported in part by Wasabi Technologies, the US DOE through the Department of Physics at the University of Illinois, Urbana-Champaign (\# 13771275), and the 2024 DOE SCGSCR program and Lawrence Berkeley National Laboratory.
This work is supported by the National Science Foundation under Cooperative Agreement PHY-2019786 (The NSF AI Institute for Artificial Intelligence and Fundamental Interactions, http://iaifi.org/).

This work was supported by National Aeronautics and Space Administration (NASA) Keck PI Data Awards, administered by the NASA Exoplanet Science Institute. 

Observations from coauthor S.A.S.\ were made with undergraduates in the spring 2024 Astronomy 331 course at Sonoma State University. These observations were made under the aegis of the ASTRAL (Astronomy/STEM Alliance with Lick) consortium, supported by a generous grant from the Gordon and Betty Moore Foundation (PI: B.\ Macintosh).

The Young Supernova Experiment (YSE) and its research infrastructure is supported by the European Research Council under the European Union's Horizon 2020 research and innovation programme (ERC Grant Agreement 101002652, PI K.\ Mandel), the Heising-Simons Foundation (2018-0913, PI R.\ Foley; 2018-0911, PI R.\ Margutti), NASA (NNG17PX03C, PI R.\ Foley), NSF (AST--1720756, AST--1815935, PI R.\ Foley; AST--1909796, AST-1944985, PI R.\ Margutti), the David \& Lucille Packard Foundation (PI R.\ Foley), VILLUM FONDEN (project 16599, PI J.\ Hjorth), and the Center for AstroPhysical Surveys (CAPS) at the National Center for Supercomputing Applications (NCSA) and the University of Illinois Urbana-Champaign.
Pan-STARRS is a project of the Institute for Astronomy of the University of Hawaii, and is supported by the NASA SSO Near Earth Observation Program under grants 80NSSC18K0971, NNX14AM74G, NNX12AR65G, NNX13AQ47G, NNX08AR22G, 80NSSC21K1572, and by the State of Hawaii.  The Pan-STARRS1 Surveys (PS1) and the PS1 public science archive have been made possible through contributions by the Institute for Astronomy, the University of Hawaii, the Pan-STARRS Project Office, the Max-Planck Society and its participating institutes, the Max Planck Institute for Astronomy, Heidelberg and the Max Planck Institute for Extraterrestrial Physics, Garching, The Johns Hopkins University, Durham University, the University of Edinburgh, the Queen's University Belfast, the Harvard-Smithsonian Center for Astrophysics, the Las Cumbres Observatory Global Telescope Network Incorporated, the National Central University of Taiwan, STScI, NASA under grant NNX08AR22G issued through the Planetary Science Division of the NASA Science Mission Directorate, NSF grant AST-1238877, the University of Maryland, Eotvos Lorand University (ELTE), the Los Alamos National Laboratory, and the Gordon and Betty Moore Foundation.

Based on observations obtained at the international Gemini Observatory, a program of NSF NOIRLab, which is managed by the Association of Universities for Research in Astronomy (AURA) under a cooperative agreement with the U.S.\ National Science Foundation on behalf of the Gemini Observatory partnership: the U.S.\ National Science Foundation (United States), National Research Council (Canada), Agencia Nacional de Investigaci\'{o}n y Desarrollo (Chile), Ministerio de Ciencia, Tecnolog\'{i}a e Innovaci\'{o}n (Argentina), Minist\'{e}rio da Ci\^{e}ncia, Tecnologia, Inova\c{c}\~{o}es e Comunica\c{c}\~{o}es (Brazil), and Korea Astronomy and Space Science Institute (Republic of Korea). Data were taken by programs GN-2024A-Q-136, GN-2024A-Q-226, and GN-2024B-Q-109. This work was enabled by observations made from the Gemini North telescope, located within the Maunakea Science Reserve and adjacent to the summit of Maunakea. We are grateful for the privilege of observing the Universe from a place that is unique in both its astronomical quality and its cultural significance.

Some of the data presented herein were obtained at Keck Observatory, which is a private 501(c)3 non-profit organization operated as a scientific partnership among the California Institute of Technology, the University of California, and the National Aeronautics and Space Administration. The Observatory was made possible by the generous financial support of the W.\ M.\ Keck Foundation.
The authors wish to recognize and acknowledge the very significant cultural role and reverence that the summit of Maunakea has always had within the indigenous Hawaiian community.  We are most fortunate to have the opportunity to conduct observations from this mountain.

The Infrared Telescope Facility, which is operated by the University of Hawaii under contract 80HQTR24DA010 with the National Aeronautics and Space Administration.

A major upgrade of the Kast spectrograph on the Shane 3~m telescope at Lick Observatory was made possible through generous gifts from the Heising-Simons Foundation as well as William and Marina Kast. Research at Lick Observatory is partially supported by a generous gift from Google.

UKIRT is owned by the University of Hawaii (UH) and operated by the UH Institute for Astronomy.

This work has made use of data from the Asteroid Terrestrial-impact Last Alert System (ATLAS) project. The Asteroid Terrestrial-impact Last Alert System (ATLAS) project is primarily funded to search for near earth asteroids through NASA grants NN12AR55G, 80NSSC18K0284, and 80NSSC18K1575; byproducts of the NEO search include images and catalogs from the survey area. This work was partially funded by Kepler/K2 grant J1944/80NSSC19K0112 and HST GO-15889, and STFC grants ST/T000198/1 and ST/S006109/1. The ATLAS science products have been made possible through the contributions of the University of Hawaii Institute for Astronomy, the Queen’s University Belfast, the Space Telescope Science Institute, the South African Astronomical Observatory, and The Millennium Institute of Astrophysics (MAS), Chile.

Based on observations obtained with the Samuel Oschin Telescope 48-inch and the 60-inch Telescope at the Palomar Observatory as part of the Zwicky Transient Facility project. ZTF is supported by the National Science Foundation under Grant Nos. AST-1440341, AST-2034437, and a collaboration including Caltech, IPAC, the Weizmann Institute for Science, the Oskar Klein Center at Stockholm University, the University of Maryland, the University of Washington, Deutsches Elektronen-Synchrotron and Humboldt University, the TANGO Consortium of Taiwan, the University of Wisconsin at Milwaukee, Trinity College Dublin, Lawrence Livermore National Laboratories, and IN2P3, France. Operations are conducted by COO, IPAC, and UW.

YSE-PZ was developed by the UC Santa Cruz Transients Team, supported in part by NASA grants NNG17PX03C, 80NSSC19K1386, and 80NSSC20K0953; NSF grants AST--1518052, AST--1815935, and AST--1911206; the Gordon \& Betty Moore Foundation; the Heising-Simons Foundation; a fellowship from the David and Lucile Packard Foundation to R.\ J.\ Foley; Gordon and Betty Moore Foundation postdoctoral fellowships and a NASA Einstein fellowship, as administered through the NASA Hubble Fellowship program and grant HST-HF2-51462.001, to D.\ O.\ Jones; and a National Science Foundation Graduate Research Fellowship, administered through grant No.\ DGE-1339067, to D.\ A.\ Coulter.

\newpage
\bibliography{references}

@ARTICLE{Hoogendam24b,
       author = {{Hoogendam}, W.~B. and {Hinkle}, J.~T. and {Shappee}, B.~J. and {Auchettl}, K. and {Kochanek}, C.~S. and {Stanek}, K.~Z. and {Maksym}, W.~P. and {Tucker}, M.~A. and {Huber}, M.~E. and {Morrell}, N. and {Burns}, C.~R. and {Hey}, D. and {Holoien}, T.~W. -S. and {Prieto}, J.~L. and {Stritzinger}, M. and {Do}, A. and {Polin}, A. and {Ashall}, C. and {Brown}, P.~J. and {DerKacy}, J.~M. and {Ferrari}, L. and {Galbany}, L. and {Hsiao}, E.~Y. and {Kumar}, S. and {Lu}, J. and {Stevens}, C.~P.},
        title = "{Discovery and follow-up of ASASSN-23bd (AT 2023clx): the lowest redshift and luminosity optically selected tidal disruption event}",
      journal = {\mnras},
         year = 2024,
        month = jun,
       volume = {530},
       number = {4},
        pages = {4501-4518},
          doi = {10.1093/mnras/stae1121},
archivePrefix = {arXiv},
       eprint = {2401.05490},
 primaryClass = {astro-ph.HE},
       adsurl = {https://ui.adsabs.harvard.edu/abs/2024MNRAS.530.4501H}
}

@ARTICLE{Hoogendam25b,
       author = {{Hoogendam}, W.~B. and {Ashall}, C. and {Jones}, D.~O. and {Shappee}, B.~J. and {Tucker}, M.~A. and {Huber}, M.~E. and {Auchettl}, K. and {Desai}, D.~D. and {Do}, A. and {Hinkle}, J.~T. and {Kong}, M.~Y. and {Romagnoli}, S. and {Shi}, J. and {Syncatto}, A. and {Kilpatrick}, C.~D.},
        title = "{Early and Extensive Ultraviolet through Near Infrared Observations of the Intermediate-luminosity Type Iax Supernovae 2024pxl}",
      journal = {\apj},
         year = 2025,
        month = aug,
       volume = {988},
       number = {2},
          eid = {209},
        pages = {209},
          doi = {10.3847/1538-4357/ade787},
       adsurl = {https://ui.adsabs.harvard.edu/abs/2025ApJ...988..209H}
}

@ARTICLE{Shingles21,
       author = {{Shingles}, L. and {Smith}, K.~W. and {Young}, D.~R. and {Smartt}, S.~J. and {Tonry}, J. and {Denneau}, L. and {Heinze}, A. and {Weiland}, H. and {Flewelling}, H. and {Stalder}, B. and {Clocchiatti}, A. and {F{\"o}rster}, F. and {Pignata}, G. and {Rest}, A. and {Anderson}, J. and {Stubbs}, C. and {Erasmus}, N.},
        title = "{Release of the ATLAS Forced Photometry server for public use}",
      journal = {Transient Name Server AstroNote},
         year = 2021,
        month = jan,
       volume = {7},
        pages = {1-7},
       adsurl = {https://ui.adsabs.harvard.edu/abs/2021TNSAN...7....1S}
}

@ARTICLE{Muller-Bravo25,
       author = {{M{\"u}ller-Bravo}, T.~E. and {Galbany}, L. and {Stritzinger}, M.~D. and {Ashall}, C. and {Baron}, E. and {Burns}, C.~R. and {H{\"o}flich}, P. and {Morrell}, N. and {Phillips}, M. and {Suntzeff}, N.~B. and {Uddin}, S.~A.},
        title = "{Analyzing type Ia supernovae near-infrared light curves with Principal Component Analysis}",
      journal = {arXiv e-prints},
         year = 2025,
        month = apr,
          eid = {arXiv:2504.05856},
        pages = {arXiv:2504.05856},
          doi = {10.48550/arXiv.2504.05856},
archivePrefix = {arXiv},
       eprint = {2504.05856},
 primaryClass = {astro-ph.SR},
       adsurl = {https://ui.adsabs.harvard.edu/abs/2025arXiv250405856M}
}

@ARTICLE{Ye24,
       author = {{Ye}, Christine and {Jones}, David O. and {Hoogendam}, Willem B. and {Shappee}, Benjamin J. and {Dhawan}, Suhail and {Sharief}, Sammy N.},
        title = "{Searching for Bumps in the Cosmological Road: Do Type Ia Supernovae with Early Excesses Have Biased Hubble Residuals?}",
      journal = {\apj},
         year = 2024,
        month = oct,
       volume = {974},
       number = {2},
          eid = {164},
        pages = {164},
          doi = {10.3847/1538-4357/ad6c3d},
archivePrefix = {arXiv},
       eprint = {2401.02926},
 primaryClass = {astro-ph.HE},
       adsurl = {https://ui.adsabs.harvard.edu/abs/2024ApJ...974..164Y}
}

@ARTICLE{Harvey25,
       author = {{Harvey}, L. and {Maguire}, K. and {Burgaz}, U. and {Dimitriadis}, G. and {Sollerman}, J. and {Goobar}, A. and {Johansson}, J. and {Nordin}, J. and {Rigault}, M. and {Smith}, M. and {Aubert}, M. and {Cartier}, R. and {Chen}, P. and {Deckers}, M. and {Dhawan}, S. and {Galbany}, L. and {Ginolin}, M. and {Kenworthy}, W.~D. and {Kim}, Y. -L. and {Liu}, C. and {Miller}, A.~A. and {Rosnet}, P. and {Senzel}, R. and {Terwel}, J.~H. and {Tomasella}, L. and {Kasliwal}, M. and {Laher}, R.~R. and {Purdum}, J. and {Rusholme}, B. and {Smith}, R.},
        title = "{ZTF SN Ia DR2: High-velocity components in the Si II $\lambda$6355}",
      journal = {arXiv e-prints},
         year = 2025,
        month = feb,
          eid = {arXiv:2502.04448},
        pages = {arXiv:2502.04448},
          doi = {10.48550/arXiv.2502.04448},
archivePrefix = {arXiv},
       eprint = {2502.04448},
 primaryClass = {astro-ph.HE},
       adsurl = {https://ui.adsabs.harvard.edu/abs/2025arXiv250204448H}
}

@ARTICLE{Cain18,
       author = {{Cain}, Christopher and {Baron}, E. and {Phillips}, M.~M. and {Contreras}, Carlos and {Ashall}, Chris and {Stritzinger}, Maximilian D. and {Burns}, Christopher R. and {Piro}, Anthony L. and {Hsiao}, Eric Y. and {Hoeflich}, P. and {Krisciunas}, Kevin and {Suntzeff}, Nicholas B.},
        title = "{Investigating the Unusual Spectroscopic Time Evolution in SN 2012fr}",
      journal = {\apj},
         year = 2018,
        month = dec,
       volume = {869},
       number = {2},
          eid = {162},
        pages = {162},
          doi = {10.3847/1538-4357/aaef34},
archivePrefix = {arXiv},
       eprint = {1810.01149},
 primaryClass = {astro-ph.SR},
       adsurl = {https://ui.adsabs.harvard.edu/abs/2018ApJ...869..162C}
}

@ARTICLE{OHora24,
       author = {{O'Hora}, J. and {Ashall}, C. and {Shahbandeh}, M. and {Hsiao}, E. and {Hoeflich}, P. and {Stritzinger}, M.~D. and {Galbany}, L. and {Baron}, E. and {DerKacy}, J. and {Kumar}, S. and {Lu}, J. and {Medler}, K. and {Shappee}, B.},
        title = "{Using nebular near-IR spectroscopy to measure asymmetric chemical distributions in 2003fg-like thermonuclear supernovae}",
      journal = {arXiv e-prints},
         year = 2024,
        month = dec,
          eid = {arXiv:2412.09352},
        pages = {arXiv:2412.09352},
          doi = {10.48550/arXiv.2412.09352},
archivePrefix = {arXiv},
       eprint = {2412.09352},
 primaryClass = {astro-ph.SR},
       adsurl = {https://ui.adsabs.harvard.edu/abs/2024arXiv241209352O}
}

@ARTICLE{Hoeflich21,
       author = {{Hoeflich}, P. and {Ashall}, C. and {Bose}, S. and {Baron}, E. and {Stritzinger}, M.~D. and {Davis}, S. and {Shahbandeh}, M. and {Anand}, G.~S. and {Baade}, D. and {Burns}, C.~R. and {Collins}, D.~C. and {Diamond}, T.~R. and {Fisher}, A. and {Galbany}, L. and {Hristov}, B.~A. and {Hsiao}, E.~Y. and {Phillips}, M.~M. and {Shappee}, B. and {Suntzeff}, N.~B. and {Tucker}, M.},
        title = "{Measuring an Off-center Detonation through Infrared Line Profiles: The Peculiar Type Ia Supernova SN 2020qxp/ASASSN-20jq}",
      journal = {\apj},
         year = 2021,
        month = dec,
       volume = {922},
       number = {2},
          eid = {186},
        pages = {186},
          doi = {10.3847/1538-4357/ac250d},
archivePrefix = {arXiv},
       eprint = {2109.03359},
 primaryClass = {astro-ph.SR},
       adsurl = {https://ui.adsabs.harvard.edu/abs/2021ApJ...922..186H}
}

@article{PypeIt_JOSS,
    doi = {10.21105/joss.02308},
    url = {https://doi.org/10.21105/joss.02308},
    year = {2020},
    publisher = {The Open Journal},
    volume = {5},
    number = {56},
    pages = {2308},
    author = {J. Xavier Prochaska and Joseph F. Hennawi and Kyle B. Westfall and Ryan J. Cooke and Feige Wang and Tiffany Hsyu and Frederick B. Davies and Emanuele Paolo Farina and Debora Pelliccia},
    title = {PypeIt: The Python Spectroscopic Data Reduction Pipeline},
    journal = {Journal of Open Source Software}
}

@ARTICLE{Galbany25,
       author = {{Galbany}, Llu{\'\i}s and {Guti{\'e}rrez}, Claudia P. and {Piscarreta}, Lara and {Alburai}, Alaa and {Ali}, Noor and {Cross}, Dane and {Gonz{\'a}lez-Ba{\~n}uelos}, Maider and {Jim{\'e}nez-Palau}, Cristina and {Kopsacheili}, Maria and {M{\"u}ller-Bravo}, Tom{\'a}s E. and {Phan}, Kim and {Sanfeliu}, Ramon and {Stritzinger}, Maximillian and {Ashall}, Chris and {Baron}, Eddie and {Folatelli}, Gast{\'o}n and {Hoogendam}, Willem and {Jha}, Saurabh and {de Jaeger}, Thomas and {Brink}, Thomas G. and {Filippenko}, Alexei V. and {Howell}, D. Andrew and {Hiramatsu}, Daichi},
        title = "{Rapid follow-up of infant supernovae with the Gran Telescopio de Canarias}",
      journal = {arXiv e-prints},
         year = 2025,
        month = jan,
          eid = {arXiv:2501.19108},
        pages = {arXiv:2501.19108},
archivePrefix = {arXiv},
       eprint = {2501.19108},
 primaryClass = {astro-ph.SR},
       adsurl = {https://ui.adsabs.harvard.edu/abs/2025arXiv250119108G}
}

@ARTICLE{Seitenzahl13,
       author = {{Seitenzahl}, Ivo R. and {Ciaraldi-Schoolmann}, Franco and {R{\"o}pke}, Friedrich K. and {Fink}, Michael and {Hillebrandt}, Wolfgang and {Kromer}, Markus and {Pakmor}, R{\"u}diger and {Ruiter}, Ashley J. and {Sim}, Stuart A. and {Taubenberger}, Stefan},
        title = "{Three-dimensional delayed-detonation models with nucleosynthesis for Type Ia supernovae}",
      journal = {\mnras},
         year = 2013,
        month = feb,
       volume = {429},
       number = {2},
        pages = {1156-1172},
          doi = {10.1093/mnras/sts402},
archivePrefix = {arXiv},
       eprint = {1211.3015},
 primaryClass = {astro-ph.SR},
       adsurl = {https://ui.adsabs.harvard.edu/abs/2013MNRAS.429.1156S}
}

@ARTICLE{Fink14,
       author = {{Fink}, Michael and {Kromer}, Markus and {Seitenzahl}, Ivo R. and {Ciaraldi-Schoolmann}, Franco and {R{\"o}pke}, Friedrich K. and {Sim}, Stuart A. and {Pakmor}, R{\"u}diger and {Ruiter}, Ashley J. and {Hillebrandt}, Wolfgang},
        title = "{Three-dimensional pure deflagration models with nucleosynthesis and synthetic observables for Type Ia supernovae}",
      journal = {\mnras},
         year = 2014,
        month = feb,
       volume = {438},
       number = {2},
        pages = {1762-1783},
          doi = {10.1093/mnras/stt2315},
archivePrefix = {arXiv},
       eprint = {1308.3257},
 primaryClass = {astro-ph.SR},
       adsurl = {https://ui.adsabs.harvard.edu/abs/2014MNRAS.438.1762F}
}

@ARTICLE{Shen10,
       author = {{Shen}, Ken J. and {Kasen}, Dan and {Weinberg}, Nevin N. and {Bildsten}, Lars and {Scannapieco}, Evan},
        title = "{Thermonuclear .Ia Supernovae from Helium Shell Detonations: Explosion Models and Observables}",
      journal = {\apj},
         year = 2010,
        month = jun,
       volume = {715},
       number = {2},
        pages = {767-774},
          doi = {10.1088/0004-637X/715/2/767},
archivePrefix = {arXiv},
       eprint = {1002.2258},
 primaryClass = {astro-ph.HE},
       adsurl = {https://ui.adsabs.harvard.edu/abs/2010ApJ...715..767S}
}

@ARTICLE{Hoeflich95,
       author = {{Hoeflich}, P. and {Khokhlov}, A.~M. and {Wheeler}, J.~C.},
        title = "{Delayed Detonation Models for Normal and Subluminous Type IA Supernovae: Absolute Brightness, Light Curves, and Molecule Formation}",
      journal = {\apj},
         year = 1995,
        month = may,
       volume = {444},
        pages = {831},
          doi = {10.1086/175656},
       adsurl = {https://ui.adsabs.harvard.edu/abs/1995ApJ...444..831H}
}

@ARTICLE{Foley18,
       author = {{Foley}, Ryan J. and {Scolnic}, Daniel and {Rest}, Armin and {Jha}, S.~W. and {Pan}, Y. -C. and {Riess}, A.~G. and {Challis}, P. and {Chambers}, K.~C. and {Coulter}, D.~A. and {Dettman}, K.~G. and {Foley}, M.~M. and {Fox}, O.~D. and {Huber}, M.~E. and {Jones}, D.~O. and {Kilpatrick}, C.~D. and {Kirshner}, R.~P. and {Schultz}, A.~S.~B. and {Siebert}, M.~R. and {Flewelling}, H.~A. and {Gibson}, B. and {Magnier}, E.~A. and {Miller}, J.~A. and {Primak}, N. and {Smartt}, S.~J. and {Smith}, K.~W. and {Wainscoat}, R.~J. and {Waters}, C. and {Willman}, M.},
        title = "{The Foundation Supernova Survey: motivation, design, implementation, and first data release}",
      journal = {\mnras},
         year = 2018,
        month = mar,
       volume = {475},
       number = {1},
        pages = {193-219},
          doi = {10.1093/mnras/stx3136},
archivePrefix = {arXiv},
       eprint = {1711.02474},
 primaryClass = {astro-ph.HE},
       adsurl = {https://ui.adsabs.harvard.edu/abs/2018MNRAS.475..193F}
}

@ARTICLE{Thielemann86,
       author = {{Thielemann}, F. -K. and {Nomoto}, K. and {Yokoi}, K.},
        title = "{Explosive nucleosynthesis in carbon deflagration models of Type I supernovae}",
      journal = {\aap},
         year = 1986,
        month = apr,
       volume = {158},
       number = {1-2},
        pages = {17-33},
       adsurl = {https://ui.adsabs.harvard.edu/abs/1986A&A...158...17T}
}

@ARTICLE{Ganeshalingam10,
       author = {{Ganeshalingam}, Mohan and {Li}, Weidong and {Filippenko}, Alexei V. and {Anderson}, Carmen and {Foster}, Griffin and {Gates}, Elinor L. and {Griffith}, Christopher V. and {Grigsby}, Bryant J. and {Joubert}, Niels and {Leja}, Joel and {Lowe}, Thomas B. and {Macomber}, Brent and {Pritchard}, Tyler and {Thrasher}, Patrick and {Winslow}, Dustin},
        title = "{Results of the Lick Observatory Supernova Search Follow-up Photometry Program: BVRI Light Curves of 165 Type Ia Supernovae}",
      journal = {\apjs},
         year = 2010,
        month = oct,
       volume = {190},
       number = {2},
        pages = {418-448},
          doi = {10.1088/0067-0049/190/2/418},
       adsurl = {https://ui.adsabs.harvard.edu/abs/2010ApJS..190..418G}
}

@ARTICLE{Tucker25,
       author = {{Tucker}, Michael A.},
        title = "{Merging white dwarf binaries produce Type Ia supernovae in elliptical galaxies}",
      journal = {\mnras},
         year = 2025,
        month = mar,
       volume = {538},
       number = {1},
        pages = {L1-L8},
          doi = {10.1093/mnrasl/slae121},
archivePrefix = {arXiv},
       eprint = {2408.00840},
 primaryClass = {astro-ph.HE},
       adsurl = {https://ui.adsabs.harvard.edu/abs/2025MNRAS.538L...1T}
}

@ARTICLE{Wang23_SBIpp,
       author = {{Wang}, Bingjie and {Leja}, Joel and {Villar}, V. Ashley and {Speagle}, Joshua S.},
        title = "{SBI$^{++}$: Flexible, Ultra-fast Likelihood-free Inference Customized for Astronomical Applications}",
      journal = {\apjl},
         year = 2023,
        month = jul,
       volume = {952},
       number = {1},
          eid = {L10},
        pages = {L10},
          doi = {10.3847/2041-8213/ace361},
archivePrefix = {arXiv},
       eprint = {2304.05281},
 primaryClass = {astro-ph.IM},
       adsurl = {https://ui.adsabs.harvard.edu/abs/2023ApJ...952L..10W}
}

@manual{Miller94,
       title = "The KAST Double Spectrograph",
       author       = "{Miller}, J.~S. and {Stone}, R.~P.~S.",
       organization = "Lick Observatory Techinical Report",
       edition      = "66",
       month        = sep,
       year         = 1994,
       note         = "Not on ADS",
}

@ARTICLE{AznarSiguan15,
       author = {{Aznar-Sigu{\'a}n}, G. and {Garc{\'\i}a-Berro}, E. and {Lor{\'e}n-Aguilar}, P. and {Soker}, N. and {Kashi}, A.},
        title = "{Smoothed particle hydrodynamics simulations of the core-degenerate scenario for Type Ia supernovae}",
      journal = {\mnras},
         year = 2015,
        month = jul,
       volume = {450},
       number = {3},
        pages = {2948-2962},
          doi = {10.1093/mnras/stv824},
archivePrefix = {arXiv},
       eprint = {1503.02444},
 primaryClass = {astro-ph.HE},
       adsurl = {https://ui.adsabs.harvard.edu/abs/2015MNRAS.450.2948A}
}

@ARTICLE{Ilkov13,
       author = {{Ilkov}, Marjan and {Soker}, Noam},
        title = "{The number of progenitors in the core-degenerate scenario for Type Ia supernovae}",
      journal = {\mnras},
         year = 2013,
        month = jan,
       volume = {428},
       number = {1},
        pages = {579-586},
          doi = {10.1093/mnras/sts053},
archivePrefix = {arXiv},
       eprint = {1208.0953},
 primaryClass = {astro-ph.SR},
       adsurl = {https://ui.adsabs.harvard.edu/abs/2013MNRAS.428..579I}
}

@ARTICLE{Masci19,
       author = {{Masci}, Frank J. and {Laher}, Russ R. and {Rusholme}, Ben and {Shupe}, David L. and {Groom}, Steven and {Surace}, Jason and {Jackson}, Edward and {Monkewitz}, Serge and {Beck}, Ron and {Flynn}, David and {Terek}, Scott and {Landry}, Walter and {Hacopians}, Eugean and {Desai}, Vandana and {Howell}, Justin and {Brooke}, Tim and {Imel}, David and {Wachter}, Stefanie and {Ye}, Quan-Zhi and {Lin}, Hsing-Wen and {Cenko}, S. Bradley and {Cunningham}, Virginia and {Rebbapragada}, Umaa and {Bue}, Brian and {Miller}, Adam A. and {Mahabal}, Ashish and {Bellm}, Eric C. and {Patterson}, Maria T. and {Juri{\'c}}, Mario and {Golkhou}, V. Zach and {Ofek}, Eran O. and {Walters}, Richard and {Graham}, Matthew and {Kasliwal}, Mansi M. and {Dekany}, Richard G. and {Kupfer}, Thomas and {Burdge}, Kevin and {Cannella}, Christopher B. and {Barlow}, Tom and {Van Sistine}, Angela and {Giomi}, Matteo and {Fremling}, Christoffer and {Blagorodnova}, Nadejda and {Levitan}, David and {Riddle}, Reed and {Smith}, Roger M. and {Helou}, George and {Prince}, Thomas A. and {Kulkarni}, Shrinivas R.},
        title = "{The Zwicky Transient Facility: Data Processing, Products, and Archive}",
      journal = {\pasp},
         year = 2019,
        month = jan,
       volume = {131},
       number = {995},
        pages = {018003},
          doi = {10.1088/1538-3873/aae8ac},
archivePrefix = {arXiv},
       eprint = {1902.01872},
 primaryClass = {astro-ph.IM},
       adsurl = {https://ui.adsabs.harvard.edu/abs/2019PASP..131a8003M}
}

@ARTICLE{York00,
       author = {{York}, Donald G. and {Adelman}, J. and {Anderson}, Jr., John E. and {Anderson}, Scott F. and {Annis}, James and {Bahcall}, Neta A. and {Bakken}, J.~A. and {Barkhouser}, Robert and {Bastian}, Steven and {Berman}, Eileen and {Boroski}, William N. and {Bracker}, Steve and {Briegel}, Charlie and {Briggs}, John W. and {Brinkmann}, J. and {Brunner}, Robert and {Burles}, Scott and {Carey}, Larry and {Carr}, Michael A. and {Castander}, Francisco J. and {Chen}, Bing and {Colestock}, Patrick L. and {Connolly}, A.~J. and {Crocker}, J.~H. and {Csabai}, Istv{\'a}n and {Czarapata}, Paul C. and {Davis}, John Eric and {Doi}, Mamoru and {Dombeck}, Tom and {Eisenstein}, Daniel and {Ellman}, Nancy and {Elms}, Brian R. and {Evans}, Michael L. and {Fan}, Xiaohui and {Federwitz}, Glenn R. and {Fiscelli}, Larry and {Friedman}, Scott and {Frieman}, Joshua A. and {Fukugita}, Masataka and {Gillespie}, Bruce and {Gunn}, James E. and {Gurbani}, Vijay K. and {de Haas}, Ernst and {Haldeman}, Merle and {Harris}, Frederick H. and {Hayes}, J. and {Heckman}, Timothy M. and {Hennessy}, G.~S. and {Hindsley}, Robert B. and {Holm}, Scott and {Holmgren}, Donald J. and {Huang}, Chi-hao and {Hull}, Charles and {Husby}, Don and {Ichikawa}, Shin-Ichi and {Ichikawa}, Takashi and {Ivezi{\'c}}, {\v{Z}}eljko and {Kent}, Stephen and {Kim}, Rita S.~J. and {Kinney}, E. and {Klaene}, Mark and {Kleinman}, A.~N. and {Kleinman}, S. and {Knapp}, G.~R. and {Korienek}, John and {Kron}, Richard G. and {Kunszt}, Peter Z. and {Lamb}, D.~Q. and {Lee}, B. and {Leger}, R. French and {Limmongkol}, Siriluk and {Lindenmeyer}, Carl and {Long}, Daniel C. and {Loomis}, Craig and {Loveday}, Jon and {Lucinio}, Rich and {Lupton}, Robert H. and {MacKinnon}, Bryan and {Mannery}, Edward J. and {Mantsch}, P.~M. and {Margon}, Bruce and {McGehee}, Peregrine and {McKay}, Timothy A. and {Meiksin}, Avery and {Merelli}, Aronne and {Monet}, David G. and {Munn}, Jeffrey A. and {Narayanan}, Vijay K. and {Nash}, Thomas and {Neilsen}, Eric and {Neswold}, Rich and {Newberg}, Heidi Jo and {Nichol}, R.~C. and {Nicinski}, Tom and {Nonino}, Mario and {Okada}, Norio and {Okamura}, Sadanori and {Ostriker}, Jeremiah P. and {Owen}, Russell and {Pauls}, A. George and {Peoples}, John and {Peterson}, R.~L. and {Petravick}, Donald and {Pier}, Jeffrey R. and {Pope}, Adrian and {Pordes}, Ruth and {Prosapio}, Angela and {Rechenmacher}, Ron and {Quinn}, Thomas R. and {Richards}, Gordon T. and {Richmond}, Michael W. and {Rivetta}, Claudio H. and {Rockosi}, Constance M. and {Ruthmansdorfer}, Kurt and {Sandford}, Dale and {Schlegel}, David J. and {Schneider}, Donald P. and {Sekiguchi}, Maki and {Sergey}, Gary and {Shimasaku}, Kazuhiro and {Siegmund}, Walter A. and {Smee}, Stephen and {Smith}, J. Allyn and {Snedden}, S. and {Stone}, R. and {Stoughton}, Chris and {Strauss}, Michael A. and {Stubbs}, Christopher and {SubbaRao}, Mark and {Szalay}, Alexander S. and {Szapudi}, Istvan and {Szokoly}, Gyula P. and {Thakar}, Anirudda R. and {Tremonti}, Christy and {Tucker}, Douglas L. and {Uomoto}, Alan and {Vanden Berk}, Dan and {Vogeley}, Michael S. and {Waddell}, Patrick and {Wang}, Shu-i. and {Watanabe}, Masaru and {Weinberg}, David H. and {Yanny}, Brian and {Yasuda}, Naoki and {SDSS Collaboration}},
        title = "{The Sloan Digital Sky Survey: Technical Summary}",
      journal = {\aj},
         year = 2000,
        month = sep,
       volume = {120},
       number = {3},
        pages = {1579-1587},
          doi = {10.1086/301513},
archivePrefix = {arXiv},
       eprint = {astro-ph/0006396},
 primaryClass = {astro-ph},
       adsurl = {https://ui.adsabs.harvard.edu/abs/2000AJ....120.1579Y}
}

@ARTICLE{Fukugita96,
       author = {{Fukugita}, M. and {Ichikawa}, T. and {Gunn}, J.~E. and {Doi}, M. and {Shimasaku}, K. and {Schneider}, D.~P.},
        title = "{The Sloan Digital Sky Survey Photometric System}",
      journal = {\aj},
         year = 1996,
        month = apr,
       volume = {111},
        pages = {1748},
          doi = {10.1086/117915},
       adsurl = {https://ui.adsabs.harvard.edu/abs/1996AJ....111.1748F}
}

@ARTICLE{Wright10,
       author = {{Wright}, Edward L. and {Eisenhardt}, Peter R.~M. and {Mainzer}, Amy K. and {Ressler}, Michael E. and {Cutri}, Roc M. and {Jarrett}, Thomas and {Kirkpatrick}, J. Davy and {Padgett}, Deborah and {McMillan}, Robert S. and {Skrutskie}, Michael and {Stanford}, S.~A. and {Cohen}, Martin and {Walker}, Russell G. and {Mather}, John C. and {Leisawitz}, David and {Gautier}, III, Thomas N. and {McLean}, Ian and {Benford}, Dominic and {Lonsdale}, Carol J. and {Blain}, Andrew and {Mendez}, Bryan and {Irace}, William R. and {Duval}, Valerie and {Liu}, Fengchuan and {Royer}, Don and {Heinrichsen}, Ingolf and {Howard}, Joan and {Shannon}, Mark and {Kendall}, Martha and {Walsh}, Amy L. and {Larsen}, Mark and {Cardon}, Joel G. and {Schick}, Scott and {Schwalm}, Mark and {Abid}, Mohamed and {Fabinsky}, Beth and {Naes}, Larry and {Tsai}, Chao-Wei},
        title = "{The Wide-field Infrared Survey Explorer (WISE): Mission Description and Initial On-orbit Performance}",
      journal = {\aj},
         year = 2010,
        month = dec,
       volume = {140},
       number = {6},
        pages = {1868-1881},
          doi = {10.1088/0004-6256/140/6/1868},
archivePrefix = {arXiv},
       eprint = {1008.0031},
 primaryClass = {astro-ph.IM},
       adsurl = {https://ui.adsabs.harvard.edu/abs/2010AJ....140.1868W}
}

@ARTICLE{Skrutskie06,
       author = {{Skrutskie}, M.~F. and {Cutri}, R.~M. and {Stiening}, R. and {Weinberg}, M.~D. and {Schneider}, S. and {Carpenter}, J.~M. and {Beichman}, C. and {Capps}, R. and {Chester}, T. and {Elias}, J. and {Huchra}, J. and {Liebert}, J. and {Lonsdale}, C. and {Monet}, D.~G. and {Price}, S. and {Seitzer}, P. and {Jarrett}, T. and {Kirkpatrick}, J.~D. and {Gizis}, J.~E. and {Howard}, E. and {Evans}, T. and {Fowler}, J. and {Fullmer}, L. and {Hurt}, R. and {Light}, R. and {Kopan}, E.~L. and {Marsh}, K.~A. and {McCallon}, H.~L. and {Tam}, R. and {Van Dyk}, S. and {Wheelock}, S.},
        title = "{The Two Micron All Sky Survey (2MASS)}",
      journal = {\aj},
         year = 2006,
        month = feb,
       volume = {131},
       number = {2},
        pages = {1163-1183},
          doi = {10.1086/498708},
       adsurl = {https://ui.adsabs.harvard.edu/abs/2006AJ....131.1163S}
}

@software{Coulter22_YSEPZ,
       author = {{Coulter}, D. A and {Jones}, D.~O. and {McGill}, P. and {Foley}, R.~J. and {Aleo}, P.~D. and {Bustamante-Rosell}, M.~J. and {Chatterjee}, D. and {Davis}, K.~W. and {Engel}, A. and {Gagliano}, A. and {Jacobson-Gal{\'a}n}, W.~V. and {Kilpatrick}, C.~D. and {Pan}, Y.-C. and {Rojas-Bravo}, C. and {Siebert}, M.~R. and {Taggart}, K.~L. and {Tinyanont}, S. and {Wang}, Q.},
        title = "{YSE-PZ: An Open-source Target and Observation Management System}",
         year = 2022,
        month = nov,
          eid = {10.5281/zenodo.7278430},
          doi = {10.5281/zenodo.7278430},
      version = {v0.3.0},
    publisher = {Zenodo},
       adsurl = {https://ui.adsabs.harvard.edu/abs/2022zndo...7278430C}
}

@ARTICLE{Collins24,
       author = {{Collins}, Christine E. and {Shingles}, Luke J. and {Sim}, Stuart A. and {Callan}, Fionntan P. and {Gronow}, Sabrina and {Hillebrandt}, Wolfgang and {Kromer}, Markus and {Pakmor}, Ruediger and {Roepke}, Friedrich K.},
        title = "{Non-LTE radiative transfer simulations: Improved agreement of the double detonation with normal Type Ia supernovae}",
      journal = {arXiv e-prints},
         year = 2024,
        month = nov,
          eid = {arXiv:2411.11643},
        pages = {arXiv:2411.11643},
          doi = {10.48550/arXiv.2411.11643},
archivePrefix = {arXiv},
       eprint = {2411.11643},
 primaryClass = {astro-ph.SR},
       adsurl = {https://ui.adsabs.harvard.edu/abs/2024arXiv241111643C}
}

@ARTICLE{Fausnaugh21,
       author = {{Fausnaugh}, M.~M. and {Vallely}, P.~J. and {Kochanek}, C.~S. and {Shappee}, B.~J. and {Stanek}, K.~Z. and {Tucker}, M.~A. and {Ricker}, George R. and {Vanderspek}, Roland and {Latham}, David W. and {Seager}, S. and {Winn}, Joshua N. and {Jenkins}, Jon M. and {Berta-Thompson}, Zachory K. and {Daylan}, Tansu and {Doty}, John P. and {F{\H{u}}r{\'e}sz}, G{\'a}bor and {Levine}, Alan M. and {Morris}, Robert and {P{\'a}l}, Andr{\'a}s and {Sha}, Lizhou and {Ting}, Eric B. and {Wohler}, Bill},
        title = "{Early-time Light Curves of Type Ia Supernovae Observed with TESS}",
      journal = {\apj},
         year = 2021,
        month = feb,
       volume = {908},
       number = {1},
          eid = {51},
        pages = {51},
          doi = {10.3847/1538-4357/abcd42},
archivePrefix = {arXiv},
       eprint = {1904.02171},
 primaryClass = {astro-ph.SR},
       adsurl = {https://ui.adsabs.harvard.edu/abs/2021ApJ...908...51F}
}

@INPROCEEDINGS{Gemini_Archive,
       author = {{Hirst}, P. and {Cardenes}, R.},
        title = "{A New Data Archive for Gemini - Fast, Cheap and in the Cloud}",
    booktitle = {Astronomical Data Analysis Software and Systems XXV},
         year = 2017,
       editor = {{Lorente}, N.~P.~F. and {Shortridge}, K. and {Wayth}, R.},
       series = {Astronomical Society of the Pacific Conference Series},
       volume = {512},
        month = dec,
        pages = {53},
       adsurl = {https://ui.adsabs.harvard.edu/abs/2017ASPC..512...53H}
}

@ARTICLE{Boos24a,
       author = {{Boos}, Samuel J. and {Townsley}, Dean M. and {Shen}, Ken J.},
        title = "{Type Ia Supernovae Can Arise from the Detonations of Both Stars in a Double Degenerate Binary}",
      journal = {\apj},
         year = 2024,
        month = sep,
       volume = {972},
       number = {2},
          eid = {200},
        pages = {200},
          doi = {10.3847/1538-4357/ad5da2},
archivePrefix = {arXiv},
       eprint = {2401.08011},
 primaryClass = {astro-ph.HE},
       adsurl = {https://ui.adsabs.harvard.edu/abs/2024ApJ...972..200B}
}

@ARTICLE{Boos24b,
       author = {{Boos}, Samuel J. and {Dessart}, Luc and {Shen}, Ken J. and {Townsley}, Dean M.},
        title = "{Non-LTE Synthetic Observables of a Multidimensional Model of Type Ia Supernovae}",
      journal = {arXiv e-prints},
         year = 2024,
        month = oct,
          eid = {arXiv:2410.22276},
        pages = {arXiv:2410.22276},
          doi = {10.48550/arXiv.2410.22276},
archivePrefix = {arXiv},
       eprint = {2410.22276},
 primaryClass = {astro-ph.HE},
       adsurl = {https://ui.adsabs.harvard.edu/abs/2024arXiv241022276B}
}

@ARTICLE{2024epr_classification,
       author = {{Sollerman}, J.},
        title = "{ZTF Transient Classification Report for 2024-03-21}",
      journal = {Transient Name Server Classification Report},
         year = 2024,
        month = mar,
       volume = {2024-779},
        pages = {1},
       adsurl = {https://ui.adsabs.harvard.edu/abs/2024TNSCR.779....1S}
}

@ARTICLE{2024epr_discovery,
       author = {{Sollerman}, J.},
        title = "{ZTF Transient Discovery Report for 2024-03-19}",
      journal = {Transient Name Server Discovery Report},
         year = 2024,
        month = mar,
       volume = {2024-746},
        pages = {1},
       adsurl = {https://ui.adsabs.harvard.edu/abs/2024TNSTR.746....1S}
}

@ARTICLE{Wang09_HV,
       author = {{Wang}, X. and {Filippenko}, A.~V. and {Ganeshalingam}, M. and {Li}, W. and {Silverman}, J.~M. and {Wang}, L. and {Chornock}, R. and {Foley}, R.~J. and {Gates}, E.~L. and {Macomber}, B. and {Serduke}, F.~J.~D. and {Steele}, T.~N. and {Wong}, D.~S.},
        title = "{Improved Distances to Type Ia Supernovae with Two Spectroscopic Subclasses}",
      journal = {\apjl},
         year = 2009,
        month = jul,
       volume = {699},
       number = {2},
        pages = {L139-L143},
          doi = {10.1088/0004-637X/699/2/L139},
archivePrefix = {arXiv},
       eprint = {0906.1616},
 primaryClass = {astro-ph.CO},
       adsurl = {https://ui.adsabs.harvard.edu/abs/2009ApJ...699L.139W}
}

@ARTICLE{Boos21,
       author = {{Boos}, Samuel J. and {Townsley}, Dean M. and {Shen}, Ken J. and {Caldwell}, Spencer and {Miles}, Broxton J.},
        title = "{Multidimensional Parameter Study of Double Detonation Type Ia Supernovae Originating from Thin Helium Shell White Dwarfs}",
      journal = {\apj},
         year = 2021,
        month = oct,
       volume = {919},
       number = {2},
          eid = {126},
        pages = {126},
          doi = {10.3847/1538-4357/ac07a2},
archivePrefix = {arXiv},
       eprint = {2101.12330},
 primaryClass = {astro-ph.HE},
       adsurl = {https://ui.adsabs.harvard.edu/abs/2021ApJ...919..126B}
}

@ARTICLE{Nomoto84,
       author = {{Nomoto}, K. and {Thielemann}, F. -K. and {Yokoi}, K.},
        title = "{Accreting white dwarf models for type I supernovae. III. Carbon deflagration supernovae.}",
      journal = {\apj},
         year = 1984,
        month = nov,
       volume = {286},
        pages = {644-658},
          doi = {10.1086/162639},
       adsurl = {https://ui.adsabs.harvard.edu/abs/1984ApJ...286..644N}
}

@ARTICLE{Iwamoto99,
       author = {{Iwamoto}, Koichi and {Brachwitz}, Franziska and {Nomoto}, Ken'ICHI and {Kishimoto}, Nobuhiro and {Umeda}, Hideyuki and {Hix}, W. Raphael and {Thielemann}, Friedrich-Karl},
        title = "{Nucleosynthesis in Chandrasekhar Mass Models for Type IA Supernovae and Constraints on Progenitor Systems and Burning-Front Propagation}",
      journal = {\apjs},
         year = 1999,
        month = dec,
       volume = {125},
       number = {2},
        pages = {439-462},
          doi = {10.1086/313278},
archivePrefix = {arXiv},
       eprint = {astro-ph/0002337},
 primaryClass = {astro-ph},
       adsurl = {https://ui.adsabs.harvard.edu/abs/1999ApJS..125..439I}
}

@ARTICLE{Thomas11,
       author = {{Thomas}, R.~C. and {Aldering}, G. and {Antilogus}, P. and {Aragon}, C. and {Bailey}, S. and {Baltay}, C. and {Bongard}, S. and {Buton}, C. and {Canto}, A. and {Childress}, M. and {Chotard}, N. and {Copin}, Y. and {Fakhouri}, H.~K. and {Gangler}, E. and {Hsiao}, E.~Y. and {Kerschhaggl}, M. and {Kowalski}, M. and {Loken}, S. and {Nugent}, P. and {Paech}, K. and {Pain}, R. and {Pecontal}, E. and {Pereira}, R. and {Perlmutter}, S. and {Rabinowitz}, D. and {Rigault}, M. and {Rubin}, D. and {Runge}, K. and {Scalzo}, R. and {Smadja}, G. and {Tao}, C. and {Weaver}, B.~A. and {Wu}, C. and {Brown}, P.~J. and {Milne}, P.~A. and {Nearby Supernova Factory}},
        title = "{Type Ia Supernova Carbon Footprints}",
      journal = {\apj},
         year = 2011,
        month = dec,
       volume = {743},
       number = {1},
          eid = {27},
        pages = {27},
          doi = {10.1088/0004-637X/743/1/27},
archivePrefix = {arXiv},
       eprint = {1109.1312},
 primaryClass = {astro-ph.CO},
       adsurl = {https://ui.adsabs.harvard.edu/abs/2011ApJ...743...27T}
}

@ARTICLE{Coulter23,
       author = {{Coulter}, D.~A. and {Jones}, D.~O. and {McGill}, P. and {Foley}, R.~J. and {Aleo}, P.~D. and {Bustamante-Rosell}, M.~J. and {Chatterjee}, D. and {Davis}, K.~W. and {Dickinson}, C. and {Engel}, A. and {Gagliano}, A. and {Jacobson-Gal{\'a}n}, W.~V. and {Kilpatrick}, C.~D. and {Kutcka}, J. and {Le Saux}, X.~K. and {Malanchev}, K. and {Pan}, Y. -C. and {Qui{\~n}onez}, P.~J. and {Rojas-Bravo}, C. and {Siebert}, M.~R. and {Taggart}, K. and {Tinyanont}, S. and {Wang}, Q.},
        title = "{YSE-PZ: A Transient Survey Management Platform that Empowers the Human-in-the-loop}",
      journal = {\pasp},
         year = 2023,
        month = jun,
       volume = {135},
       number = {1048},
          eid = {064501},
        pages = {064501},
          doi = {10.1088/1538-3873/acd662},
archivePrefix = {arXiv},
       eprint = {2303.02154},
 primaryClass = {astro-ph.IM},
       adsurl = {https://ui.adsabs.harvard.edu/abs/2023PASP..135f4501C}
}

@ARTICLE{Scolnic22,
       author = {{Scolnic}, Dan and {Brout}, Dillon and {Carr}, Anthony and {Riess}, Adam G. and {Davis}, Tamara M. and {Dwomoh}, Arianna and {Jones}, David O. and {Ali}, Noor and {Charvu}, Pranav and {Chen}, Rebecca and {Peterson}, Erik R. and {Popovic}, Brodie and {Rose}, Benjamin M. and {Wood}, Charlotte M. and {Brown}, Peter J. and {Chambers}, Ken and {Coulter}, David A. and {Dettman}, Kyle G. and {Dimitriadis}, Georgios and {Filippenko}, Alexei V. and {Foley}, Ryan J. and {Jha}, Saurabh W. and {Kilpatrick}, Charles D. and {Kirshner}, Robert P. and {Pan}, Yen-Chen and {Rest}, Armin and {Rojas-Bravo}, Cesar and {Siebert}, Matthew R. and {Stahl}, Benjamin E. and {Zheng}, WeiKang},
        title = "{The Pantheon+ Analysis: The Full Data Set and Light-curve Release}",
      journal = {\apj},
         year = 2022,
        month = oct,
       volume = {938},
       number = {2},
          eid = {113},
        pages = {113},
          doi = {10.3847/1538-4357/ac8b7a},
archivePrefix = {arXiv},
       eprint = {2112.03863},
 primaryClass = {astro-ph.CO},
       adsurl = {https://ui.adsabs.harvard.edu/abs/2022ApJ...938..113S}
}

@ARTICLE{Gronow20,
       author = {{Gronow}, Sabrina and {Collins}, Christine and {Ohlmann}, Sebastian T. and {Pakmor}, R{\"u}diger and {Kromer}, Markus and {Seitenzahl}, Ivo R. and {Sim}, Stuart A. and {R{\"o}pke}, Friedrich K.},
        title = "{SNe Ia from double detonations: Impact of core-shell mixing on the carbon ignition mechanism}",
      journal = {\aap},
         year = 2020,
        month = mar,
       volume = {635},
          eid = {A169},
        pages = {A169},
          doi = {10.1051/0004-6361/201936494},
archivePrefix = {arXiv},
       eprint = {2002.00981},
 primaryClass = {astro-ph.SR},
       adsurl = {https://ui.adsabs.harvard.edu/abs/2020A&A...635A.169G}
}

@ARTICLE{Townsley19,
       author = {{Townsley}, Dean M. and {Miles}, Broxton J. and {Shen}, Ken J. and {Kasen}, Daniel},
        title = "{Double Detonations with Thin, Modestly Enriched Helium Layers can Make Normal Type Ia Supernovae}",
      journal = {\apjl},
         year = 2019,
        month = jun,
       volume = {878},
       number = {2},
          eid = {L38},
        pages = {L38},
          doi = {10.3847/2041-8213/ab27cd},
archivePrefix = {arXiv},
       eprint = {1903.10960},
 primaryClass = {astro-ph.SR},
       adsurl = {https://ui.adsabs.harvard.edu/abs/2019ApJ...878L..38T}
}

@ARTICLE{Moore13,
       author = {{Moore}, Kevin and {Townsley}, Dean M. and {Bildsten}, Lars},
        title = "{The Effects of Curvature and Expansion on Helium Detonations on White Dwarf Surfaces}",
      journal = {\apj},
         year = 2013,
        month = oct,
       volume = {776},
       number = {2},
          eid = {97},
        pages = {97},
          doi = {10.1088/0004-637X/776/2/97},
archivePrefix = {arXiv},
       eprint = {1308.4193},
 primaryClass = {astro-ph.SR},
       adsurl = {https://ui.adsabs.harvard.edu/abs/2013ApJ...776...97M}
}

@INPROCEEDINGS{McGurk24,
       author = {{McGurk}, Rosalie C. and {Matuszewski}, Mateusz and {Neill}, James D. and {Martin}, Chris and {Bertz}, Robert and {Rockosi}, Constance and {Kassis}, Marc F.},
        title = "{The Keck Cosmic Reionization Mapper project: adding red spectroscopy to the Keck Cosmic Web Imager Integral Field Spectrograph}",
    booktitle = {Ground-based and Airborne Instrumentation for Astronomy X},
         year = 2024,
       editor = {{Bryant}, Julia J. and {Motohara}, Kentaro and {Vernet}, Jo{\"e}l. R.~D.},
       series = {Society of Photo-Optical Instrumentation Engineers (SPIE) Conference Series},
       volume = {13096},
        month = jul,
          eid = {1309647},
        pages = {1309647},
          doi = {10.1117/12.3020646},
       adsurl = {https://ui.adsabs.harvard.edu/abs/2024SPIE13096E..47M}
}

@ARTICLE{Morrissey18,
       author = {{Morrissey}, Patrick and {Matuszewski}, Matuesz and {Martin}, D. Christopher and {Neill}, James D. and {Epps}, Harland and {Fucik}, Jason and {Weber}, Bob and {Darvish}, Behnam and {Adkins}, Sean and {Allen}, Steve and {Bartos}, Randy and {Belicki}, Justin and {Cabak}, Jerry and {Callahan}, Shawn and {Cowley}, Dave and {Crabill}, Marty and {Deich}, Willian and {Delecroix}, Alex and {Doppman}, Greg and {Hilyard}, David and {James}, Ean and {Kaye}, Steve and {Kokorowski}, Michael and {Kwok}, Shui and {Lanclos}, Kyle and {Milner}, Steve and {Moore}, Anna and {O'Sullivan}, Donal and {Parihar}, Prachi and {Park}, Sam and {Phillips}, Andrew and {Rizzi}, Luca and {Rockosi}, Constance and {Rodriguez}, Hector and {Salaun}, Yves and {Seaman}, Kirk and {Sheikh}, David and {Weiss}, Jason and {Zarzaca}, Ray},
        title = "{The Keck Cosmic Web Imager Integral Field Spectrograph}",
      journal = {\apj},
         year = 2018,
        month = sep,
       volume = {864},
       number = {1},
          eid = {93},
        pages = {93},
          doi = {10.3847/1538-4357/aad597},
archivePrefix = {arXiv},
       eprint = {1807.10356},
 primaryClass = {astro-ph.IM},
       adsurl = {https://ui.adsabs.harvard.edu/abs/2018ApJ...864...93M}
}

@ARTICLE{Shen18,
       author = {{Shen}, Ken J. and {Kasen}, Daniel and {Miles}, Broxton J. and {Townsley}, Dean M.},
        title = "{Sub-Chandrasekhar-mass White Dwarf Detonations Revisited}",
      journal = {\apj},
         year = 2018,
        month = feb,
       volume = {854},
       number = {1},
          eid = {52},
        pages = {52},
          doi = {10.3847/1538-4357/aaa8de},
archivePrefix = {arXiv},
       eprint = {1706.01898},
 primaryClass = {astro-ph.HE},
       adsurl = {https://ui.adsabs.harvard.edu/abs/2018ApJ...854...52S}
}

@ARTICLE{Shen24,
       author = {{Shen}, Ken J. and {Boos}, Samuel J. and {Townsley}, Dean M.},
        title = "{Almost All Carbon/Oxygen White Dwarfs Can Host Double Detonations}",
      journal = {\apj},
         year = 2024,
        month = nov,
       volume = {975},
       number = {1},
          eid = {127},
        pages = {127},
          doi = {10.3847/1538-4357/ad7379},
archivePrefix = {arXiv},
       eprint = {2405.19417},
 primaryClass = {astro-ph.SR},
       adsurl = {https://ui.adsabs.harvard.edu/abs/2024ApJ...975..127S}
}

@ARTICLE{Mulligan19,
       author = {{Mulligan}, Brian W. and {Zhang}, Kaicheng and {Wheeler}, J. Craig},
        title = "{Exploring the shell model of high-velocity features of Type Ia supernovae using TARDIS}",
      journal = {\mnras},
         year = 2019,
        month = apr,
       volume = {484},
       number = {4},
        pages = {4785-4800},
          doi = {10.1093/mnras/stz235},
archivePrefix = {arXiv},
       eprint = {1901.08582},
 primaryClass = {astro-ph.HE},
       adsurl = {https://ui.adsabs.harvard.edu/abs/2019MNRAS.484.4785M}
}

@ARTICLE{Gerardy04,
       author = {{Gerardy}, Christopher L. and {H{\"o}flich}, Peter and {Fesen}, Robert A. and {Marion}, G.~H. and {Nomoto}, Ken'ichi and {Quimby}, Robert and {Schaefer}, Bradley E. and {Wang}, Lifan and {Wheeler}, J. Craig},
        title = "{SN 2003du: Signatures of the Circumstellar Environment in a Normal Type Ia Supernova?}",
      journal = {\apj},
         year = 2004,
        month = may,
       volume = {607},
       number = {1},
        pages = {391-405},
          doi = {10.1086/383488},
archivePrefix = {arXiv},
       eprint = {astro-ph/0309639},
 primaryClass = {astro-ph},
       adsurl = {https://ui.adsabs.harvard.edu/abs/2004ApJ...607..391G}
}

@ARTICLE{Kato18,
       author = {{Kato}, Mariko and {Saio}, Hideyuki and {Hachisu}, Izumi},
        title = "{Production of Silicon on Mass-increasing White Dwarfs: Possible Origin of High-velocity Features in Type Ia Supernovae}",
      journal = {\apj},
         year = 2018,
        month = aug,
       volume = {863},
       number = {2},
          eid = {125},
        pages = {125},
          doi = {10.3847/1538-4357/aad327},
archivePrefix = {arXiv},
       eprint = {1807.03938},
 primaryClass = {astro-ph.SR},
       adsurl = {https://ui.adsabs.harvard.edu/abs/2018ApJ...863..125K}
}

@ARTICLE{Tanaka06,
       author = {{Tanaka}, Masaomi and {Mazzali}, Paolo A. and {Maeda}, Keiichi and {Nomoto}, Ken'ichi},
        title = "{Three-dimensional Models for High-Velocity Features in Type Ia Supernovae}",
      journal = {\apj},
         year = 2006,
        month = jul,
       volume = {645},
       number = {1},
        pages = {470-479},
          doi = {10.1086/504102},
archivePrefix = {arXiv},
       eprint = {astro-ph/0603184},
 primaryClass = {astro-ph},
       adsurl = {https://ui.adsabs.harvard.edu/abs/2006ApJ...645..470T}
}

@ARTICLE{Mazzali05,
       author = {{Mazzali}, P.~A. and {Benetti}, S. and {Altavilla}, G. and {Blanc}, G. and {Cappellaro}, E. and {Elias-Rosa}, N. and {Garavini}, G. and {Goobar}, A. and {Harutyunyan}, A. and {Kotak}, R. and {Leibundgut}, B. and {Lundqvist}, P. and {Mattila}, S. and {Mendez}, J. and {Nobili}, S. and {Pain}, R. and {Pastorello}, A. and {Patat}, F. and {Pignata}, G. and {Podsiadlowski}, Ph. and {Ruiz-Lapuente}, P. and {Salvo}, M. and {Schmidt}, B.~P. and {Sollerman}, J. and {Stanishev}, V. and {Stehle}, M. and {Tout}, C. and {Turatto}, M. and {Hillebrandt}, W.},
        title = "{High-Velocity Features: A Ubiquitous Property of Type Ia Supernovae}",
      journal = {\apjl},
         year = 2005,
        month = apr,
       volume = {623},
       number = {1},
        pages = {L37-L40},
          doi = {10.1086/429874},
archivePrefix = {arXiv},
       eprint = {astro-ph/0502531},
 primaryClass = {astro-ph},
       adsurl = {https://ui.adsabs.harvard.edu/abs/2005ApJ...623L..37M}
}

@ARTICLE{Blondin13,
       author = {{Blondin}, St{\'e}phane and {Dessart}, Luc and {Hillier}, D. John and {Khokhlov}, Alexei M.},
        title = "{One-dimensional delayed-detonation models of Type Ia supernovae: confrontation to observations at bolometric maximum}",
      journal = {\mnras},
         year = 2013,
        month = mar,
       volume = {429},
       number = {3},
        pages = {2127-2142},
          doi = {10.1093/mnras/sts484},
archivePrefix = {arXiv},
       eprint = {1211.5892},
 primaryClass = {astro-ph.SR},
       adsurl = {https://ui.adsabs.harvard.edu/abs/2013MNRAS.429.2127B}
}

@ARTICLE{Kasen09,
       author = {{Kasen}, D. and {R{\"o}pke}, F.~K. and {Woosley}, S.~E.},
        title = "{The diversity of type Ia supernovae from broken symmetries}",
      journal = {\nat},
         year = 2009,
        month = aug,
       volume = {460},
       number = {7257},
        pages = {869-872},
          doi = {10.1038/nature08256},
archivePrefix = {arXiv},
       eprint = {0907.0708},
 primaryClass = {astro-ph.HE},
       adsurl = {https://ui.adsabs.harvard.edu/abs/2009Natur.460..869K}
}

@ARTICLE{Friesen14,
       author = {{Friesen}, Brian and {Baron}, E. and {Wisniewski}, John P. and {Parrent}, Jerod T. and {Thomas}, R.~C. and {Miller}, Timothy R. and {Marion}, G.~H.},
        title = "{Near-infrared Line Identification in Type Ia Supernovae during the Transitional Phase}",
      journal = {\apj},
         year = 2014,
        month = sep,
       volume = {792},
       number = {2},
          eid = {120},
        pages = {120},
          doi = {10.1088/0004-637X/792/2/120},
archivePrefix = {arXiv},
       eprint = {1407.7732},
 primaryClass = {astro-ph.SR},
       adsurl = {https://ui.adsabs.harvard.edu/abs/2014ApJ...792..120F}
}

@ARTICLE{Vallely21,
       author = {{Vallely}, P.~J. and {Kochanek}, C.~S. and {Stanek}, K.~Z. and {Fausnaugh}, M. and {Shappee}, B.~J.},
        title = "{High-cadence, early-time observations of core-collapse supernovae from the TESS prime mission}",
      journal = {\mnras},
         year = 2021,
        month = feb,
       volume = {500},
       number = {4},
        pages = {5639-5656},
          doi = {10.1093/mnras/staa3675},
archivePrefix = {arXiv},
       eprint = {2010.06596},
 primaryClass = {astro-ph.HE},
       adsurl = {https://ui.adsabs.harvard.edu/abs/2021MNRAS.500.5639V}
}

@ARTICLE{Sand16,
       author = {{Sand}, D.~J. and {Hsiao}, E.~Y. and {Banerjee}, D.~P.~K. and {Marion}, G.~H. and {Diamond}, T.~R. and {Joshi}, V. and {Parrent}, J.~T. and {Phillips}, M.~M. and {Stritzinger}, M.~D. and {Venkataraman}, V.},
        title = "{Post-maximum Near-infrared Spectra of SN 2014J: A Search for Interaction Signatures}",
      journal = {\apjl},
         year = 2016,
        month = may,
       volume = {822},
       number = {1},
          eid = {L16},
        pages = {L16},
          doi = {10.3847/2041-8205/822/1/L16},
archivePrefix = {arXiv},
       eprint = {1603.07331},
 primaryClass = {astro-ph.GA},
       adsurl = {https://ui.adsabs.harvard.edu/abs/2016ApJ...822L..16S}
}

@ARTICLE{Collins23,
       author = {{Collins}, Christine E. and {Sim}, Stuart A. and {Shingles}, Luke J. and {Gronow}, Sabrina and {R{\"o}pke}, Friedrich K. and {Pakmor}, R{\"u}diger and {Seitenzahl}, Ivo R. and {Kromer}, Markus},
        title = "{Helium as a signature of the double detonation in Type Ia supernovae}",
      journal = {\mnras},
         year = 2023,
        month = sep,
       volume = {524},
       number = {3},
        pages = {4447-4454},
          doi = {10.1093/mnras/stad2170},
archivePrefix = {arXiv},
       eprint = {2307.08660},
 primaryClass = {astro-ph.SR},
       adsurl = {https://ui.adsabs.harvard.edu/abs/2023MNRAS.524.4447C}
}

@ARTICLE{Callan24,
       author = {{Callan}, F.~P. and {Collins}, C.~E. and {Sim}, S.~A. and {Shingles}, L.~J. and {Pakmor}, R. and {Srivastav}, S. and {Pollin}, J.~M. and {Gronow}, S. and {Roepke}, F.~K. and {Seitenzahl}, I.~R.},
        title = "{Exploring the range of impacts of helium in the spectra of double detonation models for Type Ia supernovae}",
      journal = {arXiv e-prints},
         year = 2024,
        month = aug,
          eid = {arXiv:2408.03048},
        pages = {arXiv:2408.03048},
          doi = {10.48550/arXiv.2408.03048},
archivePrefix = {arXiv},
       eprint = {2408.03048},
 primaryClass = {astro-ph.HE},
       adsurl = {https://ui.adsabs.harvard.edu/abs/2024arXiv240803048C}
}

@ARTICLE{Thomas11_SYNAPPS,
       author = {{Thomas}, R.~C. and {Nugent}, P.~E. and {Meza}, J.~C.},
        title = "{SYNAPPS: Data-Driven Analysis for Supernova Spectroscopy}",
      journal = {\pasp},
         year = 2011,
        month = feb,
       volume = {123},
       number = {900},
        pages = {237},
          doi = {10.1086/658673},
       adsurl = {https://ui.adsabs.harvard.edu/abs/2011PASP..123..237T}
}

@ARTICLE{Marion15,
       author = {{Marion}, G.~H. and {Sand}, D.~J. and {Hsiao}, E.~Y. and {Banerjee}, D.~P.~K. and {Valenti}, S. and {Stritzinger}, M.~D. and {Vink{\'o}}, J. and {Joshi}, V. and {Venkataraman}, V. and {Ashok}, N.~M. and {Amanullah}, R. and {Binzel}, R.~P. and {Bochanski}, J.~J. and {Bryngelson}, G.~L. and {Burns}, C.~R. and {Drozdov}, D. and {Fieber-Beyer}, S.~K. and {Graham}, M.~L. and {Howell}, D.~A. and {Johansson}, J. and {Kirshner}, R.~P. and {Milne}, P.~A. and {Parrent}, J. and {Silverman}, J.~M. and {Vervack}, R.~J., Jr. and {Wheeler}, J.~C.},
        title = "{Early Observations and Analysis of the Type Ia SN 2014J in M82}",
      journal = {\apj},
         year = 2015,
        month = jan,
       volume = {798},
       number = {1},
          eid = {39},
        pages = {39},
          doi = {10.1088/0004-637X/798/1/39},
archivePrefix = {arXiv},
       eprint = {1405.3970},
 primaryClass = {astro-ph.GA},
       adsurl = {https://ui.adsabs.harvard.edu/abs/2015ApJ...798...39M}
}

@ARTICLE{Morrell24,
       author = {{Morrell}, N. and {Phillips}, M.~M. and {Folatelli}, G. and {Stritzinger}, M.~D. and {Hamuy}, M. and {Suntzeff}, N.~B. and {Hsiao}, E.~Y. and {Taddia}, F. and {Burns}, C.~R. and {Hoeflich}, P. and {Ashall}, C. and {Contreras}, C. and {Galbany}, L. and {Lu}, J. and {Piro}, A.~L. and {Anais}, J. and {Baron}, E. and {Burrow}, A. and {Busta}, L. and {Campillay}, A. and {Castell{\'o}n}, S. and {Corco}, C. and {Diamond}, T. and {Freedman}, W.~L. and {Gonzalez}, C. and {Krisciunas}, K. and {Kumar}, S. and {Persson}, S.~E. and {Ser{\'o}n}, J. and {Shahbandeh}, M. and {Torres}, S. and {Uddin}, S.~A. and {Anderson}, J.~P. and {Baltay}, C. and {Gall}, C. and {Goobar}, A. and {Hadjiyska}, E. and {Holmbo}, S. and {Kasliwal}, M. and {Lidman}, C. and {Marion}, G.~H. and {Mazzali}, P.~A. and {Nugent}, P. and {Perlmutter}, S. and {Pignata}, G. and {Rabinowitz}, D. and {Roth}, M. and {Ryder}, S.~D. and {Shappee}, B.~J. and {Vink{\'o}}, J. and {Wheeler}, J.~C. and {de Jaeger}, T. and {Lira}, P. and {Ruiz}, M.~T. and {Rich}, J.~A. and {Prieto}, J.~L. and {Di Mille}, F. and {Osip}, D. and {Blanc}, G. and {Palunas}, P.},
        title = "{Optical Spectroscopy of Type Ia Supernovae by the Carnegie Supernova Projects I and II}",
      journal = {\apj},
         year = 2024,
        month = may,
       volume = {967},
       number = {1},
          eid = {20},
        pages = {20},
          doi = {10.3847/1538-4357/ad38af},
archivePrefix = {arXiv},
       eprint = {2404.19208},
 primaryClass = {astro-ph.HE},
       adsurl = {https://ui.adsabs.harvard.edu/abs/2024ApJ...967...20M}
}

@ARTICLE{Brout22,
       author = {{Brout}, Dillon and {Taylor}, Georgie and {Scolnic}, Dan and {Wood}, Charlotte M. and {Rose}, Benjamin M. and {Vincenzi}, Maria and {Dwomoh}, Arianna and {Lidman}, Christopher and {Riess}, Adam and {Ali}, Noor and {Qu}, Helen and {Dai}, Mi},
        title = "{The Pantheon+ Analysis: SuperCal-fragilistic Cross Calibration, Retrained SALT2 Light-curve Model, and Calibration Systematic Uncertainty}",
      journal = {\apj},
         year = 2022,
        month = oct,
       volume = {938},
       number = {2},
          eid = {111},
        pages = {111},
          doi = {10.3847/1538-4357/ac8bcc},
archivePrefix = {arXiv},
       eprint = {2112.03864},
 primaryClass = {astro-ph.CO},
       adsurl = {https://ui.adsabs.harvard.edu/abs/2022ApJ...938..111B}
}

@ARTICLE{Taylor23,
       author = {{Taylor}, G. and {Jones}, D.~O. and {Popovic}, B. and {Vincenzi}, M. and {Kessler}, R. and {Scolnic}, D. and {Dai}, M. and {Kenworthy}, W.~D. and {Pierel}, J.~D.~R.},
        title = "{SALT2 versus SALT3: updated model surfaces and their impacts on type Ia supernova cosmology}",
      journal = {\mnras},
         year = 2023,
        month = apr,
       volume = {520},
       number = {4},
        pages = {5209-5224},
          doi = {10.1093/mnras/stad320},
archivePrefix = {arXiv},
       eprint = {2301.10644},
 primaryClass = {astro-ph.CO},
       adsurl = {https://ui.adsabs.harvard.edu/abs/2023MNRAS.520.5209T}
}

@ARTICLE{Kenworthy21,
       author = {{Kenworthy}, W.~D. and {Jones}, D.~O. and {Dai}, M. and {Kessler}, R. and {Scolnic}, D. and {Brout}, D. and {Siebert}, M.~R. and {Pierel}, J.~D.~R. and {Dettman}, K.~G. and {Dimitriadis}, G. and {Foley}, R.~J. and {Jha}, S.~W. and {Pan}, Y. -C. and {Riess}, A. and {Rodney}, S. and {Rojas-Bravo}, C.},
        title = "{SALT3: An Improved Type Ia Supernova Model for Measuring Cosmic Distances}",
      journal = {\apj},
         year = 2021,
        month = dec,
       volume = {923},
       number = {2},
          eid = {265},
        pages = {265},
          doi = {10.3847/1538-4357/ac30d8},
archivePrefix = {arXiv},
       eprint = {2104.07795},
 primaryClass = {astro-ph.CO},
       adsurl = {https://ui.adsabs.harvard.edu/abs/2021ApJ...923..265K}
}

@ARTICLE{Jones22,
       author = {{Jones}, D.~O. and {Mandel}, K.~S. and {Kirshner}, R.~P. and {Thorp}, S. and {Challis}, P.~M. and {Avelino}, A. and {Brout}, D. and {Burns}, C. and {Foley}, R.~J. and {Pan}, Y. -C. and {Scolnic}, D.~M. and {Siebert}, M.~R. and {Chornock}, R. and {Freedman}, W.~L. and {Friedman}, A. and {Frieman}, J. and {Galbany}, L. and {Hsiao}, E. and {Kelsey}, L. and {Marion}, G.~H. and {Nichol}, R.~C. and {Nugent}, P.~E. and {Phillips}, M.~M. and {Rest}, A. and {Riess}, A.~G. and {Sako}, M. and {Smith}, M. and {Wiseman}, P. and {Wood-Vasey}, W.~M.},
        title = "{Cosmological Results from the RAISIN Survey: Using Type Ia Supernovae in the Near Infrared as a Novel Path to Measure the Dark Energy Equation of State}",
      journal = {\apj},
         year = 2022,
        month = jul,
       volume = {933},
       number = {2},
          eid = {172},
        pages = {172},
          doi = {10.3847/1538-4357/ac755b},
archivePrefix = {arXiv},
       eprint = {2201.07801},
 primaryClass = {astro-ph.CO},
       adsurl = {https://ui.adsabs.harvard.edu/abs/2022ApJ...933..172J}
}

@ARTICLE{Tinyanont24,
       author = {{Tinyanont}, S. and {Foley}, R.~J. and {Taggart}, K. and {Davis}, K.~W. and {LeBaron}, N. and {Andrews}, J.~E. and {Bustamante-Rosell}, M.~J. and {Camacho-Neves}, Y. and {Chornock}, R. and {Coulter}, D.~A. and {Galbany}, L. and {Jha}, S.~W. and {Kilpatrick}, C.~D. and {Kwok}, L.~A. and {Larison}, C. and {Pierel}, J.~R. and {Siebert}, M.~R. and {Aldering}, G. and {Auchettl}, K. and {Bloom}, J.~S. and {Dhawan}, S. and {Filippenko}, A.~V. and {French}, K.~D. and {Gagliano}, A. and {Grayling}, M. and {Howell}, D.~A. and {Jacobson-Gal{\'a}n}, W.~V. and {Jones}, D.~O. and {Le Saux}, X. and {Macias}, P. and {Mandel}, K.~S. and {McCully}, C. and {Padilla Gonzalez}, E. and {Rest}, A. and {Rho}, J. and {Rojas-Bravo}, C. and {Skrutskie}, M.~F. and {Thorp}, S. and {Wang}, Q. and {Ward}, S.~M.},
        title = "{Keck Infrared Transient Survey. I. Survey Description and Data Release 1}",
      journal = {\pasp},
         year = 2024,
        month = jan,
       volume = {136},
       number = {1},
          eid = {014201},
        pages = {014201},
          doi = {10.1088/1538-3873/ad1b39},
archivePrefix = {arXiv},
       eprint = {2309.07102},
 primaryClass = {astro-ph.SR},
       adsurl = {https://ui.adsabs.harvard.edu/abs/2024PASP..136a4201T}
}

@INPROCEEDINGS{Wilson04_NIRES,
       author = {{Wilson}, John C. and {Henderson}, Charles P. and {Herter}, Terry L. and {Matthews}, Keith and {Skrutskie}, Michael F. and {Adams}, Joseph D. and {Moon}, Dae-Sik and {Smith}, Roger and {Gautier}, Nick and {Ressler}, Michael and {Soifer}, B.~T. and {Lin}, Sean and {Howard}, James and {LaMarr}, John and {Stolberg}, Todd M. and {Zink}, Jeff},
        title = "{Mass producing an efficient NIR spectrograph}",
    booktitle = {Ground-based Instrumentation for Astronomy},
         year = 2004,
       editor = {{Moorwood}, Alan F.~M. and {Iye}, Masanori},
       series = {Society of Photo-Optical Instrumentation Engineers (SPIE) Conference Series},
       volume = {5492},
        month = sep,
        pages = {1295-1305},
          doi = {10.1117/12.550925},
       adsurl = {https://ui.adsabs.harvard.edu/abs/2004SPIE.5492.1295W}
}

@ARTICLE{Jones24,
       author = {{Jones}, D.~O. and {McGill}, P. and {Manning}, T.~A. and {Gagliano}, A. and {Wang}, B. and {Coulter}, D.~A. and {Foley}, R.~J. and {Narayan}, G. and {Villar}, V.~A. and {Braff}, L. and {Engel}, A.~W. and {Farias}, D. and {Lai}, Z. and {Loertscher}, K. and {Kutcka}, J. and {Thorp}, S. and {Vazquez}, J.},
        title = "{Blast: a Web Application for Characterizing the Host Galaxies of Astrophysical Transients}",
      journal = {arXiv e-prints},
         year = 2024,
        month = oct,
          eid = {arXiv:2410.17322},
        pages = {arXiv:2410.17322},
          doi = {10.48550/arXiv.2410.17322},
archivePrefix = {arXiv},
       eprint = {2410.17322},
 primaryClass = {astro-ph.HE},
       adsurl = {https://ui.adsabs.harvard.edu/abs/2024arXiv241017322J}
}

@ARTICLE{Pearson24,
       author = {{Pearson}, Jeniveve and {Sand}, David J. and {Lundqvist}, Peter and {Galbany}, Llu{\'\i}s and {Andrews}, Jennifer E. and {Bostroem}, K. Azalee and {Dong}, Yize and {Hoang}, Emily and {Hosseinzadeh}, Griffin and {Janzen}, Daryl and {Jencson}, Jacob E. and {Lundquist}, Michael J. and {Mehta}, Darshana and {Meza Retamal}, Nicol{\'a}s and {Shrestha}, Manisha and {Valenti}, Stefano and {Wyatt}, Samuel and {Anderson}, Joseph P. and {Ashall}, Chris and {Auchettl}, Katie and {Baron}, Eddie and {Blondin}, St{\'e}phane and {Burns}, Christopher R. and {Cai}, Yongzhi and {Chen}, Ting-Wan and {Chomiuk}, Laura and {Coulter}, David A. and {Cross}, Dane and {Davis}, Kyle W. and {de Jaeger}, Thomas and {DerKacy}, James M. and {Desai}, Dhvanil D. and {Dimitriadis}, Georgios and {Do}, Aaron and {Farah}, Joseph R. and {Foley}, Ryan J. and {Gromadzki}, Mariusz and {Guti{\'e}rrez}, Claudia P. and {Haislip}, Joshua and {Gonz{\'a}lez Hern{\'a}ndez}, Jonay I. and {Hinkle}, Jason T. and {Hoogendam}, Willem B. and {Howell}, D. Andrew and {Hoeflich}, Peter and {Hsiao}, Eric and {Huber}, Mark E. and {Jha}, Saurabh W. and {Jim{\'e}nez Palau}, Cristina and {Kilpatrick}, Charles D. and {Kouprianov}, Vladimir and {Kumar}, Sahana and {Kwok}, Lindsey A. and {Larison}, Conor and {LeBaron}, Natalie and {Le Saux}, Xavier and {Lu}, Jing and {McCully}, Curtis and {Mera Evans}, Tycho and {Milne}, Peter and {Modjaz}, Maryam and {Morrell}, Nidia and {M{\"u}ller-Bravo}, Tom{\'a}s E. and {Newsome}, Megan and {Nicholl}, Matt and {Padilla Gonzalez}, Estefania and {Payne}, Anna V. and {Pellegrino}, Craig and {Phan}, Kim and {Pineda-Garc{\'\i}a}, Jonathan and {Piro}, Anthony L. and {Piscarreta}, Lara and {Polin}, Abigail and {Reichart}, Daniel E. and {Rojas-Bravo}, C{\'e}sar and {Ryder}, Stuart D. and {Salmaso}, Irene and {Schwab}, Michaela and {Shahbandeh}, Melissa and {Shappee}, Benjamin J. and {Siebert}, Matthew R. and {Smith}, Nathan and {Strader}, Jay and {Taggart}, Kirsty and {Terreran}, Giacomo and {Tinyanont}, Samaporn and {Tucker}, M.~A. and {Valerin}, Giorgio and {Young}, D.~R.},
        title = "{Strong Carbon Features and a Red Early Color in the Underluminous Type Ia SN 2022xkq}",
      journal = {\apj},
         year = 2024,
        month = jan,
       volume = {960},
       number = {1},
          eid = {29},
        pages = {29},
          doi = {10.3847/1538-4357/ad0153},
archivePrefix = {arXiv},
       eprint = {2309.10054},
 primaryClass = {astro-ph.HE},
       adsurl = {https://ui.adsabs.harvard.edu/abs/2024ApJ...960...29P}
}

@ARTICLE{Huchra99,
       author = {{Huchra}, John P. and {Vogeley}, Michael S. and {Geller}, Margaret J.},
        title = "{The CFA Redshift Survey: Data for the South Galactic CAP}",
      journal = {\apjs},
         year = 1999,
        month = apr,
       volume = {121},
       number = {2},
        pages = {287-368},
          doi = {10.1086/313194},
       adsurl = {https://ui.adsabs.harvard.edu/abs/1999ApJS..121..287H}
}

@ARTICLE{Karambelkar24,
       author = {{Karambelkar}, V. and {Earley}, N. and {Das}, K. and {Sollerman}, J. and {Fremling}, C. and {Kasliwal}, M. and {Johansson}, J. and {Miller}, A. and {Rehemtulla}, N. and {Schulze}, S. and {Harvey}, L. and {Dimitriadis}, G. and {Burgaz}, U. and {Terwel}, J.},
        title = "{ZTF discovery of a young, peculiar, thermonuclear supernova in a nearby galaxy}",
      journal = {Transient Name Server AstroNote},
         year = 2024,
        month = mar,
       volume = {81},
        pages = {1},
       adsurl = {https://ui.adsabs.harvard.edu/abs/2024TNSAN..81....1K}
}

@ARTICLE{Rest05,
       author = {{Rest}, A. and {Stubbs}, C. and {Becker}, A.~C. and {Miknaitis}, G.~A. and {Miceli}, A. and {Covarrubias}, R. and {Hawley}, S.~L. and {Smith}, R.~C. and {Suntzeff}, N.~B. and {Olsen}, K. and {Prieto}, J.~L. and {Hiriart}, R. and {Welch}, D.~L. and {Cook}, K.~H. and {Nikolaev}, S. and {Huber}, M. and {Prochtor}, G. and {Clocchiatti}, A. and {Minniti}, D. and {Garg}, A. and {Challis}, P. and {Keller}, S.~C. and {Schmidt}, B.~P.},
        title = "{Testing LMC Microlensing Scenarios: The Discrimination Power of the SuperMACHO Microlensing Survey}",
      journal = {\apj},
         year = 2005,
        month = dec,
       volume = {634},
       number = {2},
        pages = {1103-1115},
          doi = {10.1086/497060},
archivePrefix = {arXiv},
       eprint = {astro-ph/0509240},
 primaryClass = {astro-ph},
       adsurl = {https://ui.adsabs.harvard.edu/abs/2005ApJ...634.1103R}
}

@ARTICLE{Schechter93,
       author = {{Schechter}, Paul L. and {Mateo}, Mario and {Saha}, Abhijit},
        title = "{DoPHOT, A CCD Photometry Program: Description and Tests}",
      journal = {\pasp},
         year = 1993,
        month = nov,
       volume = {105},
        pages = {1342},
          doi = {10.1086/133316},
       adsurl = {https://ui.adsabs.harvard.edu/abs/1993PASP..105.1342S}
}

@ARTICLE{Holmbo23,
       author = {{Holmbo}, S. and {Stritzinger}, M.~D. and {Karamehmetoglu}, E. and {Burns}, C.~R. and {Morrell}, N. and {Ashall}, C. and {Hsiao}, E.~Y. and {Galbany}, L. and {Folatelli}, G. and {Phillips}, M.~M. and {Baron}, E. and {Guti{\'e}rrez}, C.~P. and {Leloudas}, G. and {M{\"u}ller-Bravo}, T.~E. and {Hoeflich}, P. and {Taddia}, F. and {Suntzeff}, N.~B.},
        title = "{The Carnegie Supernova Project I. Spectroscopic analysis of stripped-envelope supernovae}",
      journal = {\aap},
         year = 2023,
        month = jul,
       volume = {675},
          eid = {A83},
        pages = {A83},
          doi = {10.1051/0004-6361/202245334},
archivePrefix = {arXiv},
       eprint = {2302.11304},
 primaryClass = {astro-ph.HE},
       adsurl = {https://ui.adsabs.harvard.edu/abs/2023A&A...675A..83H}
}

@ARTICLE{Cushing04,
       author = {{Cushing}, Michael C. and {Vacca}, William D. and {Rayner}, John T.},
        title = "{Spextool: A Spectral Extraction Package for SpeX, a 0.8-5.5 Micron Cross-Dispersed Spectrograph}",
      journal = {\pasp},
         year = 2004,
        month = apr,
       volume = {116},
       number = {818},
        pages = {362-376},
          doi = {10.1086/382907},
       adsurl = {https://ui.adsabs.harvard.edu/abs/2004PASP..116..362C}
}

@ARTICLE{Hoogendam24,
       author = {{Hoogendam}, W.~B. and {Shappee}, B.~J. and {Brown}, P.~J. and {Tucker}, M.~A. and {Ashall}, C. and {Piro}, A.~L.},
        title = "{From out of the Blue: Swift Links 2002es-like, 2003fg-like, and Early Time Bump Type Ia Supernovae}",
      journal = {\apj},
         year = 2024,
        month = may,
       volume = {966},
       number = {1},
          eid = {139},
        pages = {139},
          doi = {10.3847/1538-4357/ad33ba},
archivePrefix = {arXiv},
       eprint = {2309.11563},
 primaryClass = {astro-ph.HE},
       adsurl = {https://ui.adsabs.harvard.edu/abs/2024ApJ...966..139H}
}

@ARTICLE{Chandrasekhar31,
       author = {{Chandrasekhar}, S.},
        title = "{The Maximum Mass of Ideal White Dwarfs}",
      journal = {\apj},
         year = 1931,
        month = jul,
       volume = {74},
        pages = {81},
          doi = {10.1086/143324},
       adsurl = {https://ui.adsabs.harvard.edu/abs/1931ApJ....74...81C}
}

@ARTICLE{Do25,
       author = {{Do}, Aaron and {Shappee}, Benjamin J. and {Tonry}, John L. and {Tully}, R. Brent and {de Jaeger}, Thomas and {Rubin}, David and {Ashall}, Chris and {Burns}, Christopher R. and {Desai}, Dhvanil D. and {Hinkle}, Jason T. and {Hoogendam}, Willem B. and {Huber}, Mark E. and {Jones}, David O. and {Mandel}, Kaisey S. and {Payne}, Anna V. and {Peterson}, Erik R. and {Scolnic}, Dan and {Tucker}, Michael A.},
        title = "{Hawai'i Supernova Flows: a peculiar velocity survey using over a Thousand Supernovae in the near-infrared}",
      journal = {\mnras},
         year = 2025,
        month = jan,
       volume = {536},
       number = {1},
        pages = {624-663},
          doi = {10.1093/mnras/stae2501},
archivePrefix = {arXiv},
       eprint = {2403.05620},
 primaryClass = {astro-ph.CO},
       adsurl = {https://ui.adsabs.harvard.edu/abs/2025MNRAS.536..624D}
}

@ARTICLE{Hodgkin09,
       author = {{Hodgkin}, S.~T. and {Irwin}, M.~J. and {Hewett}, P.~C. and {Warren}, S.~J.},
        title = "{The UKIRT wide field camera ZYJHK photometric system: calibration from 2MASS}",
      journal = {\mnras},
         year = 2009,
        month = apr,
       volume = {394},
       number = {2},
        pages = {675-692},
          doi = {10.1111/j.1365-2966.2008.14387.x},
archivePrefix = {arXiv},
       eprint = {0812.3081},
 primaryClass = {astro-ph},
       adsurl = {https://ui.adsabs.harvard.edu/abs/2009MNRAS.394..675H}
}

@ARTICLE{Casali07,
       author = {{Casali}, M. and {Adamson}, A. and {Alves de Oliveira}, C. and {Almaini}, O. and {Burch}, K. and {Chuter}, T. and {Elliot}, J. and {Folger}, M. and {Foucaud}, S. and {Hambly}, N. and {Hastie}, M. and {Henry}, D. and {Hirst}, P. and {Irwin}, M. and {Ives}, D. and {Lawrence}, A. and {Laidlaw}, K. and {Lee}, D. and {Lewis}, J. and {Lunney}, D. and {McLay}, S. and {Montgomery}, D. and {Pickup}, A. and {Read}, M. and {Rees}, N. and {Robson}, I. and {Sekiguchi}, K. and {Vick}, A. and {Warren}, S. and {Woodward}, B.},
        title = "{The UKIRT wide-field camera}",
      journal = {\aap},
         year = 2007,
        month = may,
       volume = {467},
       number = {2},
        pages = {777-784},
          doi = {10.1051/0004-6361:20066514},
       adsurl = {https://ui.adsabs.harvard.edu/abs/2007A&A...467..777C}
}

@ARTICLE{smith20,
       author = {{Smith}, K.~W. and {Smartt}, S.~J. and {Young}, D.~R. and {Tonry}, J.~L. and {Denneau}, L. and {Flewelling}, H. and {Heinze}, A.~N. and {Weiland}, H.~J. and {Stalder}, B. and {Rest}, A. and {Stubbs}, C.~W. and {Anderson}, J.~P. and {Chen}, T. -W. and {Clark}, P. and {Do}, A. and {F{\"o}rster}, F. and {Fulton}, M. and {Gillanders}, J. and {McBrien}, O.~R. and {O'Neill}, D. and {Srivastav}, S. and {Wright}, D.~E.},
        title = "{Design and Operation of the ATLAS Transient Science Server}",
      journal = {\pasp},
         year = 2020,
        month = aug,
       volume = {132},
       number = {1014},
          eid = {085002},
        pages = {085002},
          doi = {10.1088/1538-3873/ab936e},
archivePrefix = {arXiv},
       eprint = {2003.09052},
 primaryClass = {astro-ph.IM},
       adsurl = {https://ui.adsabs.harvard.edu/abs/2020PASP..132h5002S}
}

@ARTICLE{Aleo23,
       author = {{Aleo}, P.~D. and {Malanchev}, K. and {Sharief}, S. and {Jones}, D.~O. and {Narayan}, G. and {Foley}, R.~J. and {Villar}, V.~A. and {Angus}, C.~R. and {Baldassare}, V.~F. and {Bustamante-Rosell}, M.~J. and {Chatterjee}, D. and {Cold}, C. and {Coulter}, D.~A. and {Davis}, K.~W. and {Dhawan}, S. and {Drout}, M.~R. and {Engel}, A. and {French}, K.~D. and {Gagliano}, A. and {Gall}, C. and {Hjorth}, J. and {Huber}, M.~E. and {Jacobson-Gal{\'a}n}, W.~V. and {Kilpatrick}, C.~D. and {Langeroodi}, D. and {Macias}, P. and {Mandel}, K.~S. and {Margutti}, R. and {Matasi{\'c}}, F. and {McGill}, P. and {Pierel}, J.~D.~R. and {Ramirez-Ruiz}, E. and {Ransome}, C.~L. and {Rojas-Bravo}, C. and {Siebert}, M.~R. and {Smith}, K.~W. and {de Soto}, K.~M. and {Stroh}, M.~C. and {Tinyanont}, S. and {Taggart}, K. and {Ward}, S.~M. and {Wojtak}, R. and {Auchettl}, K. and {Blanchard}, P.~K. and {de Boer}, T.~J.~L. and {Boyd}, B.~M. and {Carroll}, C.~M. and {Chambers}, K.~C. and {DeMarchi}, L. and {Dimitriadis}, G. and {Dodd}, S.~A. and {Earl}, N. and {Farias}, D. and {Gao}, H. and {Gomez}, S. and {Grayling}, M. and {Grillo}, C. and {Hayes}, E.~E. and {Hung}, T. and {Izzo}, L. and {Khetan}, N. and {Kolborg}, A.~N. and {Law-Smith}, J.~A.~P. and {LeBaron}, N. and {Lin}, C. -C. and {Luo}, Y. and {Magnier}, E.~A. and {Matthews}, D. and {Mockler}, B. and {O'Grady}, A.~J.~G. and {Pan}, Y. -C. and {Politsch}, C.~A. and {Raimundo}, S.~I. and {Rest}, A. and {Ridden-Harper}, R. and {Sarangi}, A. and {Schr{\o}der}, S.~L. and {Smartt}, S.~J. and {Terreran}, G. and {Thorp}, S. and {Vazquez}, J. and {Wainscoat}, R.~J. and {Wang}, Q. and {Wasserman}, A.~R. and {Yadavalli}, S.~K. and {Yarza}, R. and {Zenati}, Y. and {Young Supernova Experiment}},
        title = "{The Young Supernova Experiment Data Release 1 (YSE DR1): Light Curves and Photometric Classification of 1975 Supernovae}",
      journal = {\apjs},
         year = 2023,
        month = may,
       volume = {266},
       number = {1},
          eid = {9},
        pages = {9},
          doi = {10.3847/1538-4365/acbfba},
archivePrefix = {arXiv},
       eprint = {2211.07128},
 primaryClass = {astro-ph.HE},
       adsurl = {https://ui.adsabs.harvard.edu/abs/2023ApJS..266....9A}
}

@ARTICLE{Rayner03,
       author = {{Rayner}, J.~T. and {Toomey}, D.~W. and {Onaka}, P.~M. and {Denault}, A.~J. and {Stahlberger}, W.~E. and {Vacca}, W.~D. and {Cushing}, M.~C. and {Wang}, S.},
        title = "{SpeX: A Medium-Resolution 0.8-5.5 Micron Spectrograph and Imager for the NASA Infrared Telescope Facility}",
      journal = {\pasp},
         year = 2003,
        month = mar,
       volume = {115},
       number = {805},
        pages = {362-382},
          doi = {10.1086/367745},
       adsurl = {https://ui.adsabs.harvard.edu/abs/2003PASP..115..362R}
}

@ARTICLE{Prochaska20,
       author = {{Prochaska}, J. and {Hennawi}, Joseph and {Westfall}, Kyle and {Cooke}, Ryan and {Wang}, Feige and {Hsyu}, Tiffany and {Davies}, Frederick and {Farina}, Emanuele and {Pelliccia}, Debora},
        title = "{PypeIt: The Python Spectroscopic Data Reduction Pipeline}",
      journal = {The Journal of Open Source Software},
         year = 2020,
        month = dec,
       volume = {5},
       number = {56},
          eid = {2308},
        pages = {2308},
          doi = {10.21105/joss.02308},
archivePrefix = {arXiv},
       eprint = {2005.06505},
 primaryClass = {astro-ph.IM},
       adsurl = {https://ui.adsabs.harvard.edu/abs/2020JOSS....5.2308P}
}

@INPROCEEDINGS{Elias06b,
       author = {{Elias}, Jonathan H. and {Rodgers}, Bernadette and {Joyce}, Richard R. and {Lazo}, Manuel and {Doppmann}, Gregory and {Winge}, Claudia and {Rodr{\'\i}guez-Ardila}, Alberto},
        title = "{Performance of the Gemini near-infrared spectrograph}",
    booktitle = {Ground-based and Airborne Instrumentation for Astronomy},
         year = 2006,
       editor = {{McLean}, Ian S. and {Iye}, Masanori},
       series = {Society of Photo-Optical Instrumentation Engineers (SPIE) Conference Series},
       volume = {6269},
        month = jun,
          eid = {626914},
        pages = {626914},
          doi = {10.1117/12.671765},
       adsurl = {https://ui.adsabs.harvard.edu/abs/2006SPIE.6269E..14E}
}

@INPROCEEDINGS{Elias06a,
       author = {{Elias}, Jonathan H. and {Joyce}, Richard R. and {Liang}, Ming and {Muller}, Gary P. and {Hileman}, Edward A. and {George}, James R.},
        title = "{Design of the Gemini near-infrared spectrograph}",
    booktitle = {Ground-based and Airborne Instrumentation for Astronomy},
         year = 2006,
       editor = {{McLean}, Ian S. and {Iye}, Masanori},
       series = {Society of Photo-Optical Instrumentation Engineers (SPIE) Conference Series},
       volume = {6269},
        month = jun,
          eid = {62694C},
        pages = {62694C},
          doi = {10.1117/12.671817},
       adsurl = {https://ui.adsabs.harvard.edu/abs/2006SPIE.6269E..4CE}
}

@ARTICLE{Liu23,
       author = {{Liu}, Zheng-Wei and {R{\"o}pke}, Friedrich K. and {Han}, Zhanwen},
        title = "{Type Ia Supernova Explosions in Binary Systems: A Review}",
      journal = {Research in Astronomy and Astrophysics},
         year = 2023,
        month = aug,
       volume = {23},
       number = {8},
          eid = {082001},
        pages = {082001},
          doi = {10.1088/1674-4527/acd89e},
archivePrefix = {arXiv},
       eprint = {2305.13305},
 primaryClass = {astro-ph.HE},
       adsurl = {https://ui.adsabs.harvard.edu/abs/2023RAA....23h2001L}
}

@ARTICLE{Maeda18,
       author = {{Maeda}, Keiichi and {Jiang}, Ji-an and {Shigeyama}, Toshikazu and {Doi}, Mamoru},
        title = "{Type Ia Supernovae in the First Few Days: Signatures of Helium Detonation versus Interaction}",
      journal = {\apj},
         year = 2018,
        month = jul,
       volume = {861},
       number = {2},
          eid = {78},
        pages = {78},
          doi = {10.3847/1538-4357/aac8d8},
archivePrefix = {arXiv},
       eprint = {1805.12325},
 primaryClass = {astro-ph.HE},
       adsurl = {https://ui.adsabs.harvard.edu/abs/2018ApJ...861...78M}
}

@ARTICLE{Khokhlov91,
       author = {{Khokhlov}, A.~M.},
        title = "{Delayed detonation model for type IA supernovae}",
      journal = {\aap},
         year = 1991,
        month = may,
       volume = {245},
       number = {1},
        pages = {114-128},
       adsurl = {https://ui.adsabs.harvard.edu/abs/1991A&A...245..114K}
}

@ARTICLE{Woosley11,
       author = {{Woosley}, S.~E. and {Kasen}, Daniel},
        title = "{Sub-Chandrasekhar Mass Models for Supernovae}",
      journal = {\apj},
         year = 2011,
        month = jun,
       volume = {734},
       number = {1},
          eid = {38},
        pages = {38},
          doi = {10.1088/0004-637X/734/1/38},
archivePrefix = {arXiv},
       eprint = {1010.5292},
 primaryClass = {astro-ph.HE},
       adsurl = {https://ui.adsabs.harvard.edu/abs/2011ApJ...734...38W}
}

@ARTICLE{Ropke07,
       author = {{R{\"o}pke}, F.~K. and {Niemeyer}, J.~C.},
        title = "{Delayed detonations in full-star models of type Ia supernova explosions}",
      journal = {\aap},
         year = 2007,
        month = mar,
       volume = {464},
       number = {2},
        pages = {683-686},
          doi = {10.1051/0004-6361:20066585},
archivePrefix = {arXiv},
       eprint = {astro-ph/0703378},
 primaryClass = {astro-ph},
       adsurl = {https://ui.adsabs.harvard.edu/abs/2007A&A...464..683R}
}

@ARTICLE{Jha19,
       author = {{Jha}, Saurabh W. and {Maguire}, Kate and {Sullivan}, Mark},
        title = "{Observational properties of thermonuclear supernovae}",
      journal = {Nature Astronomy},
         year = 2019,
        month = aug,
       volume = {3},
        pages = {706-716},
          doi = {10.1038/s41550-019-0858-0},
archivePrefix = {arXiv},
       eprint = {1908.02303},
 primaryClass = {astro-ph.HE},
       adsurl = {https://ui.adsabs.harvard.edu/abs/2019NatAs...3..706J}
}

@ARTICLE{Piro13,
       author = {{Piro}, Anthony L. and {Nakar}, Ehud},
        title = "{What can we Learn from the Rising Light Curves of Radioactively Powered Supernovae?}",
      journal = {\apj},
         year = 2013,
        month = may,
       volume = {769},
       number = {1},
          eid = {67},
        pages = {67},
          doi = {10.1088/0004-637X/769/1/67},
archivePrefix = {arXiv},
       eprint = {1210.3032},
 primaryClass = {astro-ph.HE},
       adsurl = {https://ui.adsabs.harvard.edu/abs/2013ApJ...769...67P}
}

@ARTICLE{Peterson23,
       author = {{Peterson}, Erik R. and {Jones}, David O. and {Scolnic}, Daniel and {S{\'a}nchez}, Bruno O. and {Do}, Aaron and {Riess}, Adam G. and {Ward}, Sam M. and {Dwomoh}, Arianna and {de Jaeger}, Thomas and {Jha}, Saurabh W. and {Mandel}, Kaisey S. and {Pierel}, Justin D.~R. and {Popovic}, Brodie and {Rose}, Benjamin M. and {Rubin}, David and {Shappee}, Benjamin J. and {Thorp}, Stephen and {Tonry}, John L. and {Tully}, R. Brent and {Vincenzi}, Maria},
        title = "{The DEHVILS survey overview and initial data release: high-quality near-infrared Type Ia supernova light curves at low redshift}",
      journal = {\mnras},
         year = 2023,
        month = jun,
       volume = {522},
       number = {2},
        pages = {2478-2494},
          doi = {10.1093/mnras/stad1077},
archivePrefix = {arXiv},
       eprint = {2301.11868},
 primaryClass = {astro-ph.CO},
       adsurl = {https://ui.adsabs.harvard.edu/abs/2023MNRAS.522.2478P}
}

@ARTICLE{Holoien17,
       author = {{Holoien}, T.~W. -S. and {Stanek}, K.~Z. and {Kochanek}, C.~S. and {Shappee}, B.~J. and {Prieto}, J.~L. and {Brimacombe}, J. and {Bersier}, D. and {Bishop}, D.~W. and {Dong}, Subo and {Brown}, J.~S. and {Danilet}, A.~B. and {Simonian}, G.~V. and {Basu}, U. and {Beacom}, J.~F. and {Falco}, E. and {Pojmanski}, G. and {Skowron}, D.~M. and {Wo{\'z}niak}, P.~R. and {{\'A}vila}, C.~G. and {Conseil}, E. and {Contreras}, C. and {Cruz}, I. and {Fern{\'a}ndez}, J.~M. and {Koff}, R.~A. and {Guo}, Zhen and {Herczeg}, G.~J. and {Hissong}, J. and {Hsiao}, E.~Y. and {Jose}, J. and {Kiyota}, S. and {Long}, Feng and {Monard}, L.~A.~G. and {Nicholls}, B. and {Nicolas}, J. and {Wiethoff}, W.~S.},
        title = "{The ASAS-SN bright supernova catalogue - I. 2013-2014}",
      journal = {\mnras},
         year = 2017,
        month = jan,
       volume = {464},
       number = {3},
        pages = {2672-2686},
          doi = {10.1093/mnras/stw2273},
archivePrefix = {arXiv},
       eprint = {1604.00396},
 primaryClass = {astro-ph.HE},
       adsurl = {https://ui.adsabs.harvard.edu/abs/2017MNRAS.464.2672H}
}

@ARTICLE{Holoien17C,
       author = {{Holoien}, T.~W. -S. and {Brown}, J.~S. and {Stanek}, K.~Z. and {Kochanek}, C.~S. and {Shappee}, B.~J. and {Prieto}, J.~L. and {Dong}, Subo and {Brimacombe}, J. and {Bishop}, D.~W. and {Bose}, S. and {Beacom}, J.~F. and {Bersier}, D. and {Chen}, Ping and {Chomiuk}, L. and {Falco}, E. and {Godoy-Rivera}, D. and {Morrell}, N. and {Pojmanski}, G. and {Shields}, J.~V. and {Strader}, J. and {Stritzinger}, M.~D. and {Thompson}, Todd A. and {Wo{\'z}niak}, P.~R. and {Bock}, G. and {Cacella}, P. and {Conseil}, E. and {Cruz}, I. and {Fernandez}, J.~M. and {Kiyota}, S. and {Koff}, R.~A. and {Krannich}, G. and {Marples}, P. and {Masi}, G. and {Monard}, L.~A.~G. and {Nicholls}, B. and {Nicolas}, J. and {Post}, R.~S. and {Stone}, G. and {Wiethoff}, W.~S.},
        title = "{The ASAS-SN bright supernova catalogue - III. 2016}",
      journal = {\mnras},
         year = 2017,
        month = nov,
       volume = {471},
       number = {4},
        pages = {4966-4981},
          doi = {10.1093/mnras/stx1544},
archivePrefix = {arXiv},
       eprint = {1704.02320},
 primaryClass = {astro-ph.HE},
       adsurl = {https://ui.adsabs.harvard.edu/abs/2017MNRAS.471.4966H}
}

@ARTICLE{Neumann23,
       author = {{Neumann}, K.~D. and {Holoien}, T.~W. -S. and {Kochanek}, C.~S. and {Stanek}, K.~Z. and {Vallely}, P.~J. and {Shappee}, B.~J. and {Prieto}, J.~L. and {Pessi}, T. and {Jayasinghe}, T. and {Brimacombe}, J. and {Bersier}, D. and {Aydi}, E. and {Basinger}, C. and {Beacom}, J.~F. and {Bose}, S. and {Brown}, J.~S. and {Chen}, P. and {Clocchiatti}, A. and {Desai}, D.~D. and {Dong}, Subo and {Falco}, E. and {Holmbo}, S. and {Morrell}, N. and {Shields}, J.~V. and {Sokolovsky}, K.~V. and {Strader}, J. and {Stritzinger}, M.~D. and {Swihart}, S. and {Thompson}, T.~A. and {Way}, Z. and {Aslan}, L. and {Bishop}, D.~W. and {Bock}, G. and {Bradshaw}, J. and {Cacella}, P. and {Castro-Morales}, N. and {Conseil}, E. and {Cornect}, R. and {Cruz}, I. and {Farfan}, R.~G. and {Fernandez}, J.~M. and {Gabuya}, A. and {Gonzalez-Carballo}, J. -L. and {Kendurkar}, M.~R. and {Kiyota}, S. and {Koff}, R.~A. and {Krannich}, G. and {Marples}, P. and {Masi}, G. and {Monard}, L.~A.~G. and {Mu{\~n}oz}, J.~A. and {Nicholls}, B. and {Post}, R.~S. and {Pujic}, Z. and {Stone}, G. and {Tomasella}, L. and {Trappett}, D.~L. and {Wiethoff}, W.~S.},
        title = "{The ASAS-SN bright supernova catalogue - V. 2018-2020}",
      journal = {\mnras},
         year = 2023,
        month = apr,
       volume = {520},
       number = {3},
        pages = {4356-4369},
          doi = {10.1093/mnras/stad355},
archivePrefix = {arXiv},
       eprint = {2210.06492},
 primaryClass = {astro-ph.HE},
       adsurl = {https://ui.adsabs.harvard.edu/abs/2023MNRAS.520.4356N}
}

@ARTICLE{Jones21,
       author = {{Jones}, D.~O. and {Foley}, R.~J. and {Narayan}, G. and {Hjorth}, J. and {Huber}, M.~E. and {Aleo}, P.~D. and {Alexander}, K.~D. and {Angus}, C.~R. and {Auchettl}, K. and {Baldassare}, V.~F. and {Bruun}, S.~H. and {Chambers}, K.~C. and {Chatterjee}, D. and {Coppejans}, D.~L. and {Coulter}, D.~A. and {DeMarchi}, L. and {Dimitriadis}, G. and {Drout}, M.~R. and {Engel}, A. and {French}, K.~D. and {Gagliano}, A. and {Gall}, C. and {Hung}, T. and {Izzo}, L. and {Jacobson-Gal{\'a}n}, W.~V. and {Kilpatrick}, C.~D. and {Korhonen}, H. and {Margutti}, R. and {Raimundo}, S.~I. and {Ramirez-Ruiz}, E. and {Rest}, A. and {Rojas-Bravo}, C. and {Siebert}, M.~R. and {Smartt}, S.~J. and {Smith}, K.~W. and {Terreran}, G. and {Wang}, Q. and {Wojtak}, R. and {Agnello}, A. and {Ansari}, Z. and {Arendse}, N. and {Baldeschi}, A. and {Blanchard}, P.~K. and {Brethauer}, D. and {Bright}, J.~S. and {Brown}, J.~S. and {de Boer}, T.~J.~L. and {Dodd}, S.~A. and {Fairlamb}, J.~R. and {Grillo}, C. and {Hajela}, A. and {Hede}, C. and {Kolborg}, A.~N. and {Law-Smith}, J.~A.~P. and {Lin}, C. -C. and {Magnier}, E.~A. and {Malanchev}, K. and {Matthews}, D. and {Mockler}, B. and {Muthukrishna}, D. and {Pan}, Y. -C. and {Pfister}, H. and {Ramanah}, D.~K. and {Rest}, S. and {Sarangi}, A. and {Schr{\o}der}, S.~L. and {Stauffer}, C. and {Stroh}, M.~C. and {Taggart}, K.~L. and {Tinyanont}, S. and {Wainscoat}, R.~J. and {Young Supernova Experiment}},
        title = "{The Young Supernova Experiment: Survey Goals, Overview, and Operations}",
      journal = {\apj},
         year = 2021,
        month = feb,
       volume = {908},
       number = {2},
          eid = {143},
        pages = {143},
          doi = {10.3847/1538-4357/abd7f5},
archivePrefix = {arXiv},
       eprint = {2010.09724},
 primaryClass = {astro-ph.HE},
       adsurl = {https://ui.adsabs.harvard.edu/abs/2021ApJ...908..143J}
}

@ARTICLE{Jones19,
       author = {{Jones}, D.~O. and {Scolnic}, D.~M. and {Foley}, R.~J. and {Rest}, A. and {Kessler}, R. and {Challis}, P.~M. and {Chambers}, K.~C. and {Coulter}, D.~A. and {Dettman}, K.~G. and {Foley}, M.~M. and {Huber}, M.~E. and {Jha}, S.~W. and {Johnson}, E. and {Kilpatrick}, C.~D. and {Kirshner}, R.~P. and {Manuel}, J. and {Narayan}, G. and {Pan}, Y. -C. and {Riess}, A.~G. and {Schultz}, A.~S.~B. and {Siebert}, M.~R. and {Berger}, E. and {Chornock}, R. and {Flewelling}, H. and {Magnier}, E.~A. and {Smartt}, S.~J. and {Smith}, K.~W. and {Wainscoat}, R.~J. and {Waters}, C. and {Willman}, M.},
        title = "{The Foundation Supernova Survey: Measuring Cosmological Parameters with Supernovae from a Single Telescope}",
      journal = {\apj},
         year = 2019,
        month = aug,
       volume = {881},
       number = {1},
          eid = {19},
        pages = {19},
          doi = {10.3847/1538-4357/ab2bec},
archivePrefix = {arXiv},
       eprint = {1811.09286},
 primaryClass = {astro-ph.CO},
       adsurl = {https://ui.adsabs.harvard.edu/abs/2019ApJ...881...19J}
}

@ARTICLE{Holoien19,
       author = {{Holoien}, T.~W. -S. and {Brown}, J.~S. and {Vallely}, P.~J. and {Stanek}, K.~Z. and {Kochanek}, C.~S. and {Shappee}, B.~J. and {Prieto}, J.~L. and {Dong}, Subo and {Brimacombe}, J. and {Bishop}, D.~W. and {Bose}, S. and {Beacom}, J.~F. and {Bersier}, D. and {Chen}, Ping and {Chomiuk}, L. and {Falco}, E. and {Holmbo}, S. and {Jayasinghe}, T. and {Morrell}, N. and {Pojmanski}, G. and {Shields}, J.~V. and {Strader}, J. and {Stritzinger}, M.~D. and {Thompson}, Todd A. and {Wo{\'z}niak}, P.~R. and {Bock}, G. and {Cacella}, P. and {Carballo}, J.~G. and {Cruz}, I. and {Conseil}, E. and {Farfan}, R.~G. and {Fernandez}, J.~M. and {Kiyota}, S. and {Koff}, R.~A. and {Krannich}, G. and {Marples}, P. and {Masi}, G. and {Monard}, L.~A.~G. and {Mu{\~n}oz}, J.~A. and {Nicholls}, B. and {Post}, R.~S. and {Stone}, G. and {Trappett}, D.~L. and {Wiethoff}, W.~S.},
        title = "{The ASAS-SN bright supernova catalogue - IV. 2017}",
      journal = {\mnras},
         year = 2019,
        month = apr,
       volume = {484},
       number = {2},
        pages = {1899-1911},
          doi = {10.1093/mnras/stz073},
archivePrefix = {arXiv},
       eprint = {1811.08904},
 primaryClass = {astro-ph.HE},
       adsurl = {https://ui.adsabs.harvard.edu/abs/2019MNRAS.484.1899H}
}

@ARTICLE{Holoien17B,
       author = {{Holoien}, T.~W. -S. and {Brown}, J.~S. and {Stanek}, K.~Z. and {Kochanek}, C.~S. and {Shappee}, B.~J. and {Prieto}, J.~L. and {Dong}, Subo and {Brimacombe}, J. and {Bishop}, D.~W. and {Basu}, U. and {Beacom}, J.~F. and {Bersier}, D. and {Chen}, Ping and {Danilet}, A.~B. and {Falco}, E. and {Godoy-Rivera}, D. and {Goss}, N. and {Pojmanski}, G. and {Simonian}, G.~V. and {Skowron}, D.~M. and {Thompson}, Todd A. and {Wo{\'z}niak}, P.~R. and {{\'A}vila}, C.~G. and {Bock}, G. and {Carballo}, J. -L.~G. and {Conseil}, E. and {Contreras}, C. and {Cruz}, I. and {And{\'u}jar}, J.~M.~F. and {Guo}, Zhen and {Hsiao}, E.~Y. and {Kiyota}, S. and {Koff}, R.~A. and {Krannich}, G. and {Madore}, B.~F. and {Marples}, P. and {Masi}, G. and {Morrell}, N. and {Monard}, L.~A.~G. and {Munoz-Mateos}, J.~C. and {Nicholls}, B. and {Nicolas}, J. and {Wagner}, R.~M. and {Wiethoff}, W.~S.},
        title = "{The ASAS-SN bright supernova catalogue - II. 2015}",
      journal = {\mnras},
         year = 2017,
        month = may,
       volume = {467},
       number = {1},
        pages = {1098-1111},
          doi = {10.1093/mnras/stx057},
archivePrefix = {arXiv},
       eprint = {1610.03061},
 primaryClass = {astro-ph.HE},
       adsurl = {https://ui.adsabs.harvard.edu/abs/2017MNRAS.467.1098H}
}

@ARTICLE{Johnson21,
    author = {{Johnson}, Benjamin D. and {Leja}, Joel and {Conroy}, Charlie and {Speagle}, Joshua S.},
        title = "{Stellar Population Inference with Prospector}",
    journal = {\apjs},
        year = 2021,
        month = jun,
    volume = {254},
    number = {2},
        eid = {22},
        pages = {22},
        doi = {10.3847/1538-4365/abef67},
archivePrefix = {arXiv},
    eprint = {2012.01426},
primaryClass = {astro-ph.GA},
    adsurl = {https://ui.adsabs.harvard.edu/abs/2021ApJS..254...22J}
}
\bibliographystyle{mnras}

\end{document}